\documentclass[twocolumn]{aastex61}
\pdfoutput=1

\usepackage{epsf}
\usepackage{graphicx}
\usepackage{natbib}
\usepackage{url}
\usepackage{mathrsfs}
\usepackage{amsmath}
\font\smcap=cmcsc10

\newcommand{\kms}{\,km~s$^{-1}$}
\newcommand{\degree}{$^\circ$}
\newcommand{\nai}{Na\,{\smcap i}}
\newcommand{\caii}{Ca\,{\smcap ii}}

\newcommand{\vio}{$(V-I)_0$}
\newcommand{\ivi}{($I,\,V-I$)}

\newcommand{\feh}{$\rm[Fe/H]$}

\newcommand{\olkhd}{$\langle L_i\rangle$}

\newcommand{\rproj}{$R_{\rm proj}$}

\newcommand\Tstrut{\rule{0pt}{2.6ex}}         

\submitjournal{ApJ}

\shorttitle{Velocity Dispersion of M31's Stellar Halo}
\shortauthors{Gilbert et~al.}

\begin{document}
\bibliographystyle{aasjournal}

\title{Global Properties of M31's Stellar Halo from the SPLASH Survey: III. Measuring the Stellar Velocity Dispersion Profile\footnote{The data presented herein were obtained at the W.M. Keck Observatory,
which is operated as a scientific partnership among the California
 Institute of Technology, the University of California and the National
Aeronautics and Space Administration. The Observatory was made
possible by the generous financial support of the W.M. Keck
Foundation.}}

\correspondingauthor{Karoline M. Gilbert}
\email{kgilbert@stsci.edu}

\author[0000-0003-0394-8377]{Karoline M. Gilbert}
\affiliation{Space Telescope Science Institute, 3700 San Martin Dr., Baltimore, MD 21218, USA} 
\affiliation{Department of Physics \& Astronomy, Bloomberg Center for Physics and Astronomy, Johns Hopkins University, 3400 N. Charles Street, Baltimore, MD 21218}

\author{Erik Tollerud}
\altaffiliation{Giacconi Fellow}
\affiliation{Space Telescope Science Institute, 3700 San Martin Dr., Baltimore, MD 21218, USA}

\author[0000-0002-1691-8217]{Rachael L. Beaton}
\altaffiliation{Hubble Fellow}
\altaffiliation{Carnegie-Princeton Fellow}
\affiliation{Department of Astrophysical Sciences, Princeton University, 4 Ivy Lane, Princeton, NJ~08544}
\affiliation{The Observatories of the Carnegie Institution for Science, 813 Santa Barbara St., Pasadena, CA ~~91101}

\author{Puragra Guhathakurta}
\affiliation{UCO/Lick Observatory, Department of Astronomy \& Astrophysics, University of California Santa Cruz, 
 1156 High Street, 
 Santa Cruz, California 95064, USA}

\author{James S. Bullock}
\affiliation{Center for Cosmology, Department of Physics and Astronomy, University of California at Irvine, Irvine, CA, 92697, USA}

\author{Masashi Chiba}
\affiliation{Astronomical Institute, Tohoku University, Aoba-ku, Sendai 980-8578, Japan}

\author{Jason S. Kalirai}
\affiliation{Space Telescope Science Institute, 3700 San Martin Dr., Baltimore, MD 21218, USA}

\author{Evan N. Kirby}
\affiliation{California Institute of Technology, 1200 E. California Boulevard, MC 249-17, Pasadena, CA 91125, USA}

\author{Steven R. Majewski}
\affiliation{Department of Astronomy, University of Virginia,
PO Box 400325,
Charlottesville, VA 22904-4325, USA}

\author{Mikito Tanaka}
\affiliation{Astronomical Institute, Tohoku University, Aoba-ku, Sendai 980-8578, Japan}

\setcounter{footnote}{11}

\begin{abstract}

We present the velocity dispersion of red giant branch (RGB) stars in M31's halo, derived by modeling the line of sight velocity distribution of over 5000 stars in 50 fields spread throughout M31's stellar halo.  The dataset was obtained as part of the SPLASH (Spectroscopic and Photometric Landscape of Andromeda's Stellar Halo) Survey, and covers projected radii of 9 to 175~kpc from M31's center.  All major structural components along the line of sight in both the Milky Way (MW) and M31 are incorporated in a Gaussian Mixture Model, including all previously identified M31 tidal debris features in the observed fields.  The probability an individual star is a constituent of M31 or the MW, based on a set of empirical photometric and spectroscopic diagnostics, is included as a prior probability in the mixture model. The velocity dispersion of stars in M31's halo is found to decrease only mildly with projected radius, from 108~\kms\ in the innermost radial bin (8.2 to 14.1~kpc) to $\sim 80$ to 90~\kms\ at projected radii of $\sim40$\,--\,130~kpc, and can be parameterized with a power-law of slope $-0.12\pm0.05$.   The quoted uncertainty on the power-law slope reflects only the precision of the method, although other sources of uncertainty we consider contribute negligibly to the overall error budget.  

\end{abstract}


\setcounter{footnote}{0}

\section{Introduction}\label{sec:intro}

The orbits of satellite dwarf galaxies, globular clusters, and halo stars trace the mass distribution of their host system to large radii and can be used to estimate the total mass of the host system.  The three-dimensional space motions of these tracers is currently out of reach for all but the Milky Way (MW) and its closest satellite galaxies.  Thus, in more distant systems the line of sight velocity distribution of these tracers is the key observable and can be used to estimate the enclosed mass
via estimators of varying complexity \citep[e.g.,][]{cappellari2006, watkins2010, wolf2010, gnedin2010, amorisco2012}.

The proximity of the Andromeda galaxy (M31) allows measurement of the line of sight velocity distributions
of all three of these tracers.  Estimates of M31's mass have previously been made based on the 
velocity distributions of M31's dwarf satellite galaxies \citep{watkins2010, tollerud2012} and globular clusters \citep{veljanoski2014}.  
The mass estimate based on M31's dwarf satellites is sensitive to which satellites are included in the
measurement: satellites that are not on virialized orbits (i.e., that are on their first infall or are not gravitationally bound)
can significantly skew the mass estimate.  Meanwhile, the globular cluster population shows significant rotation \citep{veljanoski2013}, which must be modeled, and also has a significant fraction of clusters that are statistically likely to be associated with halo substructure \citep{mackey2010,veljanoski2014}.  

M31's halo stars provide a dense network of mass tracers that can provide an independent mass estimate.  The density of halo stars allows the kinematics of distinct tidal debris features to be accounted for directly in the modeling of the velocity distribution of the underlying halo population.  A mass estimate based on the velocity distribution of halo stars can thus provide an important comparison to mass estimates from the satellite or globular cluster system.  Currently, M31 is the only external galaxy for which we can compare the velocity distributions of all three tracer systems to large radii.  This will provide an important calibration for interpreting mass estimates in more distant galaxies that are based on the galaxy's globular cluster or dwarf satellite system.  

The kinematics of M31 halo stars, especially when combined with chemical abundance measurements, can also be used to investigate the the inner regions of M31's stellar halo. In the inner regions, we expect to find a mix of {\it in situ} and accreted halo stars based on theoretical grounds \citep[e.g.,][]{abadi2006,zolotov2009,purcell2010,cooper2010,font2011}.  Observations of the inner regions of the MW have recently ignited a vigorous debate regarding the evidence for multiple halo populations \citep[e.g.,][]{majewski1992,carollo2010,schonrich2011,beers2012,schonrich2014}.  M31 provides the only external galaxy for which we can currently place observational constraints on the presence of multiple formation avenues for the stellar halo.

Large spectroscopic surveys in M31 have provided line of sight velocity measurements
for tens of thousands of stars in M31.
The majority of the analysis of the velocity measurements has been focused
on characterizing substructure \citep{ibata2004,guhathakurta2006,kalirai2006gss,gilbert2007,chapman2008,gilbert2009gss,fardal2012,mackey2014, kafle2017},
measuring the internal velocity dispersion of M31 satellite galaxies \citep{chapman2005,geha2006,letarte2009, kalirai2009,kalirai2010,geha2010, collins2010,collins2011,ho2012, tollerud2012, tollerud2013, howley2013, collins2013,collins2014,martin2014},
and measuring the dynamics of the disk \citep{ibata2005,collins2011disk, dorman2013,dorman2015}.

Existing measurements of the velocity dispersion of M31's stellar halo have been limited to
the inner regions of the halo and have primarily been made in spectroscopic fields that are dominated by
M31 disk stars.  \citet{chapman2006} used spectra obtained in fields throughout M31's disk, at projected radii of 9 to 70~kpc, to measure
the velocity dispersion of M31's halo as a function of projected radius.
To avoid contamination from M31's disk, substructure in M31, 
and MW stars, \citeauthor{chapman2006} employed window functions in velocity and 
a constrained maximum-likelihood analysis, fixing the mean velocity of
M31's stellar halo and iteratively rejecting spectroscopic fields.  
The measured gradient in the velocity dispersion 
was strongly dominated by spectroscopic fields within 40~kpc
of M31's center, and implies a central velocity dispersion of 152~\kms, decreasing at a rate of $-0.9$~\kms\ kpc$^{-1}$.  \cite{gilbert2007} modeled the velocity distribution of spectroscopically
confirmed M31 stars along the minor axis from 9 to 30~kpc, performing a maximum-likelihood analysis assuming both an M31 halo and substructure component.   \citeauthor{gilbert2007} measured a velocity dispersion 
of 129~\kms\ for the halo over this radial range.    Most recently, \citet{dorman2012} analyzed spectra in large contiguous regions in M31's inner disk ($R \lesssim 20$~kpc).  \citeauthor{dorman2012}
performed a Markov chain Monte Carlo (MCMC) analysis of spectroscopically confirmed M31 red giant branch stars, simultaneously fitting
for the disk, substructure, and halo components.  \citeauthor{dorman2012} detected significant spheroid rotation in the inner regions, and found a decrease in the line of sight velocity dispersion of M31's halo, from $\sim 140$~\kms\ at 7~kpc to $120$~\kms at 14~kpc on the major axis.  In their regions of radial overlap, the existing measurements are consistent with one another
at the $\lesssim 2\sigma$ level.

The SPLASH \citep[Spectroscopic and Photometric Landscape of Andromeda's Stellar Halo;][]{guhathakurta2005,gilbert2006} survey has 
amassed an archive of tens of thousands of spectra in lines of sight throughout 
M31's halo, disk, and dwarf galaxies \citep[e.g.,][]{kalirai2010, tollerud2012, gilbert2012, gilbert2014, dorman2012, dorman2015}. 
A large portion of the SPLASH spectroscopic fields are in the outer halo of M31, far removed from 
M31's disk.  In the first two papers of this series, we used counts of spectroscopically confirmed M31 RGB stars in 38 fields to 
measure an $\sim r^{-2}$ surface brightness profile for M31's stellar halo out to 175~kpc 
\citep{gilbert2012} and confirmed the existence of a large scale metallicity gradient in M31's halo
from 10 to 100~kpc \citep{gilbert2014}.  

In this contribution, we model the line-of-sight velocity distributions 
of more than 5000 stars in 50 fields spread throughout M31's stellar halo, with projected 
radii from M31's center of 9 to 175~kpc.  None of the fields are located on M31's disk.
We do not employ any windowing functions
on the line of sight velocity, nor do we make any cuts to the spectroscopic sample based on likely
membership in the M31 or MW populations, as has been done in previous work.  
Rather, we employ MCMC methods to sample the posterior distribution functions for the full set of model 
parameters needed to describe the kinematical components that have
been previously identified within the SPLASH survey.  

This paper is organized as follows.  Section~\ref{sec:data} provides
a brief overview of the spectroscopic dataset.  
Section~\ref{sec:halo_sample} describes the multiple stellar populations
present in the spectroscopic fields and the empirical diagnostics
available for assigning probabilities of membership.  It also explains the
motivation underlying our choice to include all observed MW and M31 halo stars
in our fitting of the velocity distribution.  Section~\ref{sec:mcmc} describes the MCMC analysis,
including the formulation of the likelihood function and the choice of priors.
Section~\ref{sec:mcmc_results} describes the results of the MCMC analysis of the 
line-of-sight velocity distributions for the primary model parameters of interest, including
the parameterization of the velocity dispersion of M31's stellar halo with radius (Section~\ref{sec:powerlaw_dispersion}).
Appendices present marginalized one- and two-dimensional 
posterior probability distributions 
for all model parameters, 
including those not discussed in Section~\ref{sec:mcmc_results}.

All radii from M31's center refer to the projected distance 
in the sky tangent plane.
For consistency with previous papers in this series, 
and to allow direct comparisons with other results,
a distance modulus of 24.47$^m$ is 
assumed for conversions of angular to physical units, which 
corresponds to a distance to M31 of 783~kpc \citep{stanek1998,mcconnachie2005}.  
However, we note that this distance is slightly greater than the recent, smaller distance modulus estimate of 24.38$^m$ based on measurements of Cepheid variables \citep{riess2012}.
Where relevant, we adopt the Planck Collaboration 2015 cosmology \citep{planck2016cosmology}.
 
\section{The SPLASH Data Set}\label{sec:data}
The majority of the SPLASH data ($\sim 90$\% of the spectroscopic masks)
have been presented in earlier contributions. The new spectroscopic data have been 
obtained with an identical observational setup and reduced using the same data reduction pipelines.
The spectroscopic masks were designed using imaging data that were obtained
and reduced in the same imaging campaigns as the fields used to design earlier spectroscopic masks.  
Thus, we only briefly summarize the spectroscopic data and its reduction below.  
Readers are referred to \citet{gilbert2012,gilbert2014}
for details of the photometric and spectroscopic data reduction. 

\subsection{Field Locations}\label{sec:fields}
\begin{figure*}[tb!]
\plotone{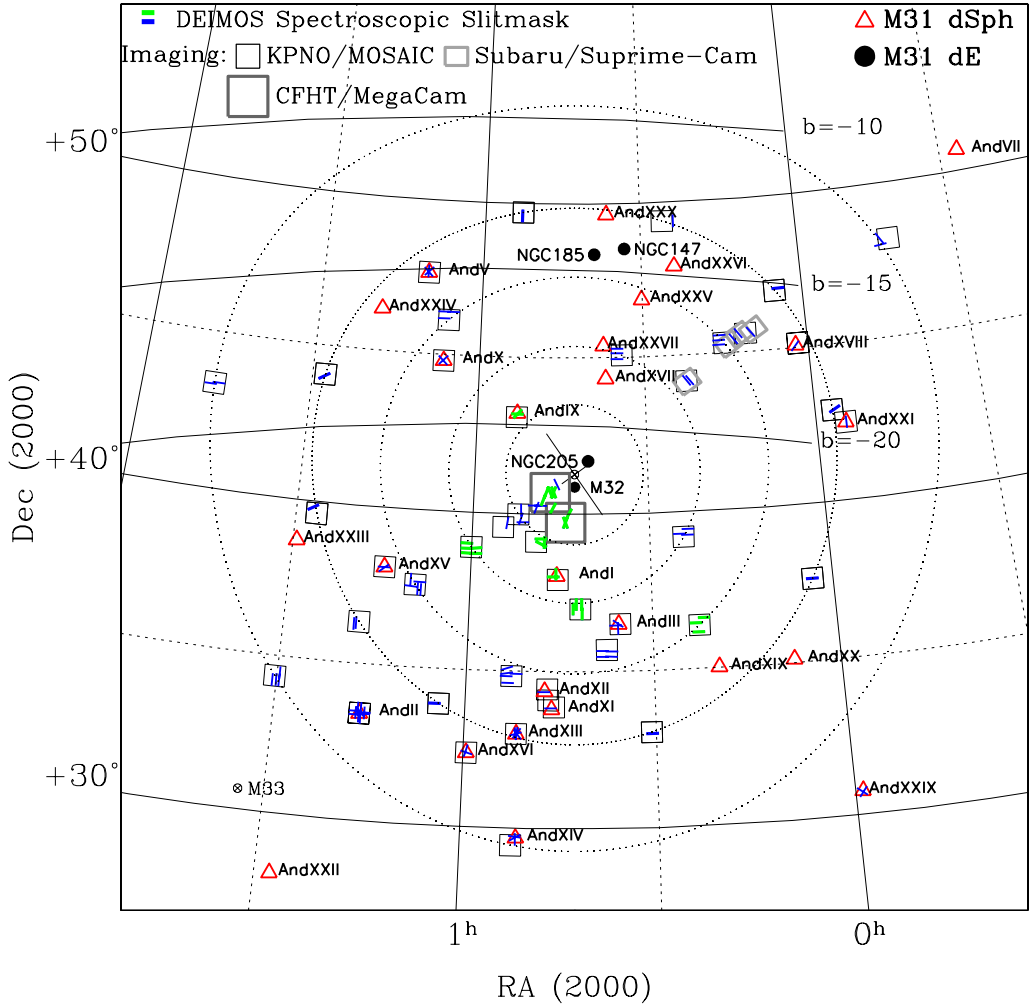}
\caption{
Locations of the spectroscopic fields.  The location and orientation of each spectroscopic mask is denoted with a small rectangle; green rectangles denote spectroscopic masks with kinematically identified substructure.  Larger rectangles denote the location and extent of the KPNO/Mosaic (black), CFHT/MegaCam (dark grey) and Subaru/Suprime-Cam (light grey) images used to design the masks.  The location of the dwarf elliptical (black circles) and dwarf spheroidal (open triangles) satellites of M31 are also shown.  M31's center is marked by an open circle, and the orientations of M31's major and minor axes are illustrated with the long and short solid lines.  The dotted circles have radii of 2, 4, 6, 8 and 11 degrees from M31's center.     
}
\label{fig:roadmap1}
\end{figure*}

\begin{figure*}[tb!]
\plotone{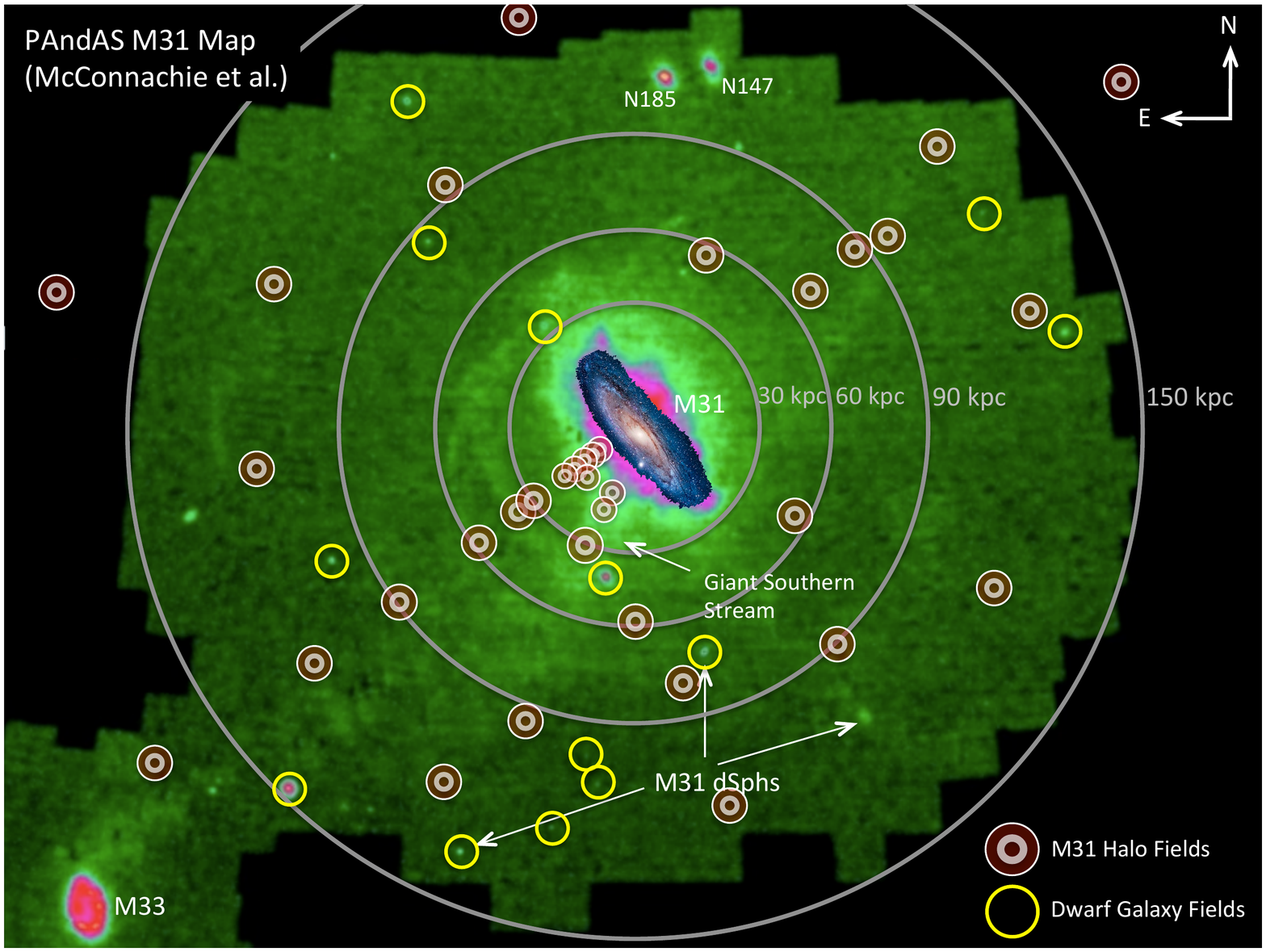}
\caption{
Approximate locations of the spectroscopic fields overlaid on the PAndAS starcount map \citep{mcconnachie2009}.  Spectroscopic fields designed primarily to target M31's dwarf galaxies are denoted by open yellow circles. The remaining fields target M31's halo, and include fields on several large tidal debris features.  Two dwarf spheroidal fields, And\,XIV and And\,XXIX, are not shown; they are to the south of M31, beyond the bounds of the figure. 
}
\label{fig:roadmap2}
\end{figure*}

The stellar spectra were obtained with 124 multi-object 
spectroscopic slitmasks in 50 fields 
spread throughout M31's stellar halo (Figure~\ref{fig:roadmap1}).
The masks span a large range in azimuth and projected radii from
the center of M31. The fields were chosen to
target relatively smooth areas of M31's halo, individual tidal debris
features, and dwarf satellites (Figure~\ref{fig:roadmap2}).

\subsection{Target Selection}\label{sec:spec_target_selection}

The majority of the multi-object spectroscopic slitmasks were designed from images in the 
Washington system $M$ and $T_2$ filters and the intermediate-width DDO51 filter, 
obtained with the Mosaic camera on the 4-m 
Mayall telescope at Kitt Peak National Observatory (KPNO).\footnote{Kitt Peak National
Observatory of the National Optical Astronomy Observatory is operated by the
Association of Universities for Research in Astronomy, Inc., under cooperative
agreement with the National Science Foundation.}  
\citep{ostheimer2003,beaton2014}.
The innermost spectroscopic slitmasks were designed using $g'$ and $i'$ band 
imaging obtained with MegaCam instrument on the 3.6-m Canada-France-Hawaii 
Telescope (CFHT)\footnote{MegaPrime/MegaCam
is a joint project of CFHT and CEA/DAPNIA,
at the Canada-France-Hawaii Telescope
which is operated by the National Research Council of Canada, the Institut
National des Science de l'Univers of the Centre National de la Recherche
Scientifique of France, and the University of Hawaii.} \citep{kalirai2006gss}.
A small number of spectroscopic masks in the outer halo of M31 were
designed from Johnson-Cousins $V$ and $I$ band imaging obtained with
the Suprime-Cam instrument on the Subaru Telescope 
\citep[fields `streamE' and `streamF';][]{tanaka2010}
and the William Herschel Telescope \citep[field `and10';][]{zucker2007}.

Stars were prioritized for inclusion on the spectroscopic masks based on their
colors and magnitudes. 
Stars with colors and magnitudes consistent with 
RGB stars at the distance of M31 were assigned high priority for inclusion 
on the spectroscopic masks, with brighter RGB stars
(within $\sim 1$ to 1.5~magnitudes of the tip of the red giant branch) given highest
priority.   When available, the surface-gravity sensitive 
intermediate-band DDO51 photometry was also used to prioritize stars with 
a high probability of being RGB stars \citep[based on their location in the 
$M-T_2$, $M-$DDO51 color-color diagram;][]{majewski2000}.

\subsection{Spectroscopic Observations}\label{sec:spec_observations}
The spectra were obtained with the DEIMOS spectrograph on the Keck~II 
10-m telescope over ten observing seasons (Fall 2002\,--\,2011).  
All of the spectra were obtained using the 1200~line~mm$^{-1}$ 
grating, which produces a dispersion of $\rm0.33~\AA$~pix$^{-1}$.  The 
survey used a slit width of 1'', which yields a resolution of 1.6\AA\ FWHM.
The wavelength range of the observed spectra is
$\lambda\lambda\sim$~6450\,--\,9150\AA, which includes the \caii\ triplet 
absorption feature ($\sim 8500$\AA) and the \nai\ absorption feature
(8190\AA).  
The Keck/DEIMOS spectra were reduced using  
the {\tt spec2d} (flat-fielding, night-sky emission line removal, and extraction
of one-dimensional spectra) and {\tt spec1d} (redshift measurement) software developed
at the University of California, Berkeley \citep{cooper2012,newman2013}.

Only stellar spectra with secure velocity measurements are 
included in the final data set \citep{gilbert2006}. 
A heliocentric correction is applied to the measured line of sight velocities
as well as a correction for imperfect centering of the star within the slit  
\citep[using the observed position of the atmospheric $A$-band absorption feature
relative to night sky emission lines;][]{simon2007,sohn2007}.

The mean (median) S/N per Angstrom of the stellar spectra with 
secure velocity measurements is 11.7 (8.2). The mean (median) S/N 
per Angstrom of the spectra of stars that are more probable to 
be M31 stars than MW stars (Section~\ref{sec:likelihoods}) is 7.8 (4.4).  
The mean (median) velocity measurement uncertainty of all stellar spectra with secure velocity 
measurements is 5.7~\kms\ (4.8~\kms), while the mean (median) velocity uncertainty of stars more
probable to be M31 stars than MW stars is 6.5~\kms\ (5.6~\kms).
The uncertainties on the velocity measurements are calculated by adding in 
quadrature the random velocity measurement uncertainty, estimated from the 
cross-correlation routine, and a systematic uncertainty of 2.2~\kms, estimated
from repeat observations of stars \citep{simon2007}. 

\section{Separation of Stellar Populations}\label{sec:halo_sample}
The final dataset of over 6600 stellar spectra 
is drawn from multiple stellar populations in M31 
and along the line of sight to M31, including 
distinct tidal debris features within M31's halo, dwarf satellites
of M31, the Milky Way disk and halo, and 
finally the relatively smooth, underlying M31 
halo whose velocity distribution we aim to measure. 
These populations all have some amount
of overlap in line-of-sight velocity space.  
However, the photometry and stellar spectra provide additional 
discriminating power beyond the line of sight velocities
for separating these populations.  
We briefly describe below the methods used to assign membership,
or probabilities of membership, among the various populations.
Each of these methods have been utilized in earlier publications, to which
readers are referred for greater detail.

\subsection{Removal of Dwarf Galaxy Members}\label{sec:dSph_members}

Almost one-third of the fields that are farther than 4\degree\ from 
M31's center 
targeted dwarf satellite galaxies 
\citep[Figure~\ref{fig:roadmap1};][]{majewski2007,kalirai2009,kalirai2010,tollerud2012}.
In these fields, a significant number of the stars observed 
on each spectroscopic mask are dSph members, rather than M31 halo 
stars or MW stars along the line of sight.

Stars that are likely to be gravitationally bound to a dwarf satellite 
galaxy are identified following the method outlined by \citet{gilbert2009gss}.
M31's dSphs are spatially compact, have small velocity dispersions, and span a limited
range of [Fe/H].  The spatial extents of the majority of the M31 dSphs are 
small enough that they cover only a portion of a DEIMOS spectroscopic slitmask.
Only stars within the King limiting radius are considered potential dSph members.
Stars 
that are outside the King limiting 
radius 
are included in our final dataset.  Thus we explicitly classify any extra-tidal dSph 
stars as M31 halo stars.  This number is small: only $\sim 5$\% of the stars beyond the
King limiting radius in dSph fields have velocities and \feh\ values consistent with
the dSph.  In addition, any stars within the King limiting radius of the dSph but well removed
from the distribution of dSph stars in line of sight velocity or \feh\ are 
classified as M31 halo stars.  The interested reader can find examples for And~I and
And~III in Figures 3 and 4 of \citet{gilbert2009gss}.

After removing stars classified as dSph members, the final dataset 
contains 5299 stars.

\subsection{Likelihood of M31 or MW Membership}\label{sec:likelihoods}
We use the method established by \citet{gilbert2006} to determine the 
likelihood that an individual star is a red giant branch star in
M31 or a foreground MW dwarf star along the line of sight to M31.  The
\citeauthor{gilbert2006}\ method determines the probability 
a star is an M31 red giant branch or MW dwarf star from multiple photometric-
and spectroscopic-based diagnostics.  The full set of diagnostics includes 
(1) line-of-sight velocity, (2) location in the ($M-T_2$, $M-$DDO51) color-color
diagram (when available), (3) strength of the \nai\ doublet absorption line as a function of
($V-I$) color\footnote{The photometrically calibrated photometry was transformed to Johnson-Cousins
$V$ and $I$ band magnitudes, using the transformation equations of
\citet{majewski2000} for the $M$ and $T_2$ magnitudes derived from the KPNO/Mosaic
imaging, and using observations of Landolt photometric standard stars for the 
$g'$ and $i'$ band magnitudes derived from the CFHT/MegaCam imaging. }, (4) location in the ($I$, $V-I$) color-magnitude diagram, 
and (5) comparison of spectroscopic and photometric metallicity estimates.  
Each diagnostic provides a (log-)likelihood that the star is an RGB or 
dwarf star: $L = {\rm log}_{10}(P_{\rm RGB}/P_{\rm dwarf}$). 
The overall likelihood, \olkhd, that a star is an M31 red giant or MW dwarf is defined 
as the sum of the individual log-likelihoods for each diagnostic (i.e., the product of the likelihoods). 

An additional factor not included in the \citeauthor{gilbert2006} method 
is projected radius from M31's center: the relative 
stellar density of M31 and MW populations changes dramatically as a function of M31-centric radius.  
For a given set of values of the stellar properties included in \olkhd, a star is in fact more likely to be an M31 star in the inner regions compared to the outer regions.  This is described in more detail in Section~\ref{sec:mcmc_lklhd_func_membershipprior}, where we explicitly include this in the model.

\begin{figure*}[tb!]
\plottwo{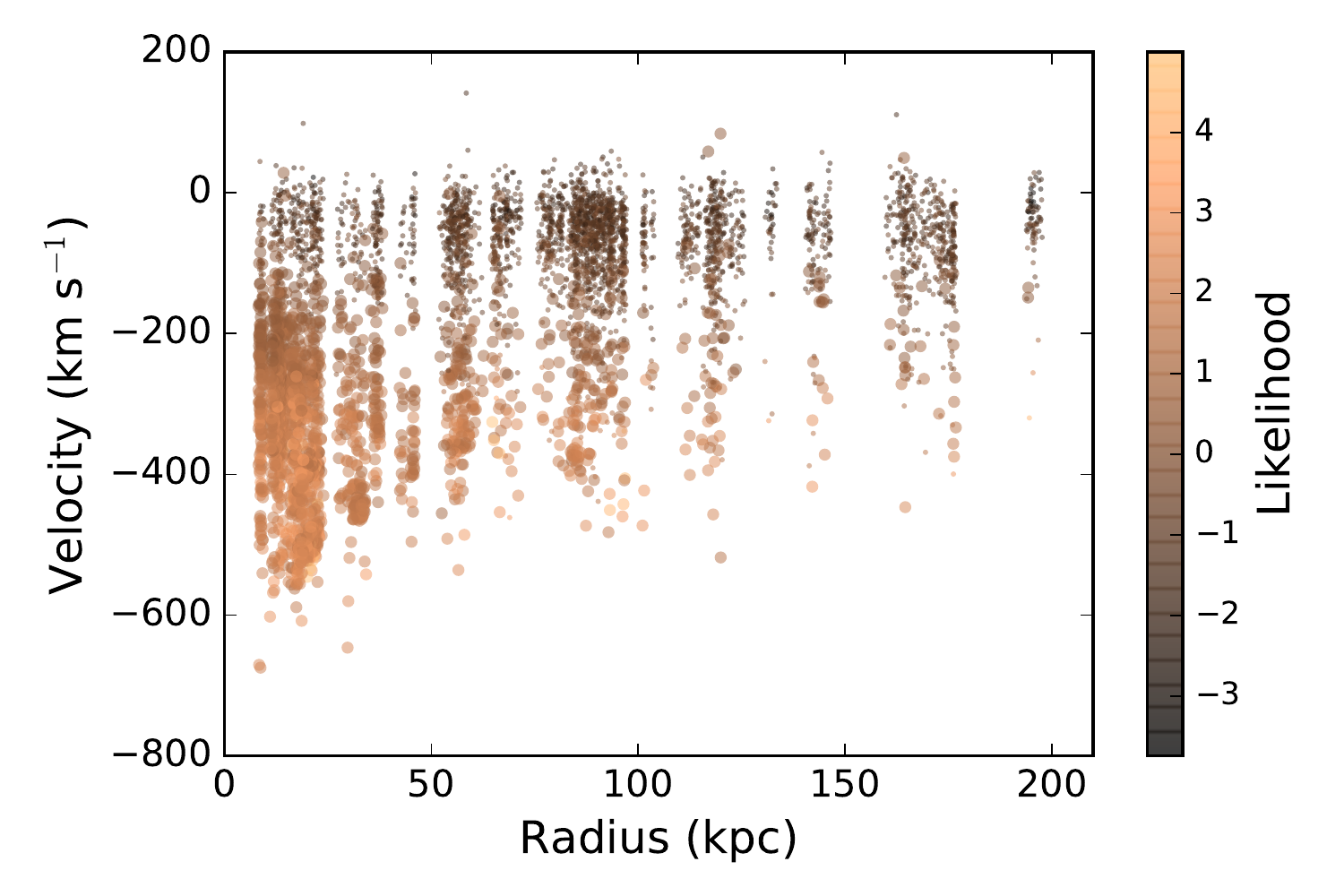}{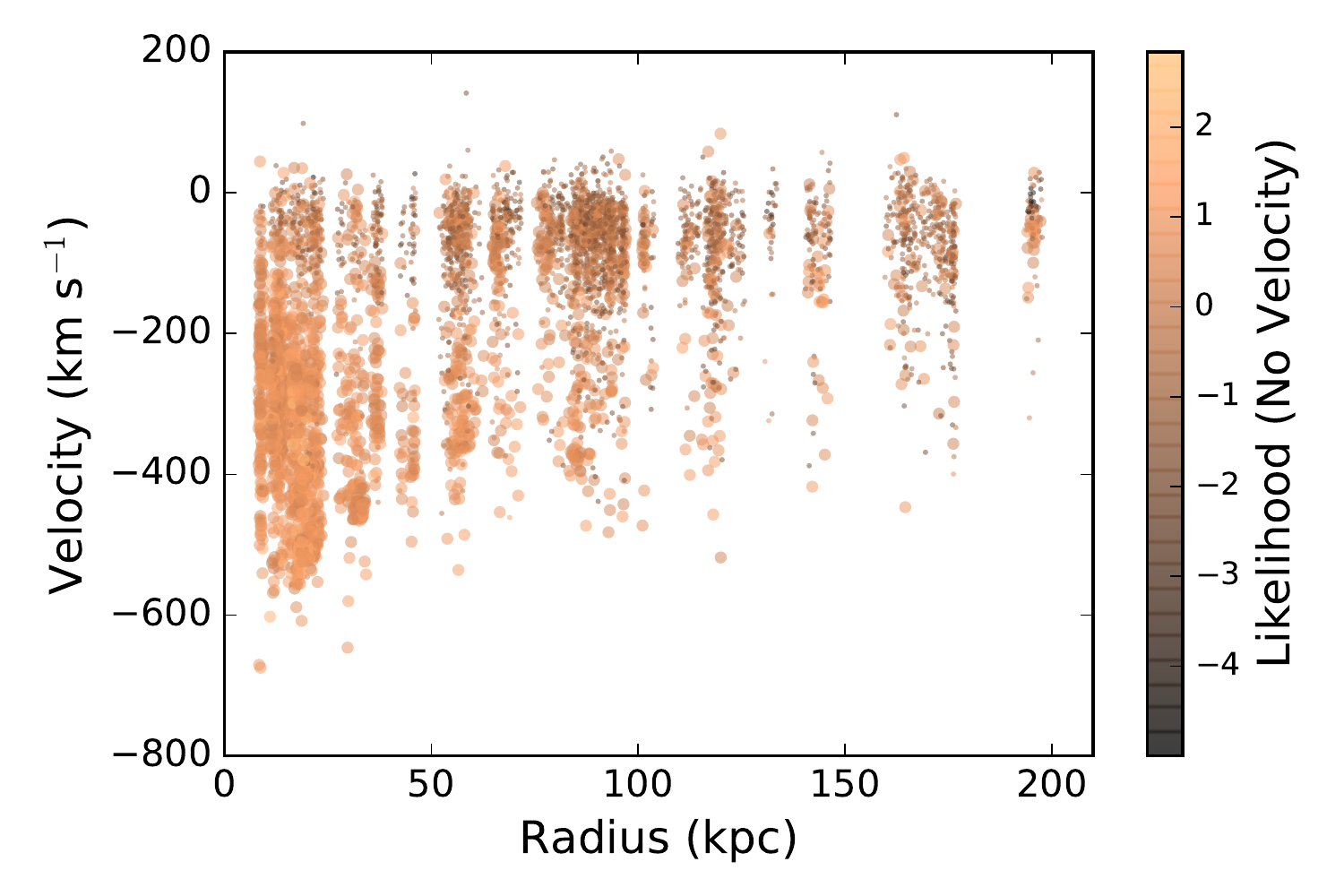}
\caption{
Line of sight, heliocentric velocity of every star in the sample as a function of projected radius, 
color-coded by the likelihood of MW or M31 membership, \olkhd, including ({\it left}) and 
excluding ({\it right}) the velocity diagnostic. Stars classified by their \olkhd\ values as
as M31 stars (\olkhd$>0$) are shown as large points, while stars classified as MW stars are shown
as small points (\olkhd$\le0$ or very blue stars; Section~\ref{sec:likelihoods}).  M31's systemic line of sight velocity is $\sim -300$~\kms.  
While including velocity in the \olkhd\ calculation results in a cleaner sample of M31 stars, it also introduces a strong bias in the velocity distribution of any sample selected based on an \olkhd\ threshold.  
}
\label{fig:data_vel_vs_rad_likelihood}
\end{figure*}

Since our aim is to model the stellar velocities, 
we do {\it not} include the velocity diagnostic in the computation of \olkhd.  Figure~\ref{fig:data_vel_vs_rad_likelihood} shows the velocity of stars as a function of projected radius from M31, 
color-coded by \olkhd\ values computed with (left) and without (right) the velocity diagnostic.
When velocity is included in the \olkhd\ computation, there is a strong trend of 
increasing \olkhd\ with decreasing line-of-sight velocity, which
is a direct result of using the velocity diagnostic in the \olkhd\ calculation.  
Although weaker, a correlation of \olkhd\ with line of sight velocity is still evident when velocity is not included
in the \olkhd\ computation, reflecting the velocity distributions of the M31 and MW populations.

In previous papers, we defined samples of M31 and MW stars based
on \olkhd\ thresholds, and stars significantly bluer than the most metal-poor, 10~Gyr
RGB isochrone have been classified as securely identified MW stars regardless 
of their \olkhd\ values \citep{gilbert2006}.  
This acknowledged that many of the empirical diagnostics have little 
discriminating power for stars with blue colors, as well as the fact that these stars
are much more likely to be MW stars than RGB stars in M31.  In this paper we use the
\olkhd\ values directly, rather than using subsets of stars classified as belonging to MW or M31.  
Thus we must choose how to treat stars bluer than the most metal-poor RGB isochrone 
in the following analysis.  
Rather than removing them from the sample altogether, we set their \olkhd\ values to $-5$.  This places these blue stars in the tail of the \olkhd\ distribution of MW stars, acknowledging that it is very unlikely for them to be M31 RGB stars. This is equivalent to what has been done in our previous papers.  
The sensitivity of the final results to this choice is explored in Section~\ref{sec:results_assumptions}.

\subsection{Probability of Belonging to Tidal Debris Features}\label{sec:kccs}

Kinematically cold tidal debris features have been identified in 
a significant fraction of the spectroscopic masks 
(Figures~\ref{fig:roadmap1} and \ref{fig:roadmap2}).
Many of the tidal debris features identified in 
individual fields are related to a single accretion event:
the Giant Southern Stream and its associated shell features
\citep[e.g.,][]{ibata2001a,fardal2007,gilbert2007,gilbert2009gss}.

Each of the tidal debris features in the dataset have been 
identified and characterized via maximum-likelihood, multi-Gaussian 
fits to the velocity distribution of M31 stars in the field,
and presented in previous papers \citep{guhathakurta2006,kalirai2006gss,
gilbert2007,gilbert2009gss,gilbert2012}.  Details of the fitting technique can be found
in the papers by \citet{gilbert2007,gilbert2012}.  Each maximum-likelihood
fit includes a Gaussian with a large velocity dispersion, representing the
underlying, kinematically hot stellar halo, and additional Gaussian 
components with small velocity dispersions, representing the kinematically
cold stellar streams.  

\begin{figure}[tb!]
\plotone{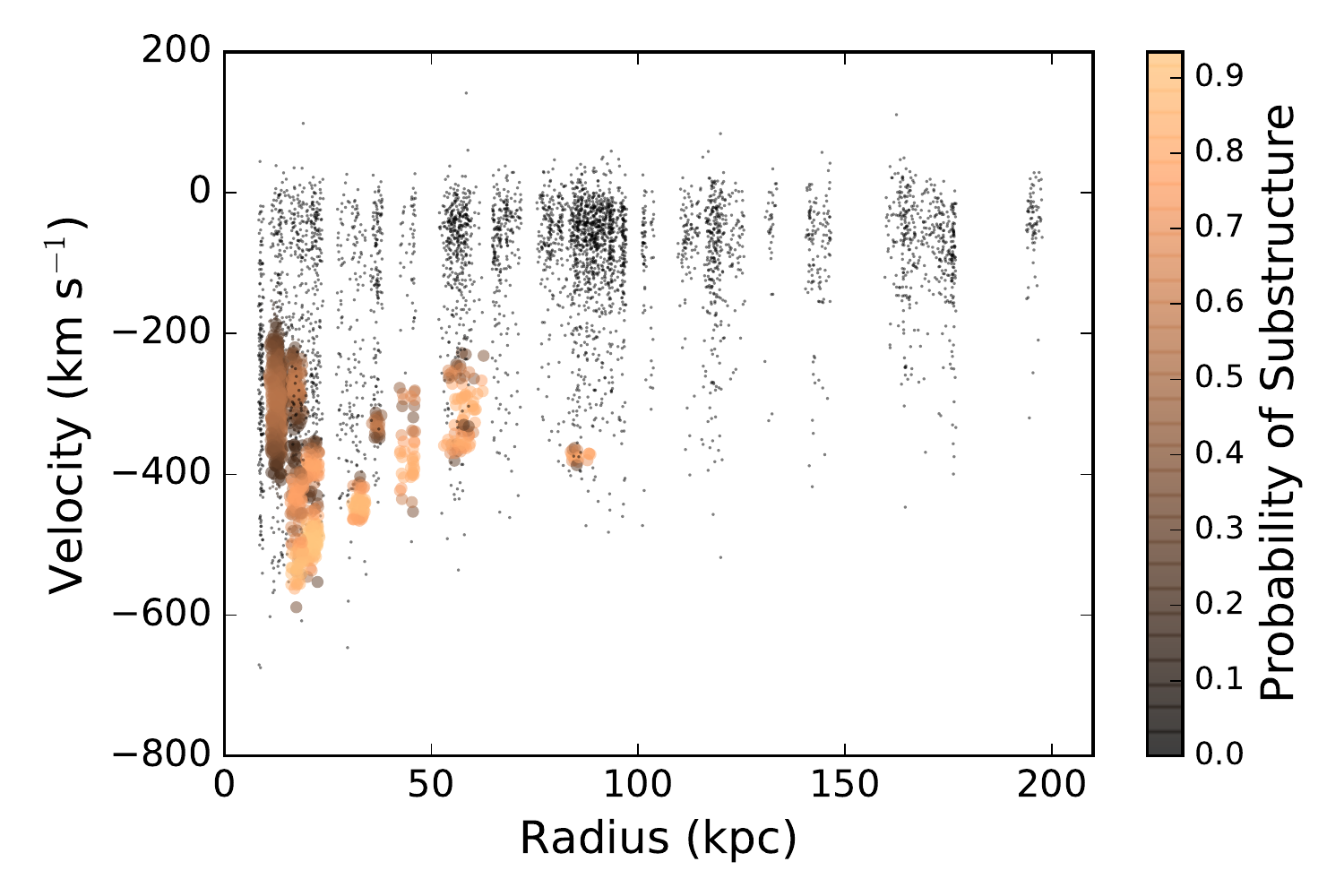}
\caption{
Line of sight, heliocentric velocity of every star in the sample as a function of projected radius, 
color-coded by the probability the star belongs to substructure.
The probabilities were computed using previously published maximum-likelihood fits to the velocity 
distribution of M31 stars (Section~\ref{sec:kccs}).
Stars with a probability greater than
0.2 of belonging to substructure are shown as large points.  The majority of identified tidal debris features 
in our dataset have velocities more negative than M31's systemic velocity ($v_{\rm sys} = -300$~\kms). 
}
\label{fig:data_vel_vs_rad_subst}
\end{figure}

The velocity distribution of the M31 halo was held fixed in all the 
published fits, with a mean heliocentric velocity of $\langle v_{\rm hel}\rangle =-300$~\kms\ (M31's systemic velocity)
and velocity dispersion of $\sigma_{\rm v} = 129$~\kms\ \citep{gilbert2007}.
The maximum-likelihood fits provide an estimate of the mean velocity, 
velocity dispersion, and fraction of M31 halo stars 
in each of the kinematically cold tidal debris features in a field.  Using the 
maximum-likelihood fits, the probability that any individual M31 star belongs to
the kinematically hot halo or a kinematically cold tidal debris feature can be
computed.  Figure~\ref{fig:data_vel_vs_rad_subst} shows the 
line of sight velocity as a function of projected radius of all M31 and MW stars in the  
dataset, color-coded by the probability the star belongs to tidal debris.
 
\subsection{Challenges of Selecting an M31 Halo Sample}\label{sec:sample_sel}

In previous studies, we successfully identified M31 and MW samples 
using cuts on the various membership criteria.  This work has included analyses of substructure 
\citep{gilbert2007,gilbert2009gss,gilbert2009a},
the surface brightness profile of M31's halo \citep{gilbert2012}, and the
metallicity profile of M31's halo \citep{kalirai2006halo,gilbert2014}. 
Depending on the analysis, the sample selection was tuned
to favor either a complete sample with some amount of contamination, or a 
clean but incomplete sample with minimal contamination.  To produce the cleanest
sample of M31 stars,  many of the analyses required that a star be 3 times 
as likely to be an M31 red giant than a foreground MW dwarf star to be 
included in the M31 sample (Section~\ref{sec:likelihoods}). 

However, simple threshold cuts on membership criteria are  
insufficient for the current analysis of the M31 halo velocity distribution,
with the exception of stars belonging to the dwarf satellites of M31.
We maintain the simple cut described above (Section~\ref{sec:dSph_members}) 
to identify and remove dSph stars from the sample.  
The small velocity dispersions, limited range of color 
(due to the limited range in [Fe/H] of the dSph stars) 
and relatively well defined spatial boundaries of the M31 dSphs 
allow us to identify and remove probable dwarf satellite members while minimally affecting 
the observed velocity distribution of M31 halo and MW stars in those fields.  
Even in dSph fields with relatively high densities of halo stars, at most 
two or three M31 halo stars are removed from the sample.  Moreover, the dSphs
in the sample have a broad range of systemic line of sight velocities. Thus,
the small amount of incompleteness introduced into the M31 and MW dataset by removing
dSph stars will not introduce a bias in the resulting velocity distribution. 

In contrast, stars belonging to tidal debris features span the full
spatial range of the spectroscopic masks.  Moreover, while the stars belonging to some 
tidal debris features have a limited range of colors and velocities, 
this is not universally true.  The tidal debris features are also significantly less 
dominant in a field than a dSph is, making the overlap of halo stars and tidal 
debris features in color and velocity space more significant.  Finally, 
the probability a given star belongs to the kinematically hot halo or a 
kinematically cold tidal debris feature, based on the fits described
in Section \ref{sec:kccs}, is calculated assuming a fixed mean velocity
and velocity dispersion of the kinematically hot M31 halo.  Thus, using these
probabilities directly in our fits would introduce an internal inconsistency in the 
velocity measurement of the M31 halo as a function of radius, and may result in a 
systematic bias in the measurements. 

MW stars and M31 stars have complete overlap in their spatial distribution, 
and large overlap in their velocity and color distributions.  
However, the velocity diagnostic cannot be included in the likelihoods,
as justified in Section \ref{sec:likelihoods}.   
Since the velocity diagnostic provides strong 
discriminating power between the M31 and MW populations, there is 
substantially more overlap of the \olkhd\ distributions of the 
M31 and MW populations when using \olkhd\ computed without the 
velocity diagnostic.  This results in an unavoidable increase in 
contamination and decrease in completeness in any sample 
defined using only an \olkhd\ threshold.  Furthermore, since the relative densities
of MW and M31 stars vary with projected radius, the level of contamination and 
completeness of any given sample will also vary with projected radius.  
Both contamination and completeness will affect the resulting 
velocity distribution 
of the final M31 sample \citep{gilbert2007}. 

The biases in the velocity distribution of any sample defined via 
an \olkhd\ threshold occur primarily in the region of significant 
overlap between the MW and M31 velocity distributions ($\sim -200 < v_{\rm hel} < -125$~\kms).
There are expected to be very few MW stars with heliocentric velocities $v_{\rm hel} < -300$~\kms\ in our sample \citep{robin2003}.  
It is thus conceivable to consider using only stars with line of sight velocities more negative
than M31's systemic velocity
to identify an M31 sample using the \citet{gilbert2006} 
likelihood technique (Section \ref{sec:likelihoods}).
However, there are several arguments against adopting this approach.  
First, it increases the sensitivity of the results to substructure, since many of the
tidal debris features observed in SPLASH have mean line of sight heliocentric velocities 
$\langle v_{\rm hel} \rangle < -300$~\kms\ (Figure~\ref{fig:data_vel_vs_rad_subst}; many of these individual features are part of the Giant Southern Stream).  
It also reduces the size of the M31 halo sample by half, and in the sparse outer halo, 
the number of observed M31 stars is already small. 
Finally, it requires assuming a mean velocity for M31's halo, removing any sensitivity to 
departures in the mean line of sight velocity of M31's halo from M31's systemic velocity. 
Hence, we do not include velocity information in the membership likelihoods, but rather explicitly model the velocity distributions of all the known populations.

\section{Modeling the Velocity Distribution of the M31 and MW Stellar Populations}\label{sec:mcmc}
The final dataset contains 5299 stars, and includes multiple MW and 
M31 components.  These stellar populations significantly overlap in all 
parameter spaces, including velocity, spatial distribution, and color and magnitude.
The challenges in defining a reasonably uncontaminated M31 sample 
that is not biased in velocity space (Section~\ref{sec:sample_sel}) 
motivates our choice to perform a Bayesian analysis: 
modeling all known components in the dataset,
incorporating our prior knowledge of the probability a star is an M31 red giant,
and comparing the full model parameter space to 
the data using Markov Chain Monte Carlo (MCMC) techniques. 
The only stellar populations (MW or M31) removed from the dataset
are members of M31's dSph satellites (Section~\ref{sec:dSph_members}), as they are
the only populations that are compact enough in parameter space to 
separate out while introducing minimal contamination or loss of completeness in the 
primary population of interest, the M31 halo.

Below, we detail the velocity transformations used to remove the effect of perspective 
motion from the line of sight velocities (Section~\ref{sec:vel_transform}), 
describe the likelihood function (Section~\ref{sec:mcmc_lklhd_func}), and 
motivate the inclusion of each of the MW and M31 
components included in the model.  
We then describe the priors applied to each of the model parameters 
(Section~\ref{sec:mcmc_priors}).  Finally, we describe our use of
MCMC techniques to efficiently sample the model parameter space 
(Section~\ref{sec:mcmc_emcee}).   

\subsection{Velocity Transformations}\label{sec:vel_transform}
The full set of fields span a significant area on the sky, with the largest angular separations between fields surpassing 20\degree.
To eliminate the effects of perspective motion, 
all line of sight velocities are 
transformed to the Galactocentric frame, and the bulk motion of M31 along the 
line of sight to each star is removed.  After this transformation, a star with no peculiar velocity  
relative to M31's bulk motion will have $v_{\rm pec} = 0$~\kms, irrespective of its position on the sky.  

To facilitate comparison with 
measurements of the velocity dispersion of M31's globular cluster 
population, we follow \citet{veljanoski2014} in using the \citet{courteau1999} relation,  
with the 
\citet{mcmillan2011} estimate of the circular speed of the Galaxy's disk at the Sun ($239$~\kms) and the \citet{schonrich2010} values for the solar peculiar motion [(U, V, W)$_{\odot}  = (11.1, 12.24, 7.25)$~\kms].  Thus, the transformation of the observed, heliocentric 
line of sight stellar velocities, $v_{\rm helio, los}$ to the Galactocentric frame, $v_{\rm Gal}$ is given by
\begin{multline}
v_{\rm Gal} = v_{\rm helio} + 251.24\sin(l)\cos(b) \\
+ 11.1\cos(l)\cos(b) + 7.25\sin(b),
\label{eqn:Gal}
\end{multline}
where $l$ and $b$ are the Galactic coordinates (longitude and latitude) of the star.  

Performing the transformation to the Galactocentric frame is vital, as it removes perspective effects in the outer halo fields that are on the same order as the dispersion we are trying to measure (several tens of \kms).  However, the analysis presented here is not sensitive to the exact values assumed in Equation~\ref{eqn:Gal} \citep[for alternate values, see ][]{bland-hawthorn2016}.  With the SPLASH dataset, the difference in Galactocentric transformations assuming alternate values, such as $218$~\kms\ for the circular speed of the disk \citep{bovy2013}, versus the transformations found using the nominal values above, is minimal: the typical effect on the measured halo dispersion would be of the order 2 to 3~\kms\ or smaller.  This is less (by a factor of two) than the typical velocity measurement error for the sample, and is more than an order of magnitude smaller than the expected halo dispersion.

To remove the bulk motion of M31 along the line of sight to each star, 
we assumed a heliocentric velocity for M31 of $v_{\rm M31, helio} = -301$~\kms, corresponding to a 
Galactocentric radial velocity of $v_{\rm M31, r} = -109$\kms\ \citep[e.g.,][]{vandermarel2008}, and 
a transverse velocity (in the Galactocentric frame) of $v_{\rm M31, t} =17$~\kms, with a position angle of $\theta_{\rm M31, t} = 287$\degree\ \citep{vandermarel2012}.  
The removal of M31's motion from the line-of-sight velocities transformed to the Galactocentric frame, resulting in peculiar line of sight velocities for each star, $v_{\rm pec}$, is calculated following \citet{vandermarel2008}:
\begin{multline}
v_{\rm pec} = v_{\rm Gal}  - v_{\rm M31, r}\cos(\rho) \\
+ v_{\rm M31, t}\sin(\rho)\cos(\phi - \theta_{\rm M31, t}),
\label{eqn:pec}
\end{multline}
where $\rho$ is the angular separation between M31's center and the star and $\phi$ is the position angle of the star with respect to M31's center.   The uncertainties in M31's tangential motion are rather large (the 1$\sigma$ confidence interval on $v_{\rm M31, t}$ is  $ \leq 34.3$~\kms).  However, as with the uncertainties in the transformation to the Galactic reference frame, we calculate the typical effect of these uncertainties on the measured dispersion to be small compared to the expected halo dispersion, on the order of $\sim 3$~\kms or less.

In the analysis that follows, all velocities have been transformed to $v_{\rm pec}$, using Equations \ref{eqn:Gal} and \ref{eqn:pec}.
Readers are referred to \citet{veljanoski2014} for a broader discussion of the above transformations in the context of M31 kinematical analyses.

\subsection{The Likelihood Function}\label{sec:mcmc_lklhd_func}
Our goal is to determine the most likely model parameters that describe
the observed velocity distribution of stars along the 
line of sight to M31, as a function of projected radius from M31's center.  
We accomplish this by inferring the probability distributions for the parameters of a 
probabilistic generative model for the data, using Bayes' theorem.  

The primary challenge in constructing the model is in determining the 
likelihood function. 
We construct the 
likelihood function by making the simplifying assumption that the line of 
sight velocity distribution of each of the individual stellar populations present 
in any line of sight can be adequately modeled by a Gaussian distribution.  
We then describe the 
likelihood function as a combination of M31 (Section~\ref{sec:mcmc_lklhd_func_m31model}) and MW (Section~\ref{sec:mcmc_lklhd_func_mwmodel}) models, each of which
is composed of a combination of normalized Gaussians.  
Each individual stellar population in M31 and the MW thus
contributes a set of parameters to the model, namely 
the mean velocity ($\mu$), velocity dispersion ($\sigma$), and a relative fraction at 
which it contributes to the M31 or MW populations ($f$).  
In the likelihood function,
each Gaussian component in the model is evaluated at the velocity of each data point, 
which is notated in Sections~\ref{sec:mcmc_lklhd_func_m31model} and  \ref{sec:mcmc_lklhd_func_mwmodel} by $\mathscr{N}(v_i | \mu, \sigma)$.  

If MW or M31 membership was known precisely for each star, 
the likelihood of the observed line of sight
velocity for an individual star given the model parameters $\Theta$ could be written as 
\begin{equation}
\mathscr{L}_{i} = \eta_i\,\mathscr{L}_{i}^{M31} + (1-\eta_i)\,\mathscr{L}_{i}^{MW}
\end{equation} 
where $\eta_i$ = 1 if the star is an RGB star in M31 and $\eta_i$ = 0 
if the star is a MW star along the line of sight, and $\mathscr{L}_{i}^{M31}$ and $\mathscr{L}_{i}^{MW}$ 
are the likelihoods for the M31 and MW models.   We do not know 
{\it a priori} which stars are M31 and MW stars.  However, we do have prior information 
on the probability of M31 membership for each star, obtained by evaluating the \citet{gilbert2006} 
photometric and spectroscopic diagnostics (Section~\ref{sec:likelihoods}).  

Thus, we combine the M31 and MW likelihood functions using a mixture model. 
For an individual star, the likelihood of the observed line of sight
velocity $v_i$ given the model parameters $\Theta$ is
\begin{equation}
\mathscr{L}_{i} = p_{\rm M31, i}\,\mathscr{L}_{i}^{M31} + (1-p_{\rm M31, i})\,\mathscr{L}_{i}^{MW}
\label{eqn:mixture_model}
\end{equation} 
with $p_{\rm M31, i}$ describing the prior probability the star is an M31 RGB star (Section~\ref{sec:mcmc_lklhd_func_membershipprior}).  
The likelihood of the observed dataset given the model parameters $\Theta$ 
is simply the product of the individual likelihoods:
\begin{equation}
\mathscr{L}_{\theta} = \prod_{i=1}^{N_{\rm stars}} \mathscr{L}_{i,\theta}.
\label{eqn:likelihood}
\end{equation}

Finally, for a set of model parameters $\Theta$, we compute 

\begin{equation}
p(\Theta | {v_i}_{i=1}^N, I) \propto p({v_i}_{i=1}^N | \Theta, I)\,p(\Theta | I)
\label{eq:bayes}
\end{equation}

where ${v_i}_{i=1}^N$ is the set of observed velocities, $I$ 
represents our prior knowledge, and $p({v_i}_{i=1}^N | \Theta, I)$ is the likelihood 
term (Equation~\ref{eqn:likelihood}). .  
Equation~\ref{eq:bayes} simply asserts that
the probability of the set of model parameters $\Theta$, given the observed data
and all prior information, is proportional to the probability of the observed 
data given the model and all prior information, multiplied by the probability 
of the model given all prior information.    

\subsubsection{M31 model}\label{sec:mcmc_lklhd_func_m31model}

We characterize the population of M31 stars as a mixture of 
all known stellar components in our spectroscopic fields: a kinematically hot
halo with multiple distinct, kinematically cold tidal debris features.  
This means that the M31 model, and thus 
the M31 likelihood function that is evaluated in Equation~\ref{eqn:mixture_model}, is
dependent on the spectroscopic field, $sf$, in which the star $i$ is observed. 

The number of kinematically cold tidal debris features is based on
our previous analyses of individual spectroscopic fields 
\citep[Section~\ref{sec:kccs};][]{gilbert2007,gilbert2009gss,gilbert2012}.
For the purposes of building the likelihood function, each kinematically cold component corresponding to an observed 
tidal debris feature is assumed to contribute only to the 
field in which it was observed. 
The dataset does contain observations of large tidal debris features
that span multiple spectroscopic fields, most notably 
the Giant Southern Stream and the Southeast Shelf \citep[both of which
are related to a single accretion event;][]{fardal2007,gilbert2007}.
However, in all cases the mean velocity, velocity dispersion, and/or
surface density of the feature is spatially dependent, 
resulting in different measured values in each spectroscopic field.  
Therefore, each kinematically cold tidal debris feature observed in a
spectroscopic field is 
included as a separate Gaussian component in the likelihood function, with
parameters that are fit independently of detections of the same 
tidal debris feature in other fields.

However, the properties of the 
underlying, dynamically hot stellar halo change much more slowly and smoothly with spatial position 
than do the presence and properties of tidal debris features.  We therefore fit a single underlying M31 halo component across all
spectroscopic fields included in a given fit. 

This results in field-independent and field-dependent (denoted with the subscript $sf$) 
M31 model parameters.  
Field-independent model parameters include 
the mean velocity and velocity dispersion of the M31 
halo components.  These model parameters are present in the M31 likelihood function
for every spectroscopic field.  

Field-dependent model parameters include the mean velocity and velocity 
dispersion of each of the M31 tidal debris features present in a field 
and the relative fractions of each of the M31 components (tidal debris features
and halo).  These model parameters are present only in the M31 likelihood function for a single
spectroscopic field.  Our spectroscopic fields have at most three kinematically distinct 
M31 components: the kinematically hot halo and two kinematically cold tidal debris features.  
Therefore, the likelihood function, $\mathscr{L}_{i,sf}^{M31}$, for a given star $i$ in a spectroscopic field $sf$ takes one of three forms.

All spectroscopic fields without tidal debris features,
and thus without any field-dependent model parameters, are described by the M31 likelihood function:
\begin{multline}
\mathscr{L}_{i}^{M31} = \mathscr{N}(v_i | \mu_{\rm M31\ halo}, \sigma_{\rm M31\ halo}).
\label{eq:nokcc_distfunc}
\end{multline}

\noindent If one kinematically cold component has been identified in a field, 

\begin{multline}
\mathscr{L}_{i}^{M31} = f_{\rm KCC 1}\mathscr{N}(v_i | \mu_{\rm KCC 1}, \sigma_{\rm KCC 1})\ \\
+ f_{\rm M31\,halo}\mathscr{N}(v_i | \mu_{\rm M31\ halo}, \sigma_{\rm M31\ halo})
\label{eq:onekcc_distfunc}
\end{multline}

\noindent If there are two kinematically cold components in a field,

\begin{multline}
\mathscr{L}_{i}^{M31} = f_{\rm KCC 1}\mathscr{N}(v_i | \mu_{\rm KCC 1}, \sigma_{\rm KCC 1})\ \\
+ f_{\rm KCC 2}\mathscr{N}(v_i | \mu_{\rm KCC 2}, \sigma_{\rm KCC 2})\  \\
+ f_{\rm M31\,halo}\mathscr{N}(v_i | \mu_{\rm M31\ halo}, \sigma_{\rm M31\ halo})
\label{eq:twokcc_distfunc}
\end{multline}

The relative fractions of the M31 components in a spectroscopic field, sf, are normalized such that
\begin{equation}
\sum_{k=1}^{N_{\rm M31,sf}} f_{k} = 1.
\label{eq:norm_m31}
\end{equation}
This is enforced by calculating the M31 halo fraction (the 
only M31 component present in every field) as 
\begin{equation}
f_{\rm M31\,halo} = 1 - \sum_{k=1}^{N_{\rm KCC}} f_{k}.
\label{eq:frac_m31halo}
\end{equation}
In fields without any kinematically cold components (KCCs), the M31 halo fraction ($f_{\rm M31\,halo}$) is equal to one.

\subsubsection{MW model}\label{sec:mcmc_lklhd_func_mwmodel}

We characterize the population of MW stars in the sample 
with three Gaussian components.  These include a 
component with a mean heliocentric line of sight velocity near zero and a relatively small velocity 
dispersion (corresponding to the MW thin disk in the direction of M31), a second component 
with a slightly more negative mean velocity and a slightly larger velocity dispersion 
(corresponding to the MW thick disk), and a third
component with a significantly more negative mean velocity and a large velocity
dispersion (the MW halo).  
In our dataset, the MW halo appears as a population of stars with blue \vio\ colors extending to 
large negative heliocentric velocities (Figure~\ref{fig:data_vel_vs_rad_likelihood}); these stars are significantly bluer than the most 
metal-poor 10~Gyr RGB isochrone. The MW disk populations 
span a range of color roughly similar to that spanned by M31 RGB stars, with
MW stars that have heliocentric velocities closest to $0$~\kms\ in general also having the reddest \ivi\ colors \citep[e.g., Figure~3 of ][]{gilbert2012}.

\begin{figure}[tb!]
\plotone{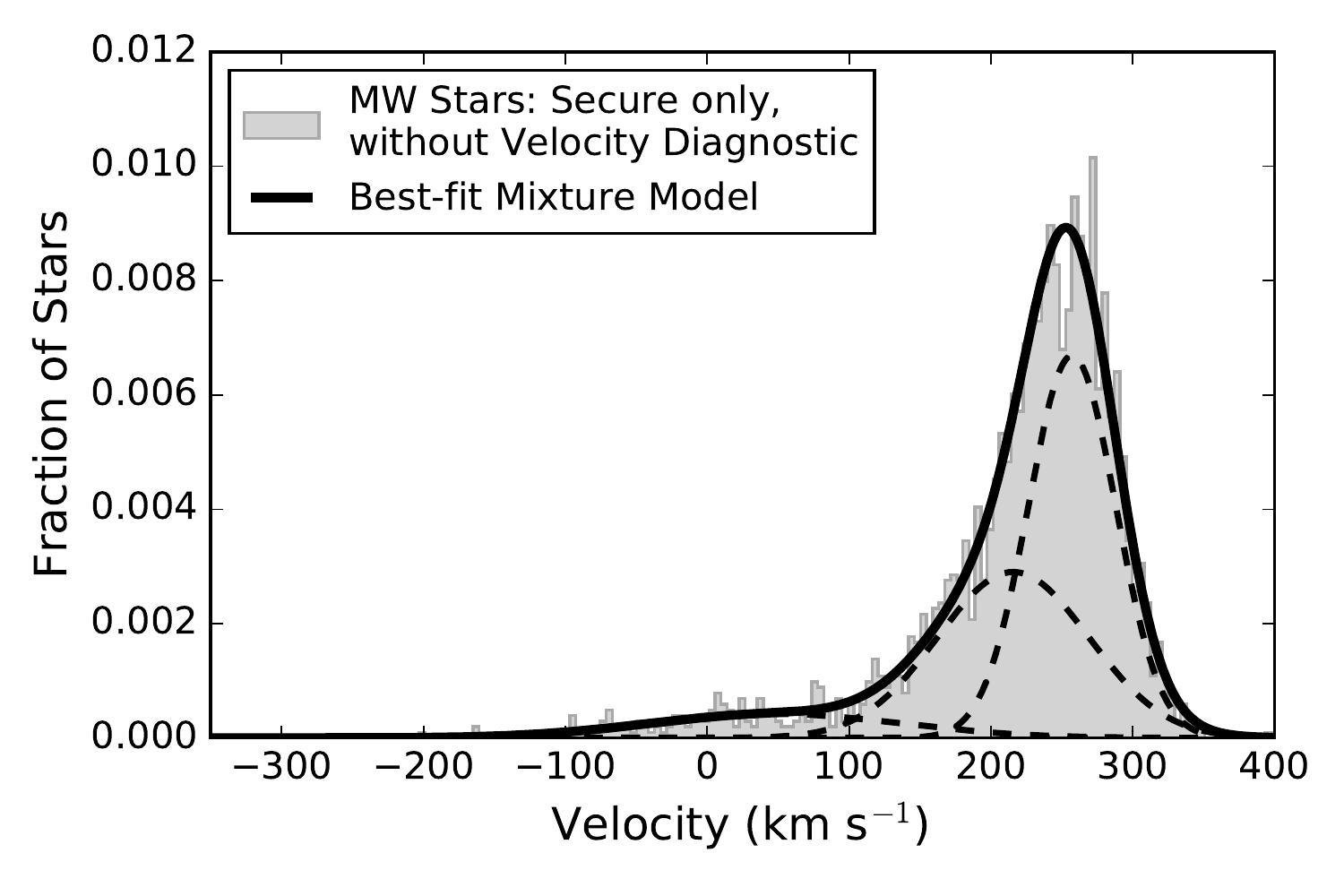}
\plotone{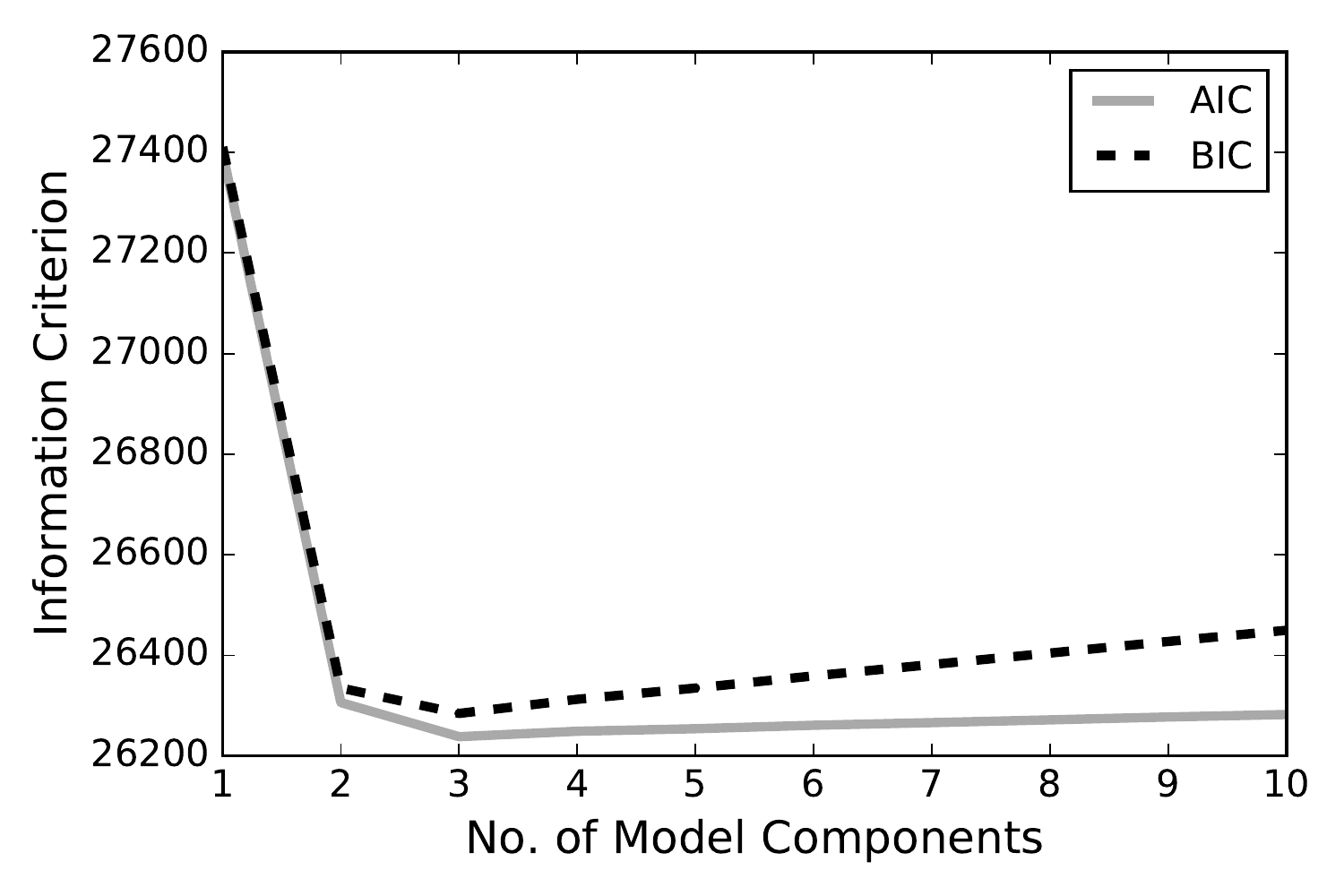}
\caption{
({\it upper panel}) Best-fit, one-dimensional Gaussian mixture model to the velocity distribution ($v_{\rm pec}$, Section~\ref{sec:vel_transform}) of 
MW stars.  The MW sample shown here includes stars securely identified as MW stars without the inclusion of the velocity diagnostic (Section~\ref{sec:likelihoods}).  
The best-fit parameter values of each of the three components is dependent on the MW sample selection method. However, three Gaussian components are statistically preferred regardless of the details of the MW sample selection (Section~\ref{sec:mcmc_lklhd_func_mwmodel}).  Observed line of sight velocities have been transformed to the Galactocentric frame, and the bulk motion of M31 has been removed (Section~\ref{sec:vel_transform}): a star with no peculiar velocity relative to M31's bulk motion will have $v = 0$~\kms.   
({\it lower panel}) The values of the Aikake information criterion (AIC) and the Bayesian information criterion (BIC) as a function of the number of Gaussian model components used to model the observed MW velocity distribution.  Three Gaussian components are preferred over two (the difference between 3 or 2 components is greater than 50 for both the AIC and BIC).  Adding additional Gaussian components does not improve either the AIC or the BIC.
}
\label{fig:MW_GMM}
\end{figure}

Figure~\ref{fig:MW_GMM} shows the velocity histogram of a sample of stars identified as MW
stars using a simple cut of \olkhd\,$<-0.5$ (Section~\ref{sec:likelihoods}), along with
the preferred Gaussian mixture model to the velocity distribution.  Based on both the Aikake information criterion (AIC) and the Bayesian information criterion (BIC), the preferred number of Gaussian components for this dataset is three. Adding additional model components beyond three improves neither the AIC nor the BIC.
The best-fit values of the mean
velocity, dispersion, and mixture fraction for each component depends on
the MW sample selection method (e.g., inclusion of the velocity diagnostic in computing \olkhd,
and/or inclusion of marginally identified MW stars, with $-0.5\le$\,\olkhd\,$< 0$). However, 
the statistical preference for three Gaussian components, with decreasing
mean velocity and increasing dispersion, 
is robust regardless of the MW sample selection: the minimum value in 
both the AIC and BIC metrics occurs with three Gaussian components for 
all MW samples tested.
 
The range in Galactic latitude spanned by the spectroscopic
fields is substantial ($\sim$20\degree, Figure~\ref{fig:roadmap1}). The 
relative fraction of the three MW components is known to change over the
full field of view of the survey \citep[e.g.,][]{martin2013pandas}, 
and it is reasonable to expect 
the mean velocity and dispersion of the disk components may change detectably 
as well.   
Ideally, the changing mixture of MW stellar populations over the field
of the survey (e.g., with galactic latitude) could be included in our model, either empirically or through use of a physical model for the Galaxy.    

A   fundamental   impediment   to   using   a   physical  MW model, or a parameterized empirical model,  is   that   the   MW   stars   displayed   in   Figure \ref{fig:MW_GMM}   are   not   a simple   and   representative   sampling   of   the   MW   components   along   the   line   of   sight   to   M31.  The MW stars with measured velocities are instead a complicated   function   of   the   target   selection   process, which was optimized   to  preferentially target   M31 stars based   on   the   available   photometric   data   in   each   field,  and   the   success   of   the   cross-correlation routine   used   to   measure   velocities   (dependent   on   both   the   observing   conditions   for   each   mask (affecting the SNR of the spectra) and   the   properties   of   the   star   itself   (affecting   the   strength   of   absorption   lines).   We nevertheless explored drawing from the Besan\c{c}on Galactic model \citep{robin2003} at   the   location   of   each   field,   limiting   the   model   results   only   to   those stars   in   the   same   apparent   magnitude   and   color   ranges   as   the   targets   in   each   field   of   the spectroscopic   survey.  We   found   that   the   velocity   distribution   of   stars   drawn   from   the   Besan\c{c}on   model   does not   match   the   observed   velocity   distribution   of   MW   stars   in   our   survey. 

Given the above considerations, we do not fix the MW component parameters based 
on fits to a selected sample of MW stars nor attempt a hierarchical fit including Galactic latitude or longitude. 
Rather, we fit all the parameters for the 
three MW components simultaneously with the parameters for the M31 components.
As was assumed for the properties of the M31 halo, the properties of each of the three MW components is assumed to change relatively slowly and smoothly with spatial position within M31's stellar halo \citep[this is consistent with the Besancon Galaxy Model;][]{robin2003}.  Thus, the MW likelihood function is not field-dependent, and is simply
\begin{equation}
\mathscr{L}_{i}^{MW} = \sum_{j=1}^{N_{\rm MW}=3}f_j \mathscr{N}(v_i | \mu_j, \sigma_j).
\end{equation}
To ensure a normalized $\mathscr{L}_{i}^{MW}$, the sum of the fractions of the MW components is constrained to equal one:
\begin{equation}
\sum_{j=1}^{N_{\rm MW}=3}f_j = 1.
\end{equation}

Enforcing this normalization reduces the number of model parameters
by one: the fraction of the third MW component is set to 
\begin{equation}
f_{\rm MW3} = 1 - f_{\rm MW1} - f_{\rm MW2}.
\end{equation}

\subsubsection{Probability of M31 Membership}\label{sec:mcmc_lklhd_func_membershipprior}
The probability of a given star $i$ being an M31 star ($p_{\rm M31}$, Equation~\ref{eqn:mixture_model}) is primarily derived 
from the overall likelihood \olkhd\ that the star is an M31 star,  
which is based on multiple spectroscopic and photometric measurements (excluding the velocity diagnostic, 
Section~\ref{sec:likelihoods}).  For each star, \olkhd\ is the logarithm of the odds ratio that the star is 
an RGB star at the distance of M31 or a MW dwarf star.  Thus, the overall likelihood can be converted into 
a probability of M31 membership via
\begin{equation}
\frac{10^{\langle L_i\rangle}}{1 + 10^{\langle L_i\rangle}}.
\label{eqn:m31probability}
\end{equation}

The distribution of \olkhd\ values of stars in a field or radial region is in reality the superposition of two independent and overlapping distributions: that of MW stars and that of M31 stars.  The MW (M31) \olkhd\ distribution has a tail that extends to positive (negative) \olkhd\ values \citep{gilbert2006, gilbert2007, gilbert2012}.  When one population (either the MW or M31) is strongly dominant, contamination (in \olkhd\ space) of the minority population by the tail of the distribution of the dominant population can be significant.  For the present analysis, this is most concerning when considering M31's outer halo, where MW stars outnumber M31 stars in the spectroscopic sample by 50:1 or more.  If this effect is not accounted for, the tail of the MW star distribution, which has M31-like \olkhd\ values (greater than zero), results in the fit being driven to include MW stars as part of the M31 model.  This will in turn drive the M31 mean velocity to larger values with increasing projected radius.

We therefore introduce a hyper-parameter $C$ to the prior probability of M31 membership to account for the possibility that a significant fraction of stars in a field may have \olkhd-based probabilities that are not in line with their actual M31 or MW membership.  Furthermore, since the density of M31 stars is the single largest driver
in the amount of contamination in M31 samples,
we add a field-dependent parameter, $\alpha_{sf}$, to the fit in fields with identified tidal debris, which typically have higher stellar densities than nearby halo fields without tidal debris features in M31's halo. Thus,
\begin{equation}
p_{\rm M31} = \alpha_{sf} C\frac{10^{\langle L_i\rangle}}{1 + 10^{\langle L_i\rangle}},
\end{equation}
and the probability that a star is an MW star is simply $1 - p_{\rm M31}$.  In fields without tidal debris features, $\alpha_{sf} \equiv 1$.

We restrict $\alpha_{sf} C$ to be less than or equal to one.  This means it can only reduce the probability of M31 membership based on the empirical photometric and spectroscopic diagnostics.  
The greatest contrast between M31 and MW stellar density is in the outer M31 halo fields, where MW stars greatly outnumber M31 stars, and improperly categorized MW stars can have a significant effect on the M31 model parameters.  In the innermost M31 halo fields included here, the converse is true: M31 stars outnumber MW stars, and improperly categorized M31 stars may affect the MW model parameters.  However, the MW model parameters are nuisance parameters which we marginalize over: the MW model parameters have no direct physical interpretation due to our experimental setup (e.g., a varying spectroscopic selection function, averaging over Galactic latitude and longitude when binning the data based on projected radius from M31, and removing M31's bulk motion from the stellar velocities).  Moreover, 
under the assumption the kinematically hot halo is a well-mixed population, the fact that the tail of M31 stars with \olkhd$ < 0$ contributes little to the M31 model will not introduce a systematic bias in the measurement of the M31 halo parameters.

\subsection{Priors}\label{sec:mcmc_priors}
As discussed above (Section~\ref{sec:mcmc_lklhd_func}), there are both field-independent and field-dependent model parameters.
We discuss our choice of priors for each of these sets of parameters in turn. 

\subsubsection{Field Independent Parameters}\label{sec:priors_findep_params}
The priors on the field-independent model parameters, which describe the M31 halo and the three MW components, are noninformative over the allowed range of the parameter with one exception.  Table~\ref{tab:primary_params_priors} lists the allowed range, choice of prior, and any additional constraints on the parameter implemented in the prior for the field-independent model parameters.  

\begin{deluxetable*}{cccccc}
\tablecolumns{6}
\tablewidth{0pc}
\tablecaption{Field-Independent Model Parameters.}
\tablehead{\multicolumn{1}{c}{Parameter} & \multicolumn{1}{c}{Description} & \multicolumn{1}{c}{Units} & \multicolumn{1}{c}{Allowed Range\tablenotemark{a}} & \multicolumn{1}{c}{Prior\tablenotemark{b}} & \multicolumn{1}{c}{Additional Constraints} 
}
\startdata
\sidehead{{\it M31 Parameters}}
$\mu_{\rm M31}$ & Mean Velocity & \kms & $-300 < \mu_{\rm M31} < 250 $ & normal & $\mu = -18$, $\sigma = 5$ \\  
$\sigma_{\rm M31}$ & Velocity Dispersion & \kms & $ 5 < \sigma_{\rm M31} < 300 $ & scale free &  \\  
\hline
\sidehead{{\it MW Disk Component 1 Parameters}}
$\mu_{\rm MW1}$ & Mean Velocity & \kms & $ 150 < \mu_{\rm MW1} < 350 $ & uniform &  $\mu_{\rm MW3} < \mu_{\rm MW2} <  \mu_{\rm MW1}$ \\  
$\sigma_{\rm MW1}$ & Velocity Dispersion & \kms & $ 5 < \sigma_{\rm MW1} < 350 $ & scale free & $\sigma_{\rm MW3} < \sigma_{\rm MW2} <  \sigma_{\rm MW1}$ \\  
$f_{\rm MW1}$ & Fraction & &  $ 0 < f_{\rm MW1} < 1 $ & uniform & $ f_{\rm MW1} + f_{\rm MW2} < 1  $ \\
\hline
\sidehead{{\it MW Disk Component 2 Parameters}}
$\mu_{\rm MW2}$ & Mean Velocity & \kms & $ 50 < \mu_{\rm MW2} < 350 $ & uniform &  $\mu_{\rm MW3} < \mu_{\rm MW2} <  \mu_{\rm MW1}$ \\  
$\sigma_{\rm MW2}$ & Velocity Dispersion & \kms & $ 5 < \sigma_{\rm MW2} < 350 $ & scale free & $\sigma_{\rm MW3} < \sigma_{\rm MW2} <  \sigma_{\rm MW1}$ \\  
$f_{\rm MW2}$ & Fraction & &  $ 0 < f_{\rm MW2} < 1 $ & uniform & $ f_{\rm MW1} + f_{\rm MW2} < 1  $ \\
\hline
\sidehead{{\it MW Halo Parameters}}
$\mu_{\rm MW3}$ & Mean Velocity & \kms & $ -200 < \mu_{\rm MW3} < 350 $ & uniform &  $\mu_{\rm MW3} < \mu_{\rm MW2} <  \mu_{\rm MW1}$ \\  
$\sigma_{\rm MW3}$ & Velocity Dispersion & \kms & $ 5 < \sigma_{\rm MW3} < 350 $ & scale free & $\sigma_{\rm MW3} < \sigma_{\rm MW2} <  \sigma_{\rm MW1}$  \\ 
\hline
\sidehead{{\it Hyper-Parameter}}
$C$ & affects prior probability & & $0 < C \le 1$ & uniform & \\
 &  of M31 membership & & & & 
\enddata
\tablenotetext{a}{Mean velocities are in the Galactocentric frame, with the bulk motion of M31 removed (Section~\ref{sec:mcmc}).
}
\tablenotetext{b}{The designated prior is applied over the allowed range of the model parameter.
}
\label{tab:primary_params_priors}
\end{deluxetable*}

The prior is uniform for the mean velocities and fractions of the total stellar population in the three MW components.  The priors on the velocity dispersions of the MW components and the M31 halo go as $1/\sigma$, which is scale-free.  A uniform prior is used for the hyper-parameter $C$ (Section~\ref{sec:mcmc_lklhd_func_membershipprior})  

The only informative prior on a field-independent model parameter is placed on the mean velocity of the M31 halo, for which we implement a normal prior with $\mu = -18$~\kms\ and $\sigma = 5$~\kms.  This is based on the marginalized one-dimensional posterior distribution found for the M31 halo mean velocity resulting from performing a fit to all stars within 40~kpc of M31's center, using the same fitting procedure described above, but assuming a uniform prior for the mean velocity of M31's halo.  This result is insensitive to the exact choice of projected radius used; fits including all stars within 30~kpc or 50~kpc of M31's center return similar results.   
This is consistent with what has been found for the mean velocity of M31's halo by Beaton et al. (in prep), who have used a separate kinematical analysis of the same spectroscopic dataset to measure the proper motion of M31.  

We implement reasonable yet conservative minimum and maximum bounds on the range of each parameter, allowing considerable freedom for exploration of parameter space by the MCMC chains while eliminating clearly unphysical values.  
We also implement a hierarchy of mean velocities and velocity dispersions for the three MW components.  This choice is informed by our physical understanding of the origin of the three MW components, as well as the results of Gaussian mixture model fits to samples of likely MW stars (Section~\ref{sec:mcmc_lklhd_func_mwmodel}).  

\subsubsection{Field Dependent Parameters}\label{sec:priors_fdep_params}

The field-dependent parameters describe the kinematically cold tidal debris features identified in individual fields.  
Table~\ref{tab:subst_params} lists the allowed range, choice of prior, and any additional constraints on the parameter implemented in the prior for the field-dependent model parameters.  

The priors for the mean velocity and velocity dispersion of each tidal debris feature are set by the literature values 
for each component (Section~\ref{sec:kccs}, Table~\ref{tab:subst_params}).  
We implement a normal prior on each mean velocity and velocity dispersion, using the published maximum-likelihood 
values and uncertainties on the maximum likelihood value of the parameter.  The median RA and Dec of all stars in the field was used to transform the published mean velocity for each tidal debris feature to the Galactocentric frame, and to remove the bulk motion of M31.  The published uncertainties on the mean velocities and velocity dispersions are in general not symmetric, and this is included in the prior.  The range of allowed values has been bounded to be within 5 times the published upper and lower error on the maximum-likelihood value.

As for the field-independent parameters, additional constraints on the parameters are implemented in the prior.  
By construction, the sum of the fractions computed for the M31 components (halo and all tidal debris features present) 
in an individual field can not exceed unity. All components must have a fractional contribution greater than zero.   
In fields with more than one tidal debris feature, a hierarchy of values is enforced for the mean velocity of each 
tidal debris feature: the second kinematically cold component must have a mean velocity more negative than the first.   

The parameter $\alpha$ (Section~\ref{sec:mcmc_lklhd_func_membershipprior}) is given a uniform prior and 
the freedom to vary over a significant range of values (0 to 10) to account for increased M31 stellar density due to tidal debris features, subject to the additional enforced constraint that $\alpha_{sf} C <=1$ (Section~\ref{sec:mcmc_lklhd_func_membershipprior}).  In practice, $\alpha_{sf}$ is very close to one in fields with \rproj\ $< 40$~kpc, and less than $\sim 2.5$ for fields with \rproj\,$> 40$~kpc.  

\startlongtable
\begin{deluxetable*}{cccccccc}
\tablecolumns{8}
\tablewidth{0pc}
\tablecaption{Field-Dependent Model Parameters\,-- Substructure Parameters and Priors.}
\tablehead{\multicolumn{1}{c}{Field} & \multicolumn{1}{c}{Parameter} & \multicolumn{1}{c}{Allowed Range\tablenotemark{a}} & \multicolumn{3}{c}{Prior} & \multicolumn{1}{c}{Additional} & \multicolumn{1}{c}{Reference} \\ 
\cline{4-6}
\multicolumn{1}{c}{and Component} & & & \multicolumn{1}{c}{Form} & \multicolumn{1}{c}{Mean\tablenotemark{b}} & \multicolumn{1}{c}{Standard} &\multicolumn{1}{c}{Constraints} &  \\ 
 & & & & & \multicolumn{1}{c}{Deviation\tablenotemark{b}} & &  
}
\startdata
f115 KCC 1 & $\mu_{\rm KCC1}$ & $ -49.00 < \mu_{\rm KCC1} <   77.00$ &  normal & $   13.00$  & $   12.60$  & ... & 1  \Tstrut\\ 
 & $\sigma_{\rm KCC1}$ & $    0.00 < \sigma_{\rm KCC1} <  104.70$ &  normal & $  42.20$ & $  13.40$ &  ... &\\ 
 & $f_{\rm KCC1}$ &                         $ 0 < f_{\rm KCC1} < 1 $ & uniform & ... & ...                         &  ... & \\ 
\hline 
f116 KCC 1 & $\mu_{\rm KCC1}$ & $ -98.39 < \mu_{\rm KCC1} <   85.11$ &  normal & $  -10.89$  & $   18.35$  & ... & 1  \Tstrut\\ 
 & $\sigma_{\rm KCC1}$ & $    0.00 < \sigma_{\rm KCC1} <  173.20$ &  normal & $  51.20$ & $  19.70$ &  ... & \\ 
 & $f_{\rm KCC1}$ &                         $ 0 < f_{\rm KCC1} < 1 $ & uniform & ... & ...                         &  ... & \\ 
\hline 
H11 KCC 1 & $\mu_{\rm KCC1}$ & $ -83.54 < \mu_{\rm KCC1} <   90.96$ &  normal & $    4.46$  & $   17.45$  & ... & 1  \Tstrut\\ 
 & $\sigma_{\rm KCC1}$ & $    0.00 < \sigma_{\rm KCC1} <  133.50$ &  normal & $  55.50$ & $  14.15$ &  ... & \\ 
 & $f_{\rm KCC1}$ &                         $ 0 < f_{\rm KCC1} < 1 $ & uniform & ... & ...                         & ... &  \\ 

\hline
f207 KCC 1 & $\mu_{\rm KCC1}$ & $-168.18 < \mu_{\rm KCC1} <  -76.68$ &  normal & $ -126.68$  & $    9.15$  & ...  & 2 \Tstrut\\ 
 & $\sigma_{\rm KCC1}$ & $    0.00 < \sigma_{\rm KCC1} <   85.30$ &  normal & $  20.80$ & $  10.25$ & ... & \\ 
 & $f_{\rm KCC1}$ &                         $ 0 < f_{\rm KCC1} < 1 $ & uniform & ... & ...                         & ... &  \\ 
\hline 
f207 KCC 2 & $\mu_{\rm KCC2}$ & $ -262.48 < \mu_{\rm KCC2} < -186.98$ &  normal & $-225.48$ & $   7.55$ & $\mu_{\rm KCC1} > \mu_{\rm KCC2} $ & 2 \\ 
 & $\sigma_{\rm KCC2}$ & $   0.00 < \sigma_{\rm KCC2} <   59.20$ &  normal & $  23.20$ & $   6.10$ & ... & \\ 
 & $f_{\rm KCC2}$ &                     $ 0 < f_{\rm KCC2} < 1 $ & uniform & ... & ...                     & $f_{\rm KCC1} + f_{\rm KCC2} < 1$ \\ 
\hline 
f123 KCC 1 & $\mu_{\rm KCC1}$ & $  -4.71 < \mu_{\rm KCC1} <   43.79$ &  normal & $   18.29$  & $    4.85$  & ...   & 1 \Tstrut\\ 
 & $\sigma_{\rm KCC1}$ & $    0.00 < \sigma_{\rm KCC1} <   45.10$ &  normal & $  10.60$ & $   5.95$ & ... & \\ 
 & $f_{\rm KCC1}$ &                         $ 0 < f_{\rm KCC1} < 1 $ & uniform & ... & ...                         & ... &  \\ 
\hline 
H13s KCC 1 & $\mu_{\rm KCC1}$ & $-116.99 < \mu_{\rm KCC1} <  -62.99$ &  normal & $  -89.99$  & $    5.40$  & ...  & 2 \Tstrut\\ 
 & $\sigma_{\rm KCC1}$ & $    0.00 < \sigma_{\rm KCC1} <   50.00$ &  normal & $  17.00$ & $   6.75$ & ... & \\ 
 & $f_{\rm KCC1}$ &                         $ 0 < f_{\rm KCC1} < 1 $ & uniform & ... & ...                         & ... &  \\ 
\hline 
H13s KCC 2 & $\mu_{\rm KCC2}$ & $ -212.39 < \mu_{\rm KCC2} < -168.89$ &  normal & $-191.39$ & $   4.35$ & $\mu_{\rm KCC1} > \mu_{\rm KCC2} $  & 2 \\ 
 & $\sigma_{\rm KCC2}$ & $   5.30 < \sigma_{\rm KCC2} <   41.30$ &  normal & $  21.30$ & $   3.60$ & ... & \\ 
 & $f_{\rm KCC2}$ &                     $ 0 < f_{\rm KCC2} < 1 $ & uniform & ... & ...                     & $f_{\rm KCC1} + f_{\rm KCC2} < 1$ \\ 
\hline 
f115 KCC 1 & $\mu_{\rm KCC1}$ & $ -49.24 < \mu_{\rm KCC1} <   76.76$ &  normal & $   12.76$  & $   12.60$  & ... & 1 \Tstrut\\ 
 & $\sigma_{\rm KCC1}$ & $    0.00 < \sigma_{\rm KCC1} <  104.70$ &  normal & $  42.20$ & $  13.40$ & ... & \\ 
 & $f_{\rm KCC1}$ &                         $ 0 < f_{\rm KCC1} < 1 $ & uniform & ... & ...                         & ... &  \\ 
\hline 
f135 KCC 1 & $\mu_{\rm KCC1}$ & $ -59.97 < \mu_{\rm KCC1} <   85.03$ &  normal & $   25.03$  & $   14.50$  & ... & 1 \Tstrut\\ 
 & $\sigma_{\rm KCC1}$ & $    0.00 < \sigma_{\rm KCC1} <  103.60$ &  normal & $  30.10$ & $  11.85$ & ... & \\ 
 & $f_{\rm KCC1}$ &                         $ 0 < f_{\rm KCC1} < 1 $ & uniform & ... & ...                         & ... &  \\ 
\hline 
f135 KCC 2 & $\mu_{\rm KCC2}$ & $ -320.97 < \mu_{\rm KCC2} <   -5.97$ &  normal & $-150.97$ & $  31.50$ & $\mu_{\rm KCC1} > \mu_{\rm KCC2} $ & 1 \\ 
 & $\sigma_{\rm KCC2}$ & $   0.00 < \sigma_{\rm KCC2} <  191.00$ &  normal & $  55.50$ & $  23.80$ & ... & \\ 
 & $f_{\rm KCC2}$ &                     $ 0 < f_{\rm KCC2} < 1 $ & uniform & ... & ...                     & $f_{\rm KCC1} + f_{\rm KCC2} < 1$ \\ 

\hline
a3 KCC 1 & $\mu_{\rm KCC1}$ & $-169.25 < \mu_{\rm KCC1} < -118.75$ &  normal & $ -144.75$  & $    5.05$  & ... & 2  \Tstrut\\ 
 & $\sigma_{\rm KCC1}$ & $    0.30 < \sigma_{\rm KCC1} <   39.80$ &  normal & $  16.80$ & $   3.95$ & ... & \\ 
 & $f_{\rm KCC1}$ &                         $ 0 < f_{\rm KCC1} < 1 $ & uniform & ... & ...                         & ... &  \\ 
\hline 
and9 KCC 1 & $\mu_{\rm KCC1}$ & $ -63.57 < \mu_{\rm KCC1} <   20.93$ &  normal & $  -30.57$  & $    8.45$  & ... & 2  \Tstrut\\ 
 & $\sigma_{\rm KCC1}$ & $    0.00 < \sigma_{\rm KCC1} <   67.60$ &  normal & $  12.60$ & $   7.90$ & ... & \\ 
 & $f_{\rm KCC1}$ &                         $ 0 < f_{\rm KCC1} < 1 $ & uniform & ... & ...                         &  ... & \\ 

\hline
and1 KCC 1 & $\mu_{\rm KCC1}$ & $ -31.34 < \mu_{\rm KCC1} <   39.66$ &  normal & $    5.66$  & $    7.10$  & ... & 2  \Tstrut\\ 
 & $\sigma_{\rm KCC1}$ & $    0.00 < \sigma_{\rm KCC1} <   51.20$ &  normal & $   8.20$ & $   6.55$ & ... & \\ 
 & $f_{\rm KCC1}$ &                         $ 0 < f_{\rm KCC1} < 1 $ & uniform & ... & ...                         & ... &  \\ 
\hline 
and1 KCC 2 & $\mu_{\rm KCC2}$ & $ -145.14 < \mu_{\rm KCC2} <  -25.14$ &  normal & $ -87.14$ & $  12.00$ & $\mu_{\rm KCC1} > \mu_{\rm KCC2} $  & 2 \\ 
 & $\sigma_{\rm KCC2}$ & $   0.00 < \sigma_{\rm KCC2} <   96.80$ &  normal & $  30.30$ & $  10.95$ & ... & \\ 
 & $f_{\rm KCC2}$ &                     $ 0 < f_{\rm KCC2} < 1 $ & uniform & ... & ...                     & $f_{\rm KCC1} + f_{\rm KCC2} < 1$ \\ 
\hline 
a13 KCC 1 & $\mu_{\rm KCC1}$ & $ -63.67 < \mu_{\rm KCC1} <   54.33$ &  normal & $   -5.67$  & $   11.80$  & ... & 2  \Tstrut\\ 
 & $\sigma_{\rm KCC1}$ & $    0.00 < \sigma_{\rm KCC1} <   91.20$ &  normal & $  32.20$ & $  11.20$ & ... & \\ 
 & $f_{\rm KCC1}$ &                         $ 0 < f_{\rm KCC1} < 1 $ & uniform & ... & ...                         & ... &  \\ 
\hline 
m4 KCC 1 & $\mu_{\rm KCC1}$ & $  10.38 < \mu_{\rm KCC1} <   64.38$ &  normal & $   34.88$  & $    5.40$  & ... & 2  \Tstrut\\ 
 & $\sigma_{\rm KCC1}$ & $    0.00 < \sigma_{\rm KCC1} <   36.60$ &  normal & $   6.60$ & $   4.45$ & ... & \\ 
 & $f_{\rm KCC1}$ &                         $ 0 < f_{\rm KCC1} < 1 $ & uniform & ... & ...                         & ... &  \\ 
\hline 
m4 KCC 2 & $\mu_{\rm KCC2}$ & $  -88.12 < \mu_{\rm KCC2} <  -40.12$ &  normal & $ -65.62$ & $   4.80$ & $\mu_{\rm KCC1} > \mu_{\rm KCC2} $  & 2 \\ 
 & $\sigma_{\rm KCC2}$ & $   0.00 < \sigma_{\rm KCC2} <   37.40$ &  normal & $  11.40$ & $   4.65$ & ... & \\ 
 & $f_{\rm KCC2}$ &                     $ 0 < f_{\rm KCC2} < 1 $ & uniform & ... & ...                     & $f_{\rm KCC1} + f_{\rm KCC2} < 1$ \\ 

\hline
R06A220 KCC 1 & $\mu_{\rm KCC1}$ & $ -86.33 < \mu_{\rm KCC1} <  -56.33$ &  normal & $  -71.33$  & $    3.00$  & ... & 3 \Tstrut\\ 
 & $\sigma_{\rm KCC1}$ & $    0.00 < \sigma_{\rm KCC1} <   19.60$ &  normal & $   6.10$ & $   2.20$ & ... & \\ 
 & $f_{\rm KCC1}$ &                         $ 0 < f_{\rm KCC1} < 1 $ & uniform & ... & ...                         & ... &  \\ 

\hline
each field, & $\alpha_{sf}$ & $ 0 < \alpha_{sf} < 10$ &  uniform & ...  & ... & $\alpha_{sf}C \le 1$ & ... \Tstrut\\ 
$sf$, above & & & & & & 

\enddata
\tablenotetext{a}{Mean velocities are in the Galactocentric frame, with the bulk motion of M31 removed (Section~\ref{sec:mcmc}).}
\tablenotetext{b}{Mean and Standard Deviation used in the case of normal priors.} 
\tablenotetext{}{{\bf References.} (1) \citet{gilbert2007}; (2) \citet{gilbert2009gss}; (3) \citet{gilbert2012} }
\label{tab:subst_params}
\end{deluxetable*}

\subsection{Sampling the Posterior Probability Distribution with MCMC}\label{sec:mcmc_emcee} 

Following Bayes' theorem, we multiplied the likelihood of the data given the model (Section~\ref{sec:mcmc_lklhd_func}) 
with the prior distribution for each parameter in the model (Section~\ref{sec:mcmc_priors}) to compute the likelihood of the 
model given the data.  
We used MCMC methods to efficiently sample the parameter space, and marginalized over all model parameters to obtain posterior probability distributions for each parameter of interest. 

We sampled the posterior probability distributions for all of the model parameters 
described above using the open-source {\tt emcee} python package \citep{foreman-mackey_emcee, emcee-sw}, 
which provides an efficient implementation of the \citet{goodman&weare} 
Affine Invariant MCMC Ensemble sampler.  
The number of parameters in the model is dependent on the number of 
tidal debris features (Section~\ref{sec:mcmc_lklhd_func}), and thus varies 
based on the spectroscopic fields included in the fit.  In order to balance the required computational resources 
with the final number of independent samples, the number of MCMC chains was 
set to be at least ten times, and no more than 20 times, the number of model parameters included in the fit.  
However, if that value was less than 300, 
we instead ran the MCMC analysis with a minimum threshold of 300 chains (Table~\ref{tab:results_m31}).

Chains were initialized using a random, 
uniform distribution over the valid parameter space implemented in the prior 
for each parameter.  The chains were run for a minimum of 12,000 steps.  
The marginalized one- and two-dimensional projections of the 
posterior probability distributions are computed by drawing 
the values of the model parameters from the last half of each MCMC chain.  
The choice to use the first half of the chains as the burn-in 
period was made as a conservative choice that could be applied uniformly to all chains, for all radial bins.
The autocorrelation time of the chains stabilized 
by half way through the chain for all the parameters considered in the following analysis.  In most cases, the autocorrelation time stabilized well before the half way point.

Post-burn in, the chains for all parameters were inspected to ensure they had settled to an equilibrium.  
The parameter values and confidence limits as a function of the number of steps in the chain 
were also inspected to ensure stability had been achieved.  The chains were run long enough to 
supply a sufficient number of independent samples (estimated using the autocorrelation length) 
to estimate the {\it uncertainties} on the parameters of interest to a level of $\sim 1$\%.  

The results of the MCMC analysis presented below define the best estimate for each parameter as the 50th percentile of the one-dimensional posterior probability distribution, marginalized over all other model parameters. The associated uncertainties on the best estimates are the 16th and 84th percentiles of the posterior probability distributions. 

\section{The M31 Halo Velocity Dispersion as a Function of Radius}\label{sec:mcmc_results} 
To characterize the change of the velocity dispersion of M31's halo with projected radius from M31's center, the full spectroscopic dataset was split into multiple radial bins.  The MCMC analysis described in Section~\ref{sec:mcmc} was completed for each radial bin.  The bounds of each radial bin were chosen to minimize the number of fields that span two radial bins, while ensuring that at least 100 stars more likely to be M31 stars than MW stars based on their \olkhd\ values (Section~\ref{sec:likelihoods}, not including the velocity diagnostic) were present in each bin.  Table~\ref{tab:results_m31} lists the total number of stars in each bin, as well as the number of stars likely to be M31 stars (Section~\ref{sec:likelihoods}).  A final consideration was including at least one field without identified tidal debris features in each bin; this removed degeneracies between the field-independent and field-dependent 
model hyper-parameters $C$ and $\alpha_{sf}$ (Section~\ref{sec:mcmc_lklhd_func_membershipprior}).

\begin{figure*}[tb!]
\includegraphics[width=3.0in]{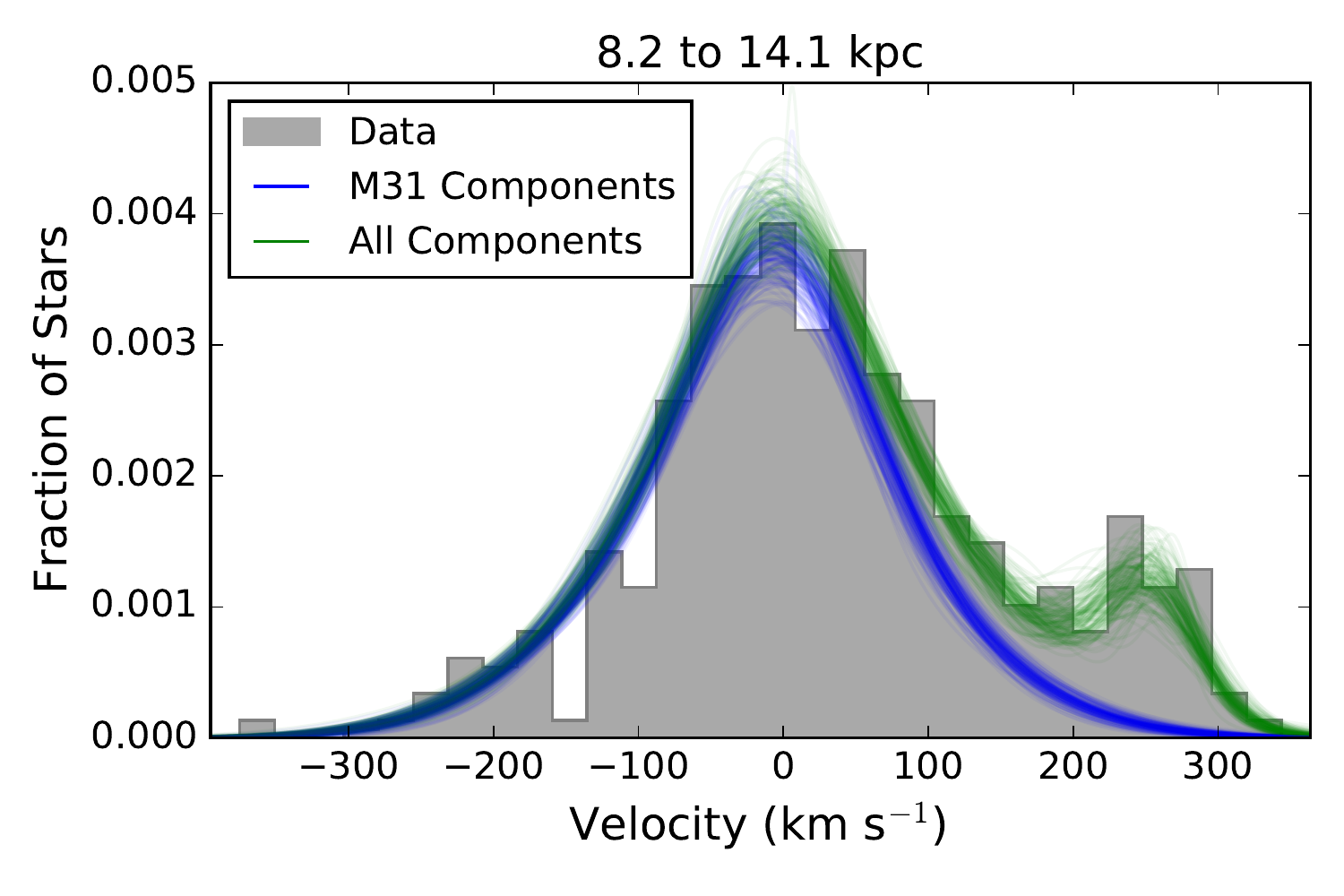}
\includegraphics[width=3.0in]{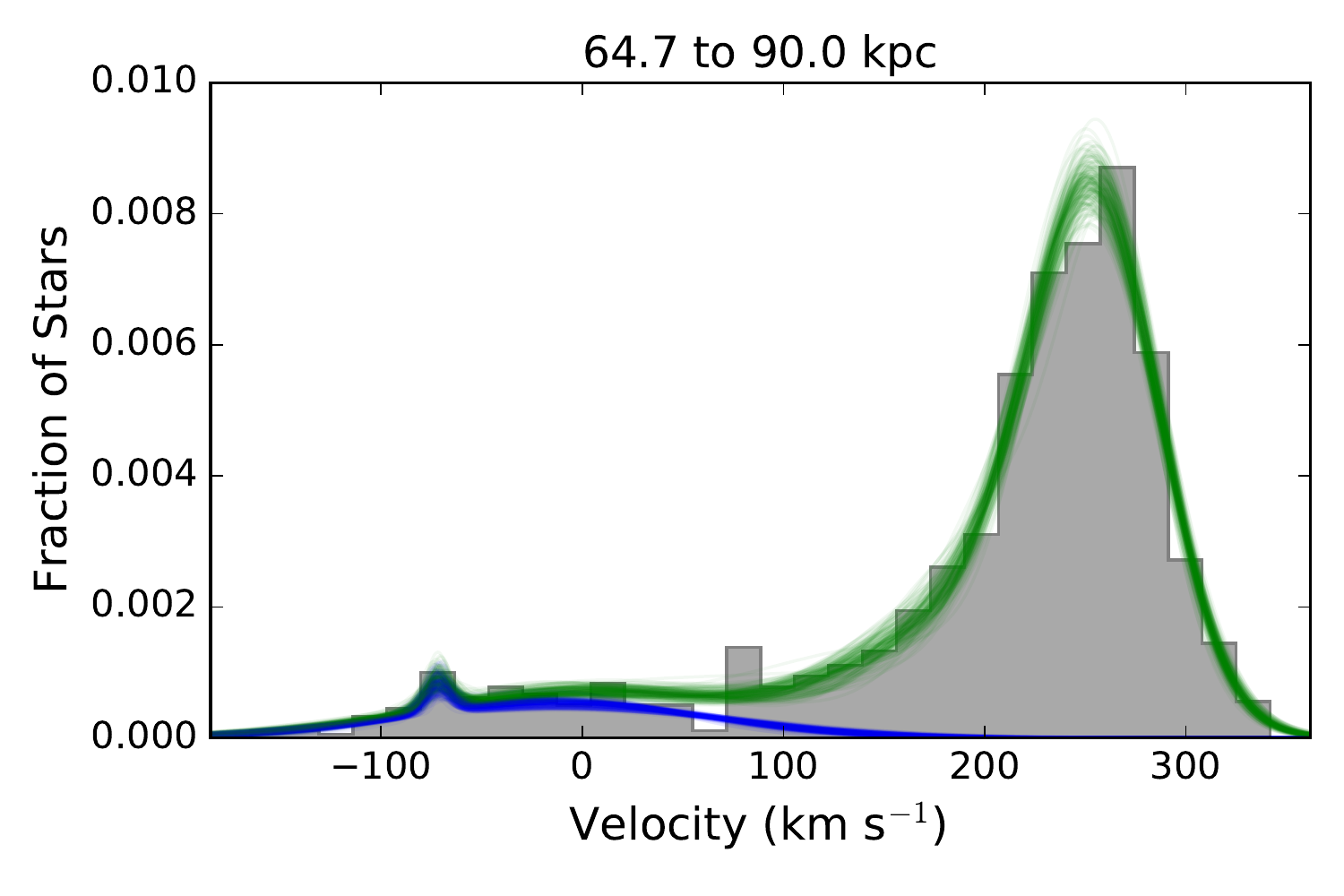}
\includegraphics[width=3.0in]{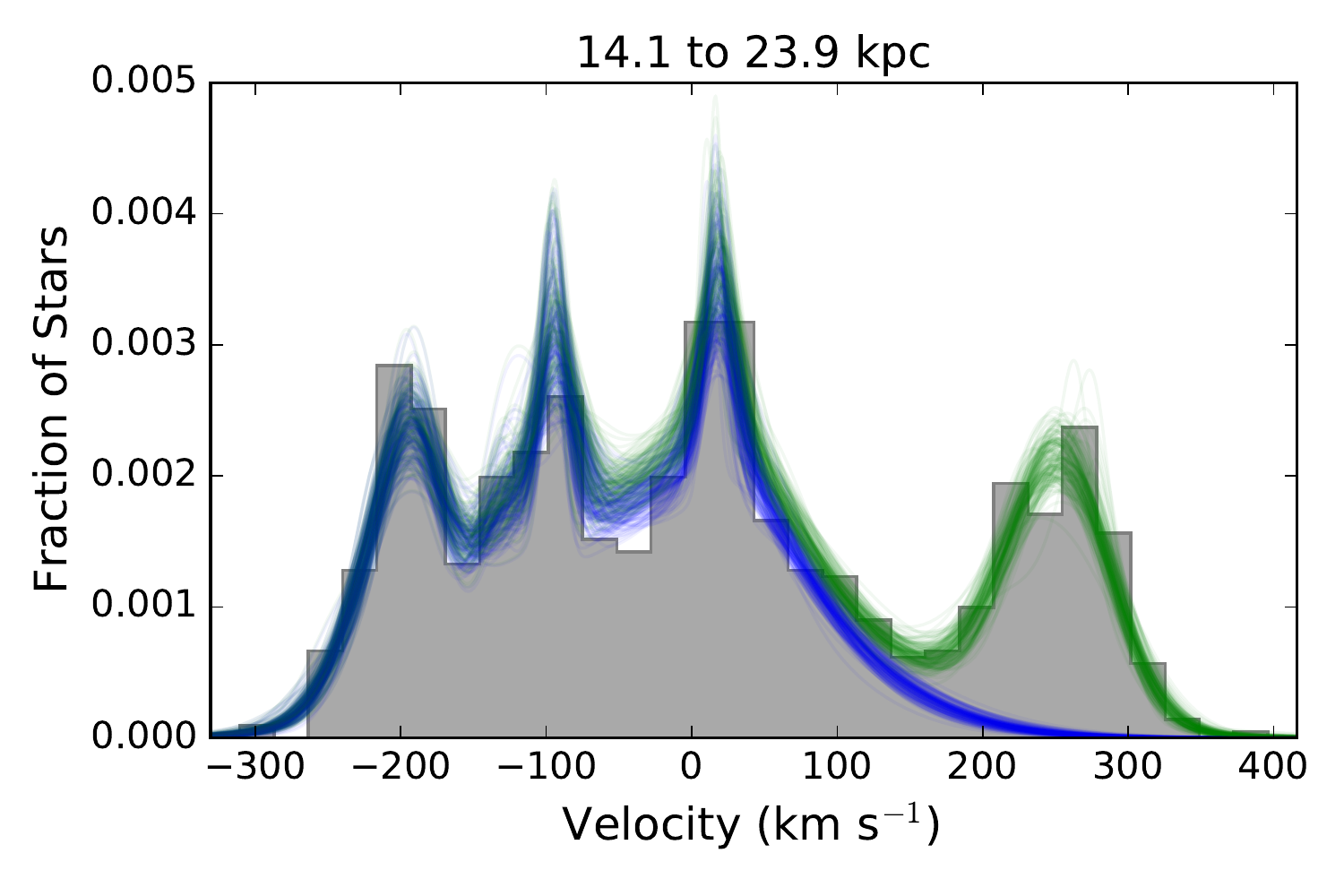}
\includegraphics[width=3.0in]{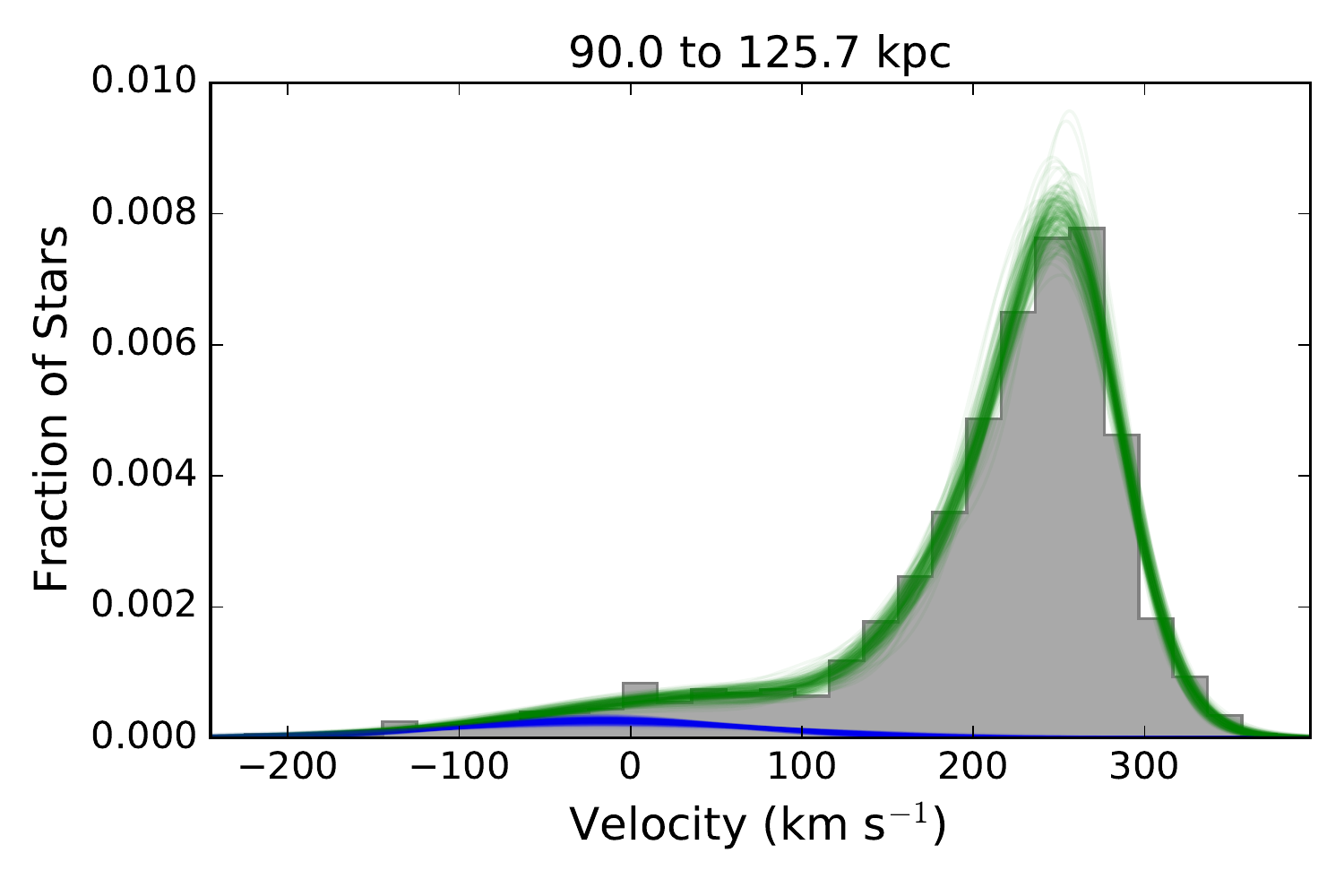}
\includegraphics[width=3.0in]{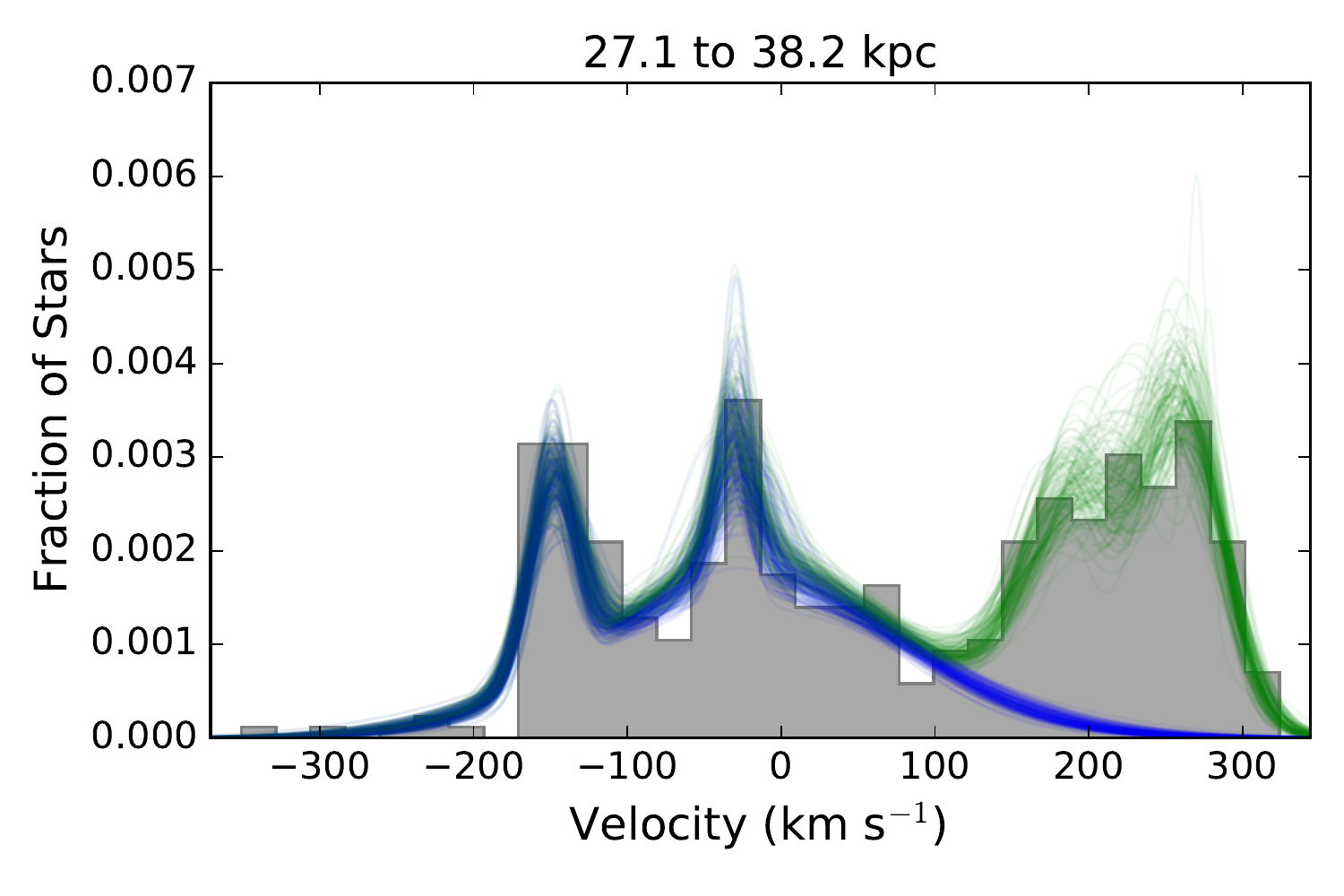}
\includegraphics[width=3.0in]{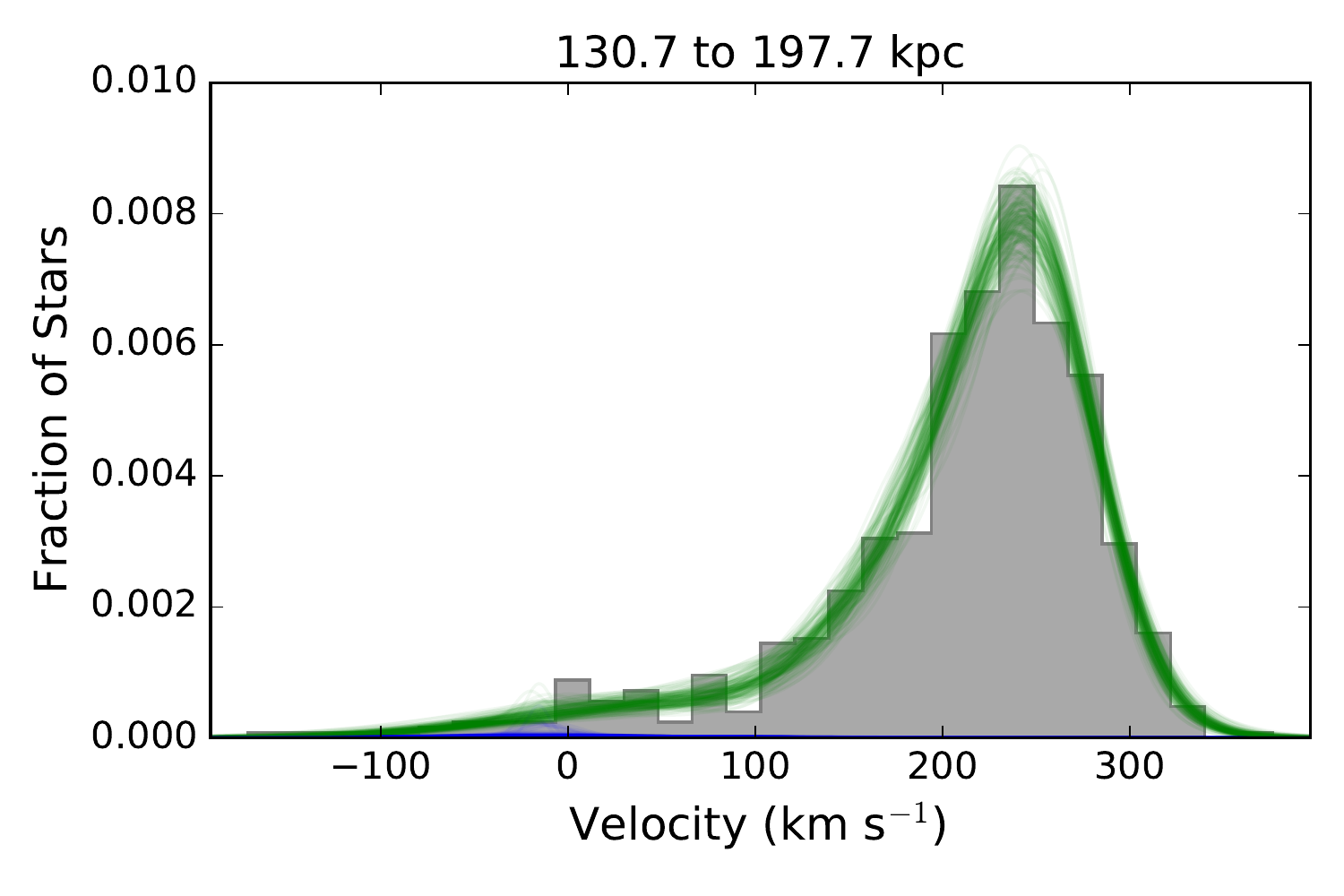}
\includegraphics[width=3.0in]{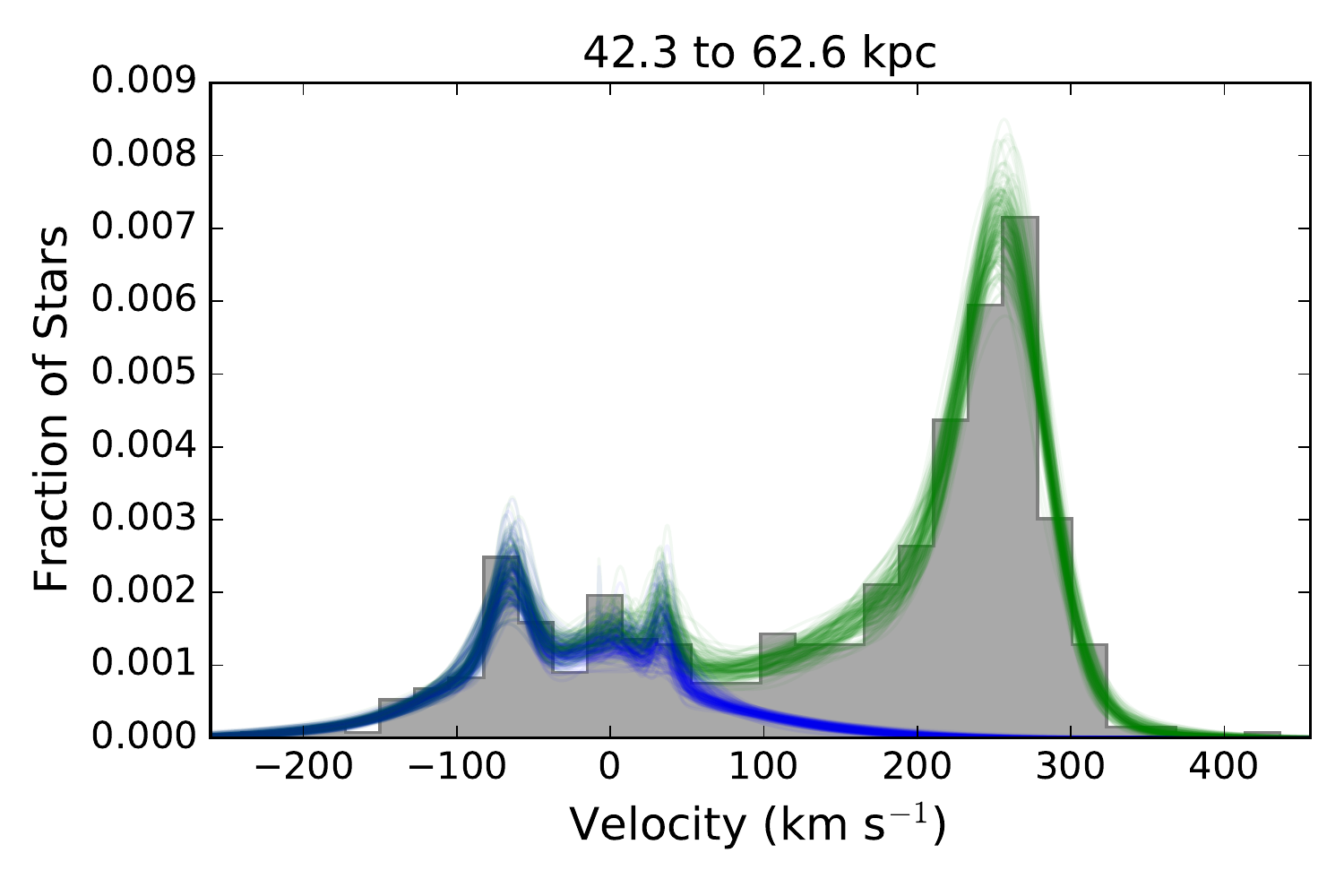}
\caption{
The stellar velocity distribution ($v_{\rm pec}$, Section~\ref{sec:vel_transform}) in each of the seven radial bins.  Overlaid are 150 samples of the parameterized velocity distribution, drawn from the MCMC chain.  The blue curves include only the M31 components, while the green curves include all M31 and MW components.  Observed line of sight velocities have been transformed to the Galactocentric frame, and the bulk motion of M31 has been removed (Section~\ref{sec:mcmc}): a star with no peculiar velocity relative to M31's bulk motion will have $v = 0$~\kms.   
}
\label{fig:mcmc_vel_hist}
\end{figure*}

\begin{deluxetable*}{rrrrrrrrc}
\tablecolumns{9}
\tablewidth{0pc}
\tablecaption{Parameters Describing the Velocity Distribution of M31 Halo Stars.}
\tablehead{\multicolumn{1}{c}{$R_{\rm min}$} & \multicolumn{1}{c}{$R_{\rm max}$} & \multicolumn{1}{c}{No.} & \multicolumn{1}{c}{No. M31} & \multicolumn{1}{c}{No.} & \multicolumn{1}{c}{No.} & \multicolumn{1}{c}{Mean} & \multicolumn{1}{c}{Velocity} & \multicolumn{1}{c}{Fields with} \\
\multicolumn{1}{c}{(kpc)} & \multicolumn{1}{c}{(kpc)} & \multicolumn{1}{c}{Stars} & \multicolumn{1}{c}{Stars\tablenotemark{a}} & \multicolumn{1}{c}{Model} & \multicolumn{1}{c}{MCMC} & \multicolumn{1}{c}{Velocity\tablenotemark{b}} &  \multicolumn{1}{c}{Dispersion\tablenotemark{b}} &  \multicolumn{1}{c}{Substructure}  \\
& & & & \multicolumn{1}{c}{Parameters} & \multicolumn{1}{c}{Chains} & \multicolumn{1}{c}{$\mu_{\rm M31}$} & \multicolumn{1}{c}{$\sigma_{\rm M31}$}  & \\
& &  & & & & \multicolumn{1}{c}{(\kms)} & \multicolumn{1}{c}{(\kms)} & 
}
\startdata
8.2    & 14.1 & 617  & 525 & 23 & 460 & $ -16.5^{+   4.3}_{-   4.5}$ & $108.2^{+  6.8}_{-  6.6}$ & f115, f116, H11 \Tstrut \\
14.1 & 24   & 896  & 697 & 40 & 400 & $ -20.6^{+   4.4}_{-   4.2}$ & $ 98.1^{+  5.3}_{-  5.0}$ & f207, f123, H13s, f115, f135 \Tstrut \\
24   & 40   & 382  & 240 & 19 & 380 & $ -18.9^{+   4.4}_{-   4.4}$ & $ 98.0^{+  7.2}_{-  6.6}$ & a3, and9 \Tstrut \\
40   & 63   & 589  & 202 & 29 & 436 & $ -17.1^{+   4.9}_{-   4.8}$ & $ 93^{+ 11}_{-  10}$ & and1, a13, m4 \Tstrut \\
63   & 90   & 1068 & 247 & 15 & 300 & $ -11.8^{+   4.7}_{-   4.7}$ & $ 76.0^{+  8.6}_{-  7.6}$ & R06A220 \Tstrut \\
90   & 130  & 1013 & 202 & 11 & 300 & $ -17.4^{+   4.8}_{-   4.8}$ & $ 88^{+ 13}_{- 10}$ \Tstrut \\
130  & 200  & 684  & 104 & 11 & 300 & $ -17.2^{+   5.0}_{-   5.0}$ & $ 92^{+ 40}_{- 31}$ \Tstrut \\
\enddata
\tablenotetext{a}{Number of stars more likely to be M31 stars than MW stars based on their \olkhd\ values, without inclusion of velocity in the calculation of the M31/MW probabilities (Section~\ref{sec:likelihoods}).
}
\tablenotetext{b}{Results are the 50th percentile of the one-dimensional posterior probability distribution, marginalized over all other model parameters. The quoted errors are the 16th and 84th percentiles of the posterior probability distribution. Mean velocities are in the Galactocentric frame, with the bulk motion of M31 removed (Section~\ref{sec:mcmc}).
}
\label{tab:results_m31}
\end{deluxetable*}

Figure~\ref{fig:mcmc_vel_hist} shows the velocity distribution of stars in each radial bin, along with a visualization of 150 randomly drawn samples from the MCMC chain overlaid.  As the overall ratio of MW and M31 stars is not a parameter in the mixture model, the fraction of stars with posterior probabilities of M31 membership greater than 0.5 is used as the M31 fraction.  This is done purely to allow the visualization of the models in Figure~\ref{fig:mcmc_vel_hist}.  

\begin{figure}[tb!]
\plotone{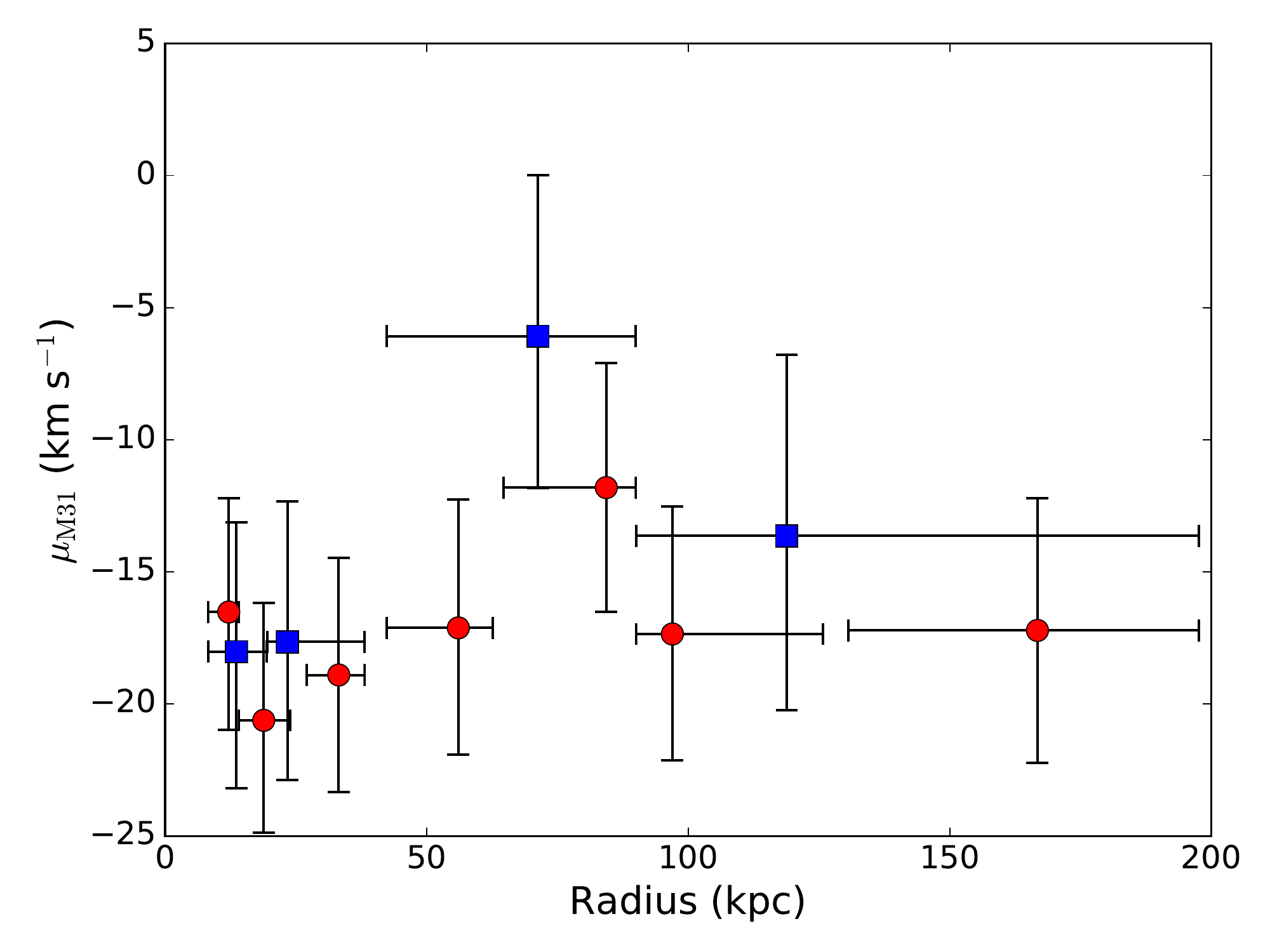}
\plotone{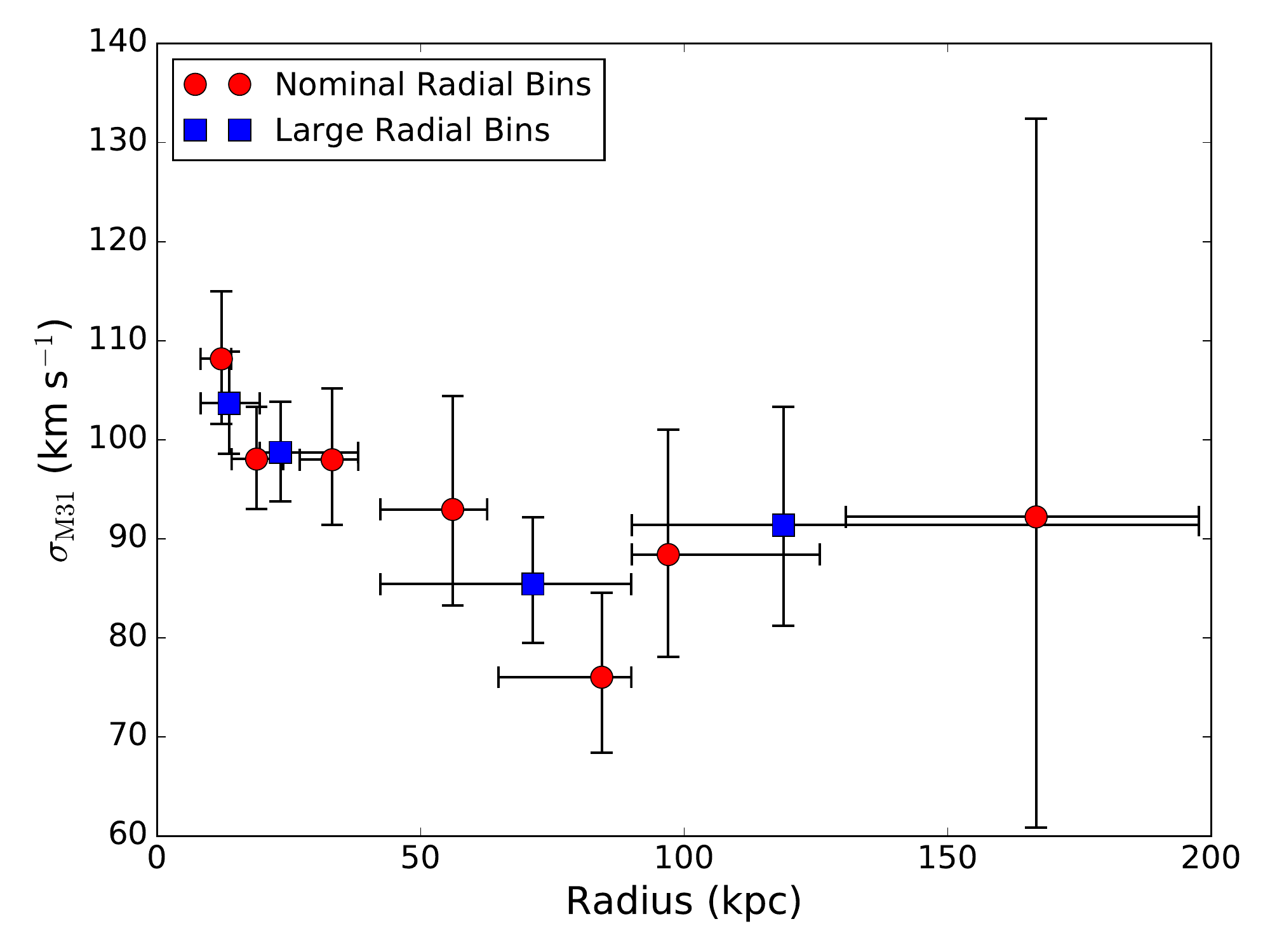}
\caption{
Field independent M31 model parameters as a function of projected radius from M31's center [halo mean velocity ({\it upper panel}; $v_{\rm pec}$, Section~\ref{sec:vel_transform}) and halo velocity dispersion ({\it lower panel})].  Each point is placed at the median projected radius of all stars in the radial bin, with the error bar showing the full range of radii of the stars in the bin.  The model parameter value is the 50th percentile of the marginalized one-dimensional posterior probability distribution, and the errorbar on the parameter value shows the span of the 16th and 84th percentiles.  The velocity dispersion of M31's halo decreases only mildly with radius. In addition to the 7 nominal radial bins, this figure also shows the results when the data is binned into 4 large radial bins (large blue squares); the results are consistent with the smaller radial bins used in the analysis. 
}
\label{fig:mcmc_modelparams_radius}
\end{figure}

The velocity dispersion of the kinematically hot M31 halo component is well constrained in all but the outermost radial bin.  However, only a weak gradient in the M31 halo velocity dispersion with projected radius is seen (Figure~\ref{fig:mcmc_modelparams_radius}).  For each radial bin, Table~\ref{tab:results_m31} lists the 50th percentile of the marginalized one-dimensional posterior probability distribution of the field-independent M31 model parameters, along with the estimated uncertainties based on the 16th and 84th percentiles.  Marginalized one- and two- dimensional posterior probability distributions for all field-independent and field-dependent model parameters are shown in the Appendix.  

The uncertainties listed in Table~\ref{tab:results_m31} are based solely on the results of the MCMC analysis, and do not include velocity measurement (Section~\ref{sec:spec_observations}) or transformation (Section~\ref{sec:vel_transform}) uncertainties.  Assuming the measured velocity dispersion is a combination of the intrinsic velocity dispersion of halo stars, the velocity measurement uncertainties, and the uncertainties in the velocity transformations (all added in quadrature), we can estimate the impact of these uncertainties on our measurement of the velocity dispersion.  For most radial bins, the typical values of these uncertainties will inflate the measured dispersion by only a few tenths of a~\kms, which is only $\sim 10$\% of the estimated {\it uncertainties} from the MCMC method.  The uncertainties in the transformations are expected to lead to larger effects in the outer radial bins than in the inner radial bins, as fields in the outer bins are often separated by many degrees on the sky (Figure~\ref{fig:roadmap1}).  The maximum effect from the velocity transformations is estimated to introduce an uncertainty on the order of $8$\,--\,$10$~\kms; this would be the maximum effect for the outermost radial bin.  Nevertheless, we can use this to provide a conservative upper bound: using these values, we find that the measured dispersion is inflated by only $\sim 1$\,--\,1.3~\kms, which is $<2$\% of the measured dispersion in the three outermost radial bins, and is still a small fraction ($\lesssim 15$\%) of the estimated {\it uncertainties} from the MCMC method in these bins.

\subsection{Parameterization of the Velocity Dispersion of M31's Halo with Radius}\label{sec:powerlaw_dispersion}

\begin{figure}[tb!]
\plotone{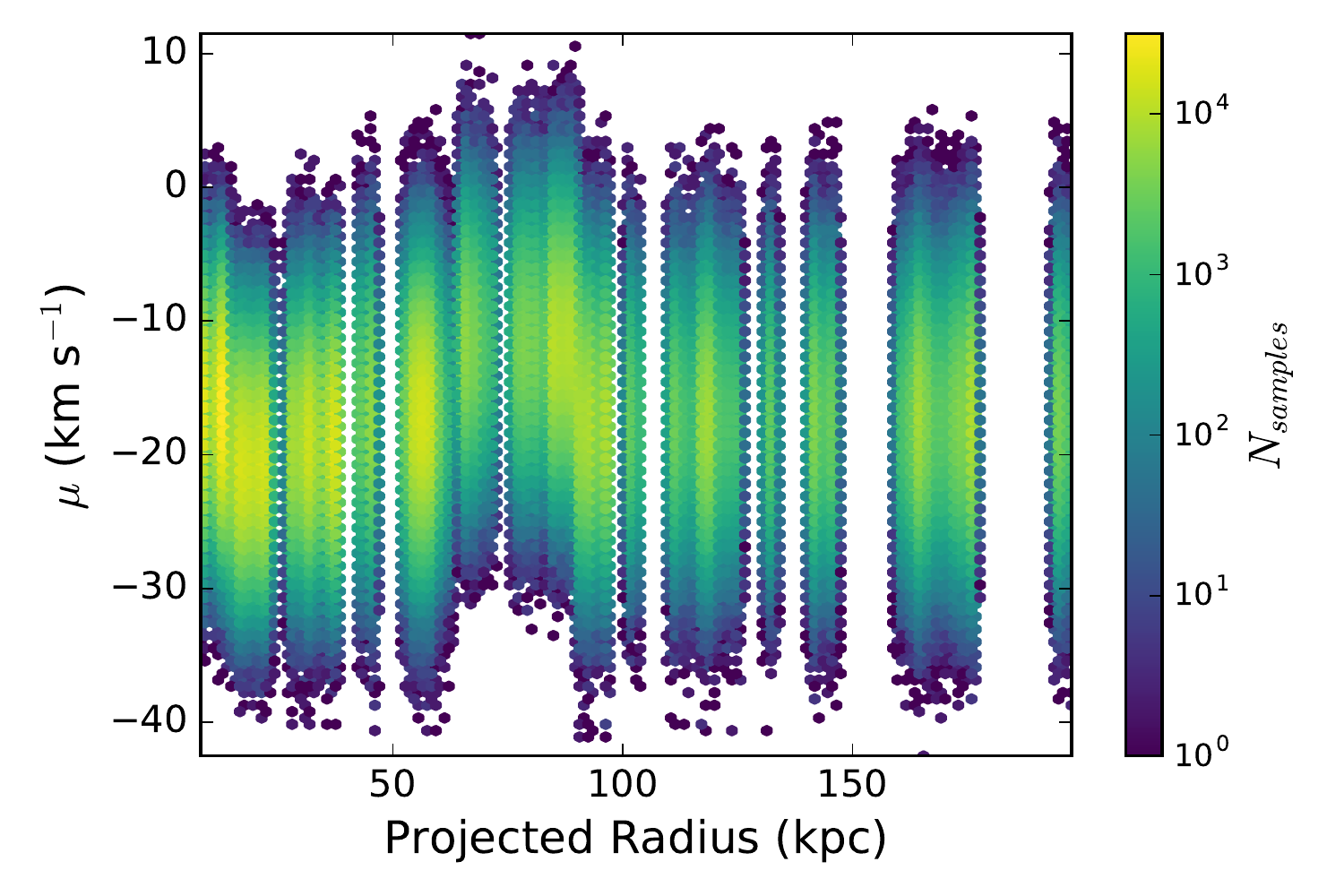}
\plotone{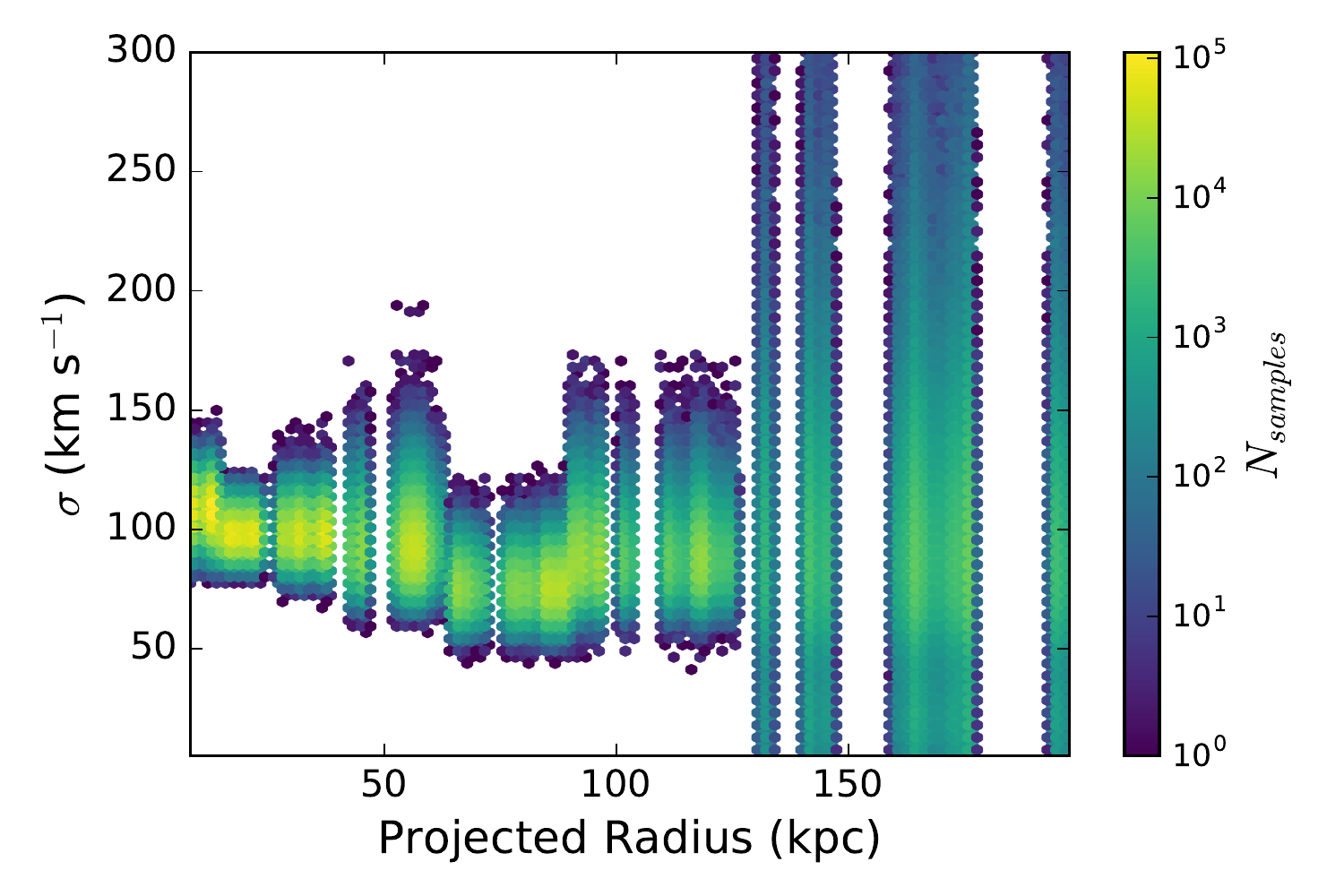}
\caption{
A representation of the probability distribution of the halo mean velocity ({\it upper panel}; $v_{\rm pec}$, Section~\ref{sec:vel_transform}) and velocity dispersion ({\it lower panel}) of M31's stellar halo as a function of projected radius.  The M31 halo velocity mean and dispersion from each MCMC sample was randomly assigned to the radius of a star in that radial bin.    
The resulting data were then randomly sampled, drawing a total of $1.5\times10^6$ samples for each radial bin.
Purple represents the regions of low probability density while yellow represents the regions of highest probability density. White regions of the plot indicate where the probability density is zero, as well as where there are gaps in the radial coverage of the spectroscopic dataset.   
}
\label{fig:mcmc_hexbin}
\end{figure}

To parameterize the change in the velocity dispersion of M31's halo with projected radius, we leverage the one-dimensional posterior probability distributions, i.e., those that are marginalized over all other model parameters.  A visualization of the posterior probability distributions for the M31 halo velocity dispersion as a function of radius is shown in Figure~\ref{fig:mcmc_hexbin}.   In this representation, it is very clear that the halo dispersion in outermost radial bins is poorly constrained: the probability distribution extends over a large range of values, and does not have a prominent peak.  

We randomly sample the marginalized, one-dimensional posterior probability distributions of the M31 halo dispersion in each radial bin to parameterize the change in the velocity dispersion of M31's halo with radius.  We fit a power-law of the form
\begin{equation}
\sigma = \sigma_0\,(R_{\rm med}/R_{\rm 0})^{-\gamma},
\end{equation} 
where the power-law is normalized at the projected radius $R_{\rm 0}$.  We set $R_{\rm 0}=30$~kpc, which is the scale radius chosen for the analysis of the M31 GC dispersion profile in \citet{veljanoski2014}.  $R_{\rm med}$ is the median value of the projected radii of all stars in each radial bin.  

\begin{figure}[tb!]
\plotone{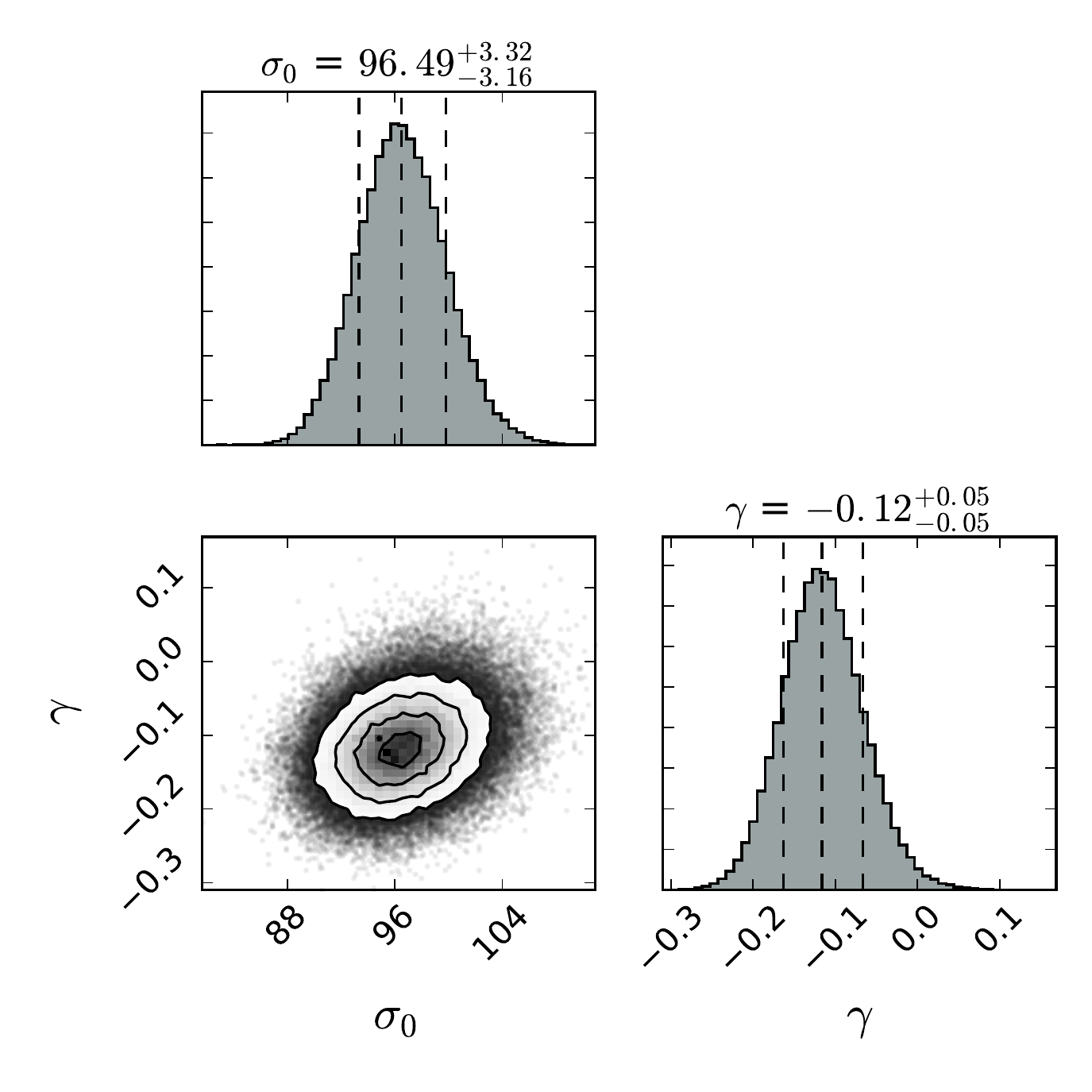}
\caption{
One- and two-dimensional distributions of the best-fit power-law parameters describing the change in the velocity dispersion of M31's halo with projected radius.  Each point comes from fitting a power-law to random draws from the M31 halo velocity dispersion posterior probability distributions in all but the outermost radial bin (Figure~\ref{fig:mcmc_hexbin}).  
}
\label{fig:powerlaw_params}
\end{figure}

\begin{figure}[tb!]
\plotone{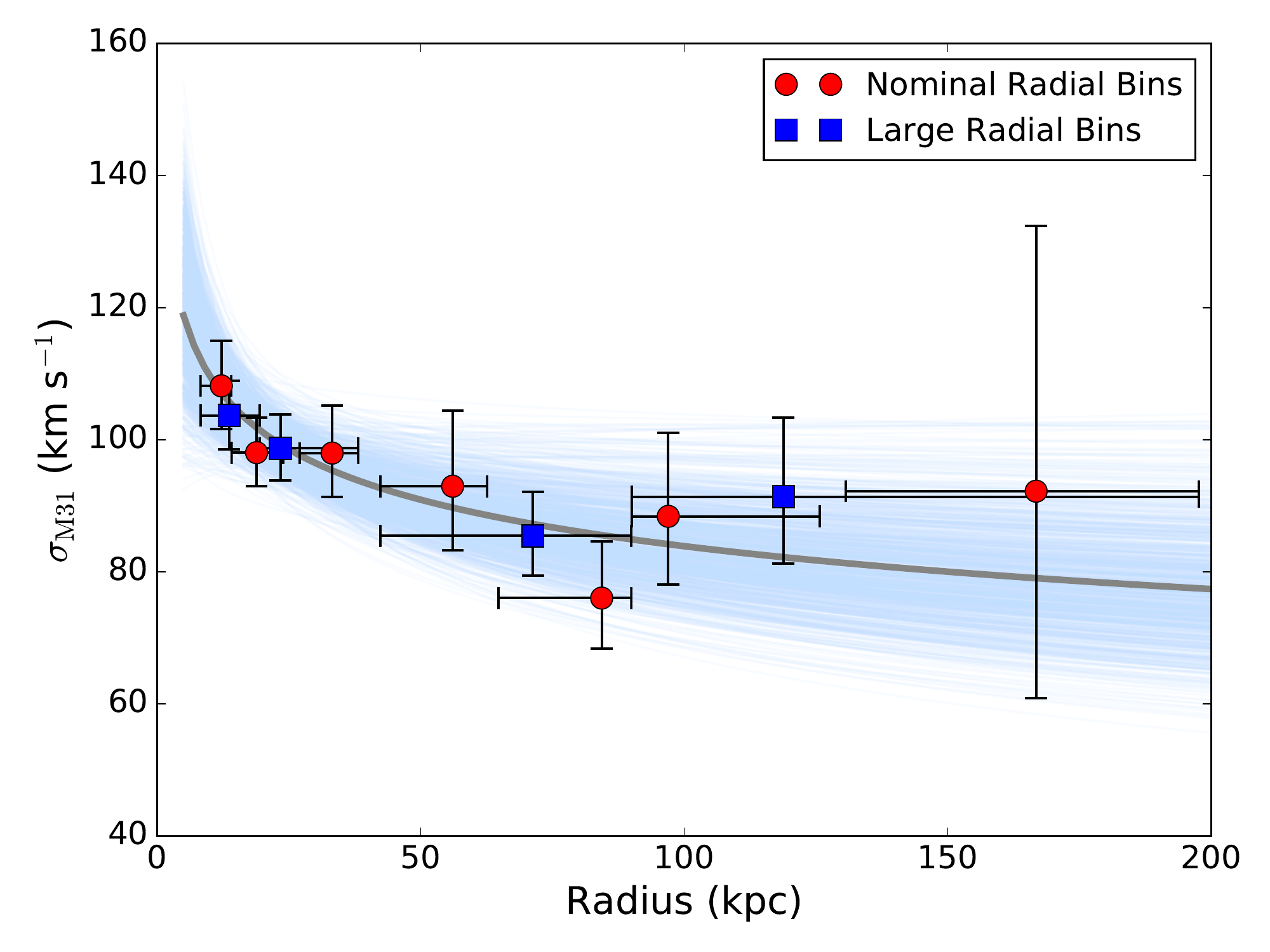}
\caption{
Same as the lower panel of Figure~\ref{fig:mcmc_modelparams_radius}, with the power-law fits of the velocity dispersion of M31's halo as a function of projected radius overlaid.  The light blue curves show a subset of the power-law fits to 10000 random draws from the marginalized one-dimensional posterior distribution functions for the M31 halo velocity dispersion in each of the first six radial bins. The dark gray curve shows a power-law composed of the 50th percentile values of the normalization and slope distributions.  
}
\label{fig:powerlaw_fits}
\end{figure}

We make 10,000 random draws (with replacement) from the one-dimensional posterior probability distributions for the M31 halo velocity dispersion in the first 6 radial bins, and perform a least-squares fit on each draw to determine the power-law parameters $\sigma_0$ and $\gamma$.  The resulting distribution functions of $\sigma_0$ and $\gamma$ provide estimates of the most likely values and uncertainty on each parameter (Figure~\ref{fig:powerlaw_params}).  The 50th percentile values, with uncertainties estimated from the 16th and 84th percentiles of the distributions, are $\gamma = -0.12\pm 0.05$ and $\sigma_0 = 96.5^{+3.3}_{-3.2}$~\kms.  A random drawing of 1000 of these power-law fits is shown in Figure~\ref{fig:powerlaw_fits}.  This parameterization confirms the likely existence of a weak gradient in the velocity dispersion with projected radius:  only a small percentage of the fits are consistent with a flat or increasing dispersion with radius.   

The uncertainties in the power-law slope are estimated based on the method, and do not include formal propagation of uncertainties such as those in the velocity measurements or transformations.  However, as discussed in Section~\ref{sec:mcmc_results}, these uncertainties are estimated to have a minimal effect on the measured velocity dispersions, and this will propagate to a minimal effect on the power-law slope uncertainties.  We discuss the effect of modeling choices on the power-law slope below (Section~\ref{sec:results_assumptions}).

\subsection{Sensitivity to Modeling Choices}\label{sec:results_assumptions}

We made several choices regarding both the data and the modeling with the potential to influence the results.  We performed alternative fits to explore the impact of these choices on the measured dispersion profile.   

We instituted a strong prior probability that stars significantly bluer than the most metal-poor red giant branch isochrone are MW stars along the line of sight, rather than M31 red giant branch stars, by assigning them very negative \olkhd\ values of $-5$.  This places them on the tail end of the distribution of MW stars in \olkhd\ space, and is consistent with the approach taken in previous papers.  However, we also ran the above analysis after assigning less negative \olkhd\ values to these stars, based on the \olkhd\ distribution of stars classified as MW stars: they were assigned an \olkhd\ value one sigma below the mean \olkhd\ of all MW stars.  This still places them firmly as having a high probability of being MW stars, but is a less draconian \olkhd\ assignment.  This resulted in no appreciable difference in the power-law parameters found for the stellar halo dispersion profile ($\sigma_0 = 96.5^{+3.3}_{-3.2}$~\kms, $\gamma = -0.12\pm 0.05$). 

We also explored fits with variations on the M31 halo mean velocity parameter, including a delta prior on the mean velocity at M31's systemic velocity (0~\kms).  This also made little difference in the power-law parameters, resulting in values consistent with those found in the nominal fit ($\sigma_0 = 100.0^{+3.4}_{-3.3}$~\kms, $\gamma = -0.09\pm 0.05$).  Using a normal prior on the M31 halo mean velocity parameter, but centered at M31's systemic velocity and with a width of $\sigma=10$\kms, resulted in a slightly flatter halo velocity dispersion profile, but with power-law parameters statistically consistent (within one sigma) with the nominal fit ($\sigma_0 = 99.3^{+3.5}_{-3.3}$~\kms, $\gamma = -0.07\pm 0.05$).

Relaxing the boundary constraints on the substructure prior (allowing values within $\pm10$ times the uncertainties on the published best-fit values) also did not result in a significant change in the power-law parameters ($\sigma_0 = 96.3^{+3.4}_{-3.2}$ ~\kms, $\gamma = -0.11\pm 0.05$). 

Finally, we have measured the power-law based on projected radii calculated using a distance modulus to M31 of $24.47^m$ (Section~\ref{sec:intro}).  If instead we use a distance modulus of $24.38^m$, this results in an $\sim 4$\% difference in the estimated projected radii in kpc.  The primary effect of assuming a different distance modulus on the estimated power-law slope is easy to compute by simply recalculating the power-law after recomputing the median projected radius for each bin.  The resulting power-law parameters ($\sigma_0 = 96.1\pm 3.3$~\kms, $\gamma = -0.12\pm 0.05$) are fully consistent with the nominal fit.

\subsection{Discussion}\label{sec:results_discussion}

Only one other measurement of the dispersion profile of stars in M31's halo over a large range of radii has been attempted, by \citet[Section~\ref{sec:intro}]{chapman2006}. 
The \citeauthor{chapman2006} fields were primarily within 40~kpc of M31's center and almost exclusively located on M31's disk. This required the application of strict window functions to the velocity distribution to remove M31 disk stars and MW stars.  This early analysis could not account for recent discoveries of rotation in M31's inner spheroid \citep[$v_{\rm rot}\sim 50$~\kms;][]{dorman2012} and the relatively large line of sight velocity dispersion of disk RGB stars \citep[$\sigma_{v} = 90$\kms;][]{dorman2015}, both of which will affect the interpretation of the relative mix of stellar populations assumed to fall within the \citeauthor{chapman2006} velocity window functions.  Nevertheless, the results are consistent with the recent measurements of \citet{veljanoski2014}, who measured the dispersion profile of M31 halo globular clusters (GCs) over a large radial range.  The study by \citet{veljanoski2014} sampled projected radii of 25\,--\,145~kpc, and while it includes many less tracer objects (72 beyond 30~kpc), there are fewer populations from which these objects are drawn: all are firmly in the M31 system.  After accounting for the rotation observed in the GC system, \citeauthor{veljanoski2014} measured a power-law dispersion profile with both a larger value of $\sigma_0$ and a steeper power-law slope, $-0.45$, with only a 1\% posterior probability of $\gamma = 0$.  While \citeauthor{veljanoski2014} find clear evidence of coherent velocity patterns amongst groups of GCs that are spatially correlated with tidal debris features, the analysis did not account for this.  The ability to do so is an advantage of using much more abundant halo RGB stars as tracers of the halo.   
 
Measurements of the velocity dispersion of MW halo stars have found a sharp decrease in the velocity dispersion in the inner regions of the MW halo from $\sim 150$ to 100\kms\ over the inner 20~kpc, settling to a relatively flat dispersion profile at large radii, with measurements of $\sigma_{r} \sim 100$~\kms\ from $\sim 20$\,--\,80~kpc \citep{battaglia2005,xue2008,bond2010,brown2010,kafle2012,deason2012,kafle2014}.  A graphical summary of the MW's velocity dispersion profile can be found in \citet{bland-hawthorn2016}.  

M31 also appears to have a sharply decreasing velocity dispersion in the inner regions (\citet{dorman2012} measured a velocity dispersion of $140$~\kms\ at \rproj$=7~kpc$ in M31), followed by a relatively flat dispersion to large radii.  However, the reader should note that the MW profiles measure primarily the radial velocity of MW halo stars. Given the large spread of the SPLASH spectroscopic fields on the sky, the M31 line of sight velocity dispersion profile measures a combination of the stars' tangential and radial velocities in the M31 coordinate frame, with the relative contributions changing with field position.  

To date, there have been few analyses of the velocity dispersion profiles of MW- or M31-like stellar halos in $\Lambda$CDM simulations \citep[one example is][]{abadi2006}.   The stellar density profiles, substructure characteristics, and metallicity profiles of the M31 and MW halos have proven to be useful constraints and checks on $\Lambda$CDM simulations of stellar halo formation, and comparisons of the simulations to observations have provided insight into the physical origins of the stellar halos of M31 and the MW \citep[e.g.,][]{font2006,font2008,zolotov2010,font2011,mccarthy2012,gilbert2012,gilbert2014}.  We expect future comparisons of the observed MW and M31 velocity dispersion profiles with simulated halos will yield further insights into the origins of stellar halos.

\section{Conclusion} \label{sec:conc}

We modeled the velocity distribution of more than 5000 stars observed in M31 halo fields as part of the SPLASH survey, including all major MW and M31 components in the halo fields.  Photometric and spectroscopic information on likely MW or M31 membership for each star was incorporated into the Gaussian mixture model as a prior probability.  Tidal debris features in M31 halo fields were included in the model, and the marginalized posterior distributions for each are presented in the appendix.  

Marginalizing over all model parameters, we parameterized the dispersion of stars in M31's halo as a function of projected radius.  The dispersion of M31's halo stars is found to decrease only mildly with projected radius, over a radial range of 9 to 100~kpc.  Our measurement finds a significantly flatter profile with radius than that measured for M31's globular cluster population.  The measurement of the velocity dispersion profile is the first step towards using halo stars as tracers of M31's mass.  In future work, the dispersion M31's halo stars will be used to model M31's total mass distribution.  

\acknowledgments
The authors recognize and acknowledge the very significant cultural role and reverence that the summit of Mauna Kea has always had within the indigenous Hawaiian community. We are most fortunate to have the opportunity to conduct observations from this mountain.  The authors thank Laura Watkins for useful discussions during the preparation of the manuscript.
This research made use of Astropy, a community-developed core Python package for Astronomy  \citep{astropy2013}\footnote{http://www.astropy.org}, as well as the following python packages: {\tt astroML} \citep{astroML,astroMLtext}, {\tt emcee}  \citep{foreman-mackey_emcee, emcee-sw}, and {\tt corner} \citep{corner-sw_v1, corner}. 
  
Support for this work was provided by NASA through a Giacconi Fellowship and Hubble Fellowship 
grant 51316.01 awarded to E.J.T. by the Space Telescope 
Science Institute, which is operated by the Association of Universities for 
Research in Astronomy, Inc., for NASA, under contract NAS 5-26555.
K.M.G., P.G., J.S.B., S.R.M., and R.L.B. acknowledge support from collaborative NSF grants AST-1614569, AST-1412648, AST-1010039, AST-1009973, AST-1009882, and AST-0607726. This project was also supported by NSF grants AST03-07842, AST03-07851, AST06-07726, AST08-07945, and AST10-09882, NASA grant HST-GO-12105.03 through STScI, NASA/JPL contract 1228235, the David and Lucile Packard Foundation, and the F. H. Levinson Fund of the Peninsula Community Foundation (S.R.M., R.J.P., and R.L.B.). E.N.K acknowledges support from the Southern California Center for Galaxy Evolution, a multicampus research program funded by the University of California Office of Research, and partial support from NSF grant AST-1009973. R.L.B. acknowledges receipt of the Mark C. Pirrung Family Graduate Fellowship from the Jefferson Scholars Foundation and a Fellowship Enhancement for Outstanding Doctoral Candidates from the Office of the Vice President of Research at the University of Virginia. M.T. acknowledges support from Grant-in-Aid for Scientific Research (25800098) of the Ministry of
Education, Culture, Sports, Science, and Technology of Japan.  The analysis pipeline used to reduce the DEIMOS data was developed at UC Berkeley with support from NSF grant AST-0071048.

\appendix
\section{Posterior Probability Distributions of the Field-Independent Parameters}\label{app:findep_params}

\begin{figure}
\plotone{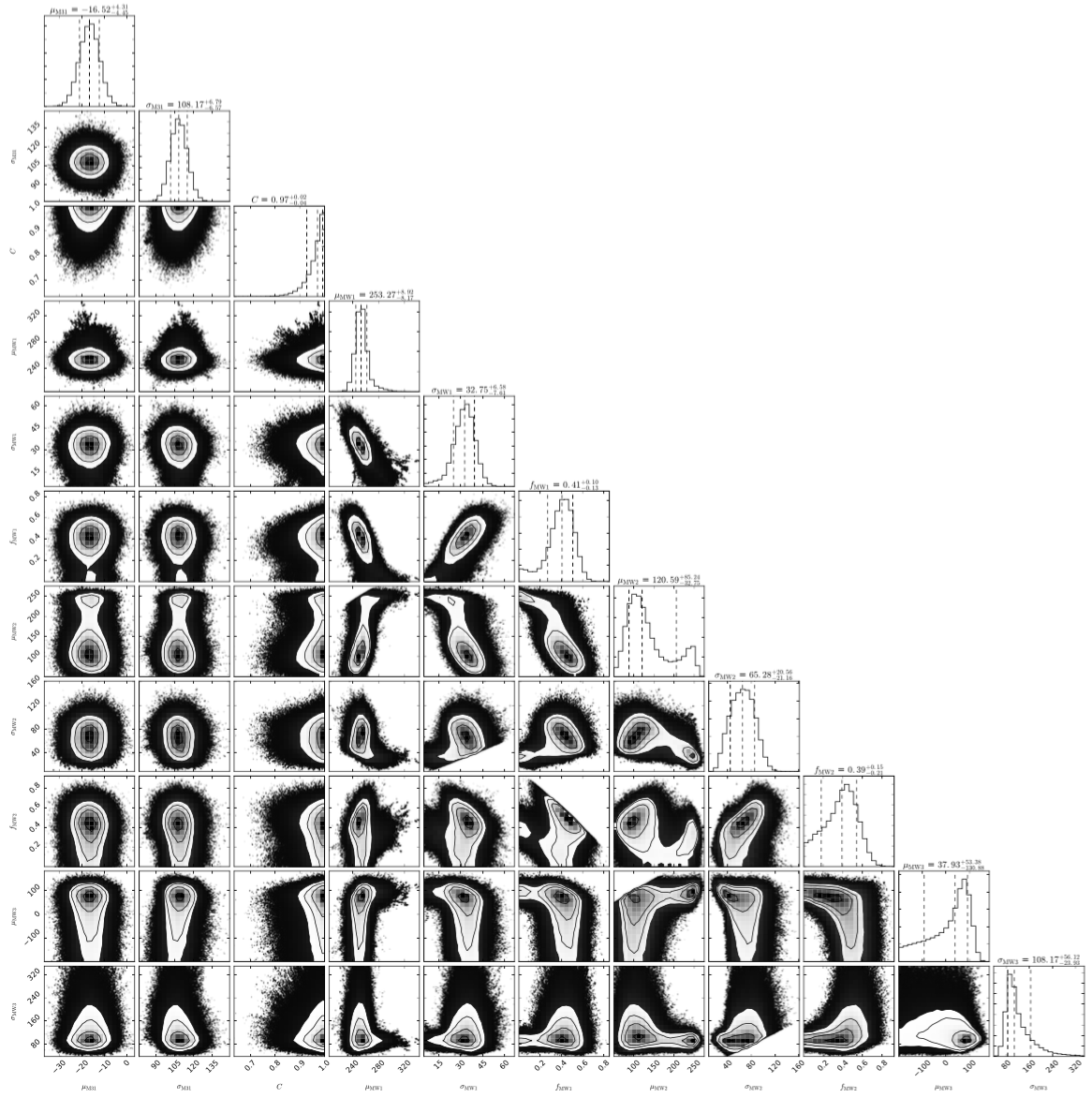}
\caption{Marginalized one- and two-dimensional posterior probability distributions for the field-independent model parameters (Section~A\ref{app:findep_params}) in the first radial bin ($8.2\le$\,\rproj\,$<14.1$~kpc).   The distributions are computed using the values of the model parameters in the half of each of the MCMC chains.  The 16th, 50th, and 84th percentiles of the marginalized one-dimensional distributions are marked by dashed lines, and these values are 
listed for each parameter at the top of each column.  In the two-dimensional distribution panels, contours are displayed at levels of 0.5, 1, 1.5, and 2$\sigma$. This figure and those that follow were created using the open source python package {\tt corner} \citep{corner-sw_v1, corner}.
}
\label{fig:primary_params_corner_start}
\end{figure}

Figures \ref{fig:primary_params_corner_start} through \ref{fig:primary_params_corner_end} show the marginalized one- and two-dimensional posterior probability distributions for the field-independent model parameters in each radial bin.  The model parameters describing the M31 halo component are well constrained in all but the outermost radial bin, where the number of M31 stars in the dataset is very low.  Conversely, in the inner radial bins, the number of MW stars in the dataset is small, and the constraints on the model parameters describing the MW components is poor.  The constraints on the model parameters describing the MW components increase in the fits at large projected radius, where the dataset contains significantly more MW stars.  The results for the MW component model parameters reflect only the distribution of spectroscopic targets, and are dependent on the spectroscopic target selection functions.  Thus, they should not be used as general measurements of the physical properties of the MW velocity distribution towards M31.

\begin{figure}
\plotone{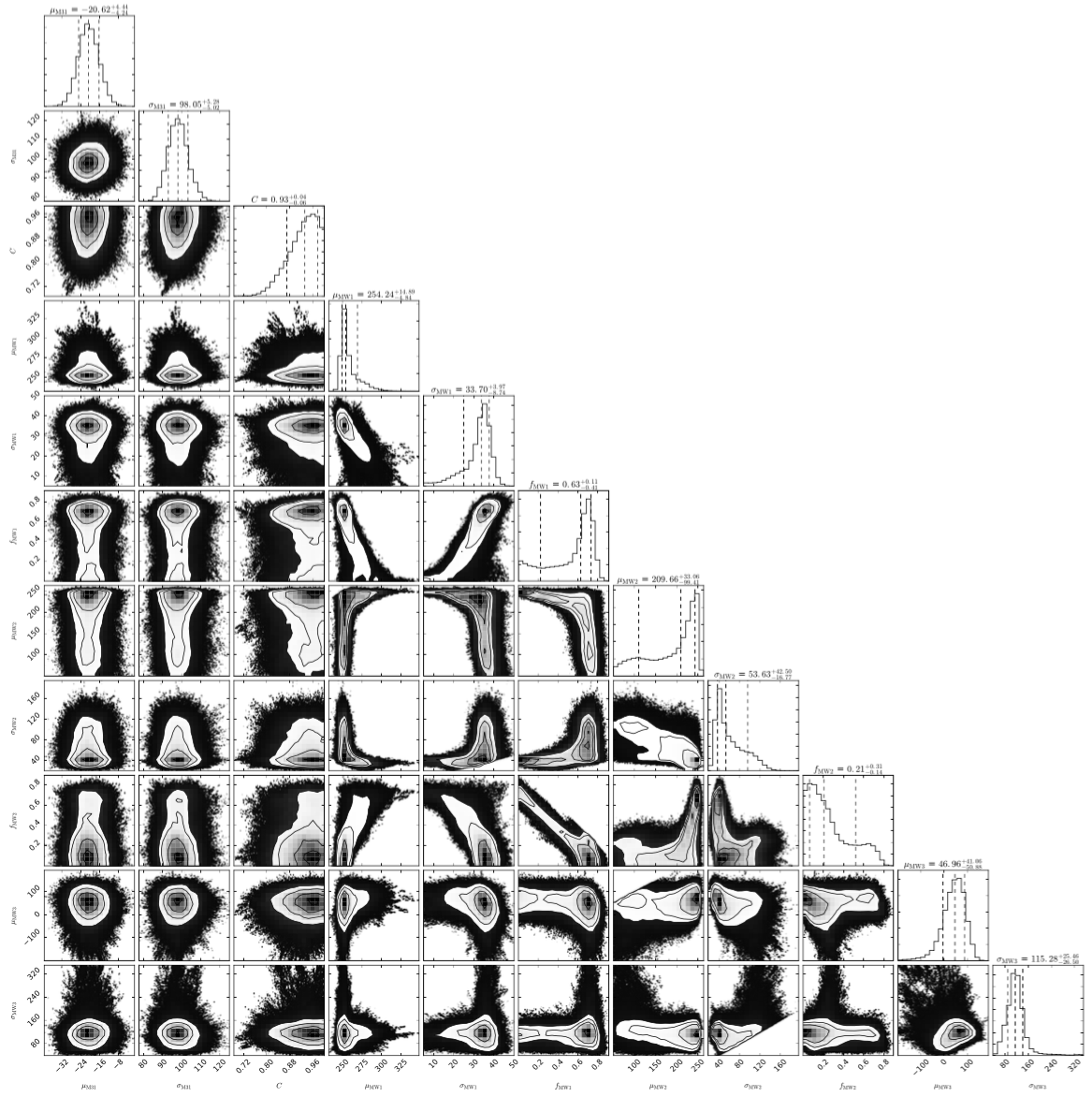}
\caption{Same as Figure~\ref{fig:primary_params_corner_start} for the second radial bin ($14.1\le$\,\rproj\,$<24$~kpc).}
\end{figure}

\begin{figure}
\plotone{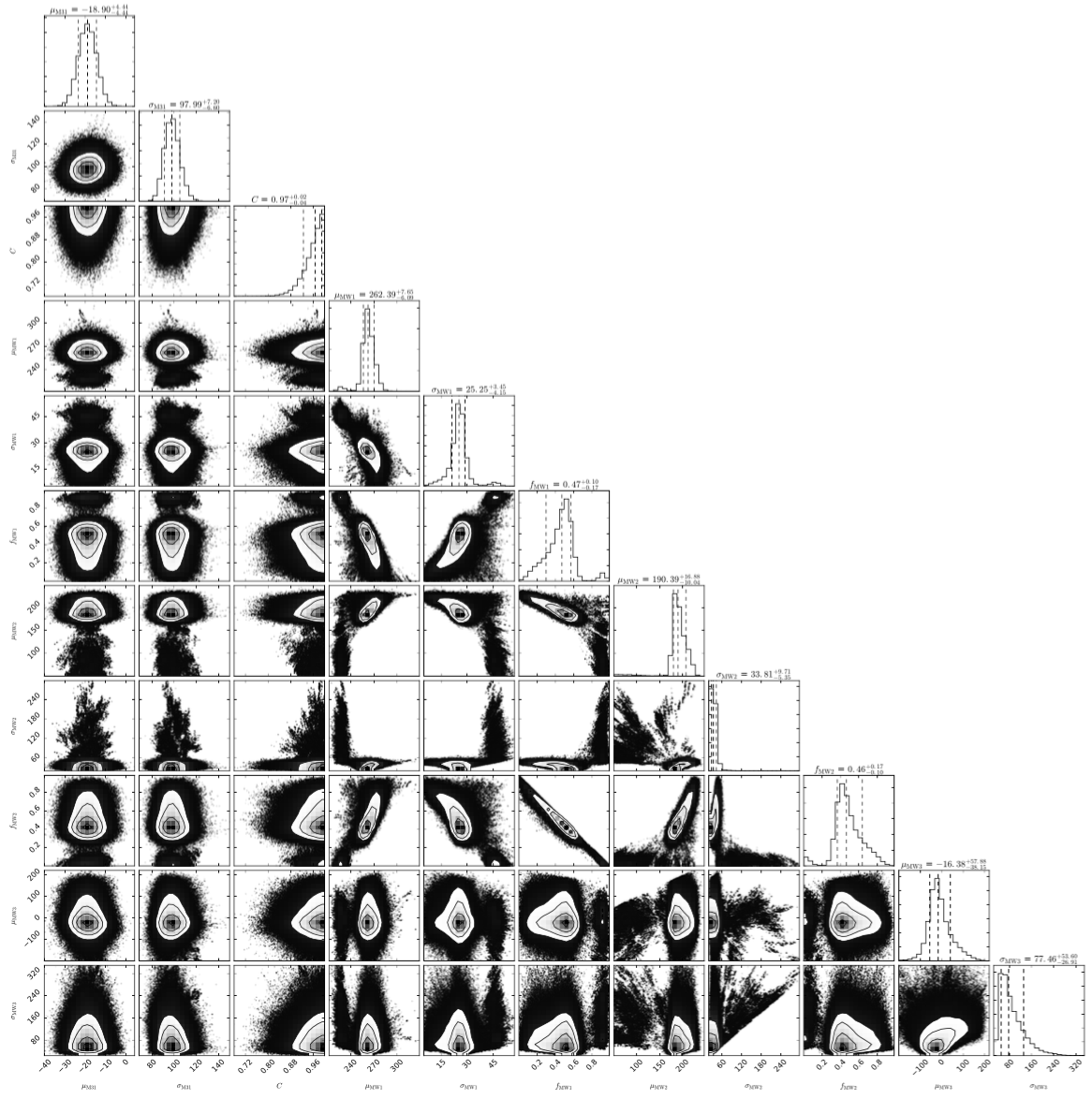}
\caption{Same as Figure~\ref{fig:primary_params_corner_start} for the third radial bin ($24\le$\,\rproj\,$<40$~kpc).}
\end{figure}

\begin{figure}
\plotone{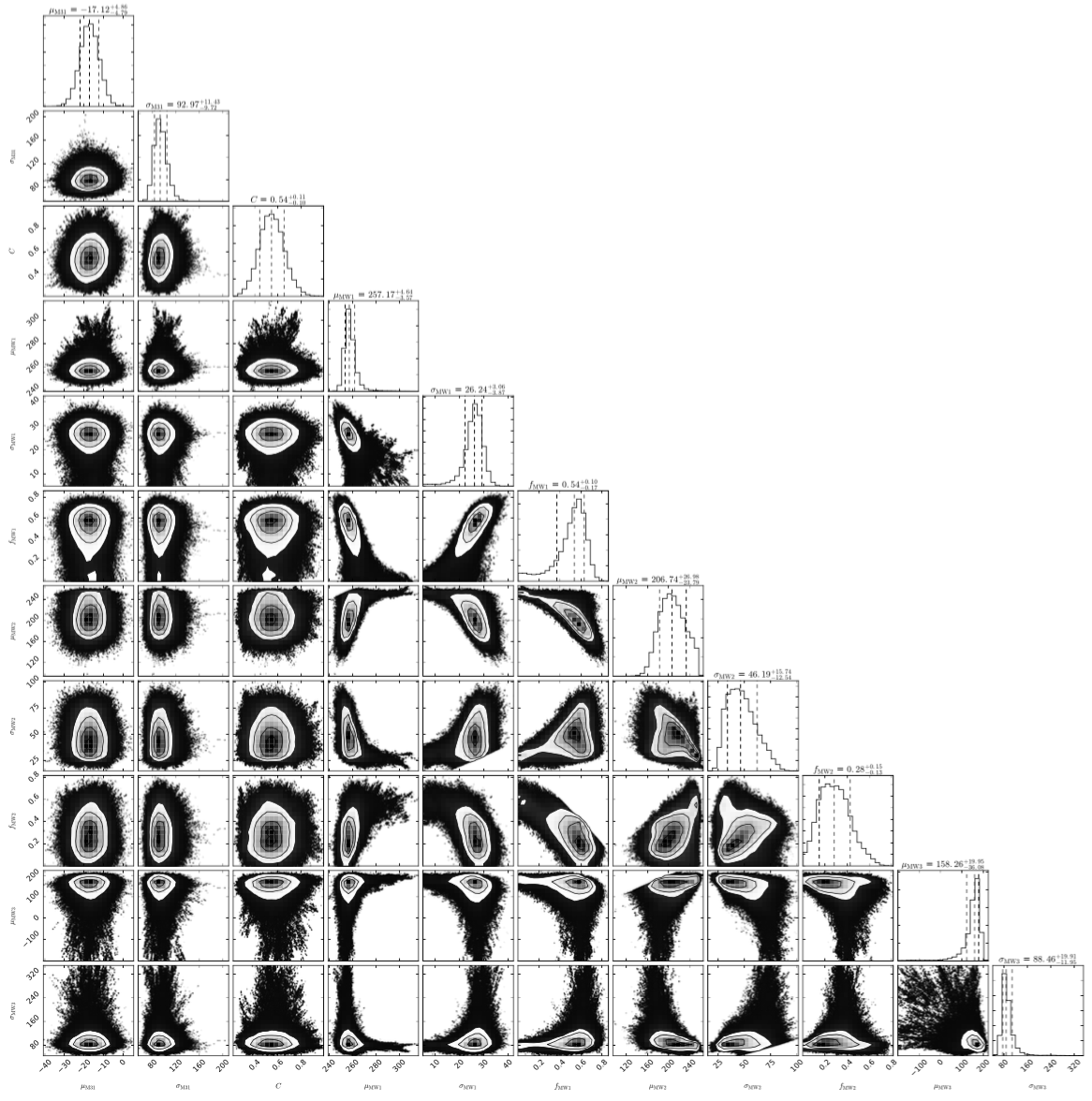}
\caption{Same as Figure~\ref{fig:primary_params_corner_start} for the fourth radial bin ($40\le$\,\rproj\,$<63$~kpc).}
\end{figure}

\begin{figure}
\plotone{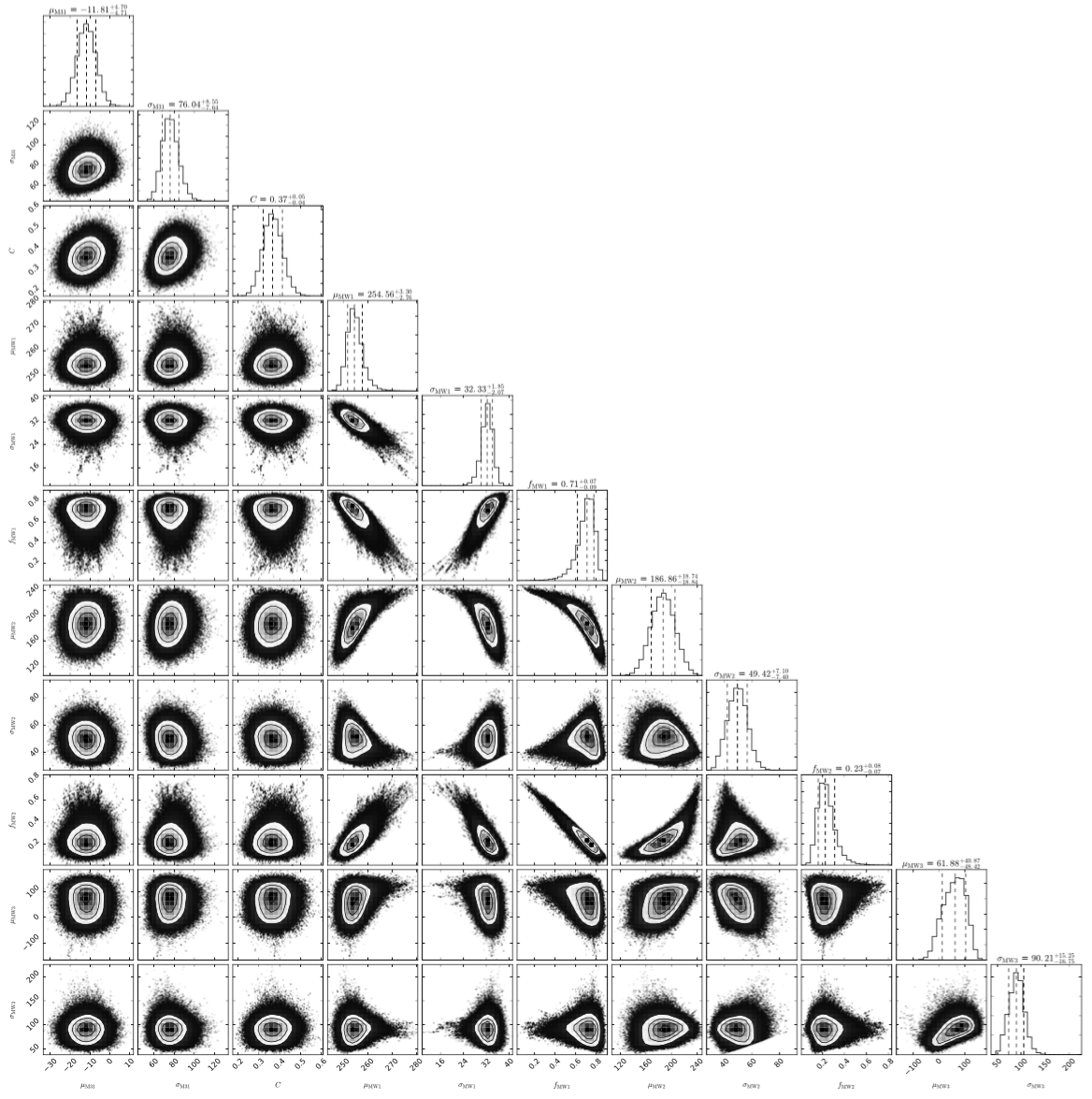}
\caption{Same as Figure~\ref{fig:primary_params_corner_start} for the fifth radial bin ($63\le$\,\rproj\,$<90$~kpc).}
\end{figure}

\begin{figure}
\plotone{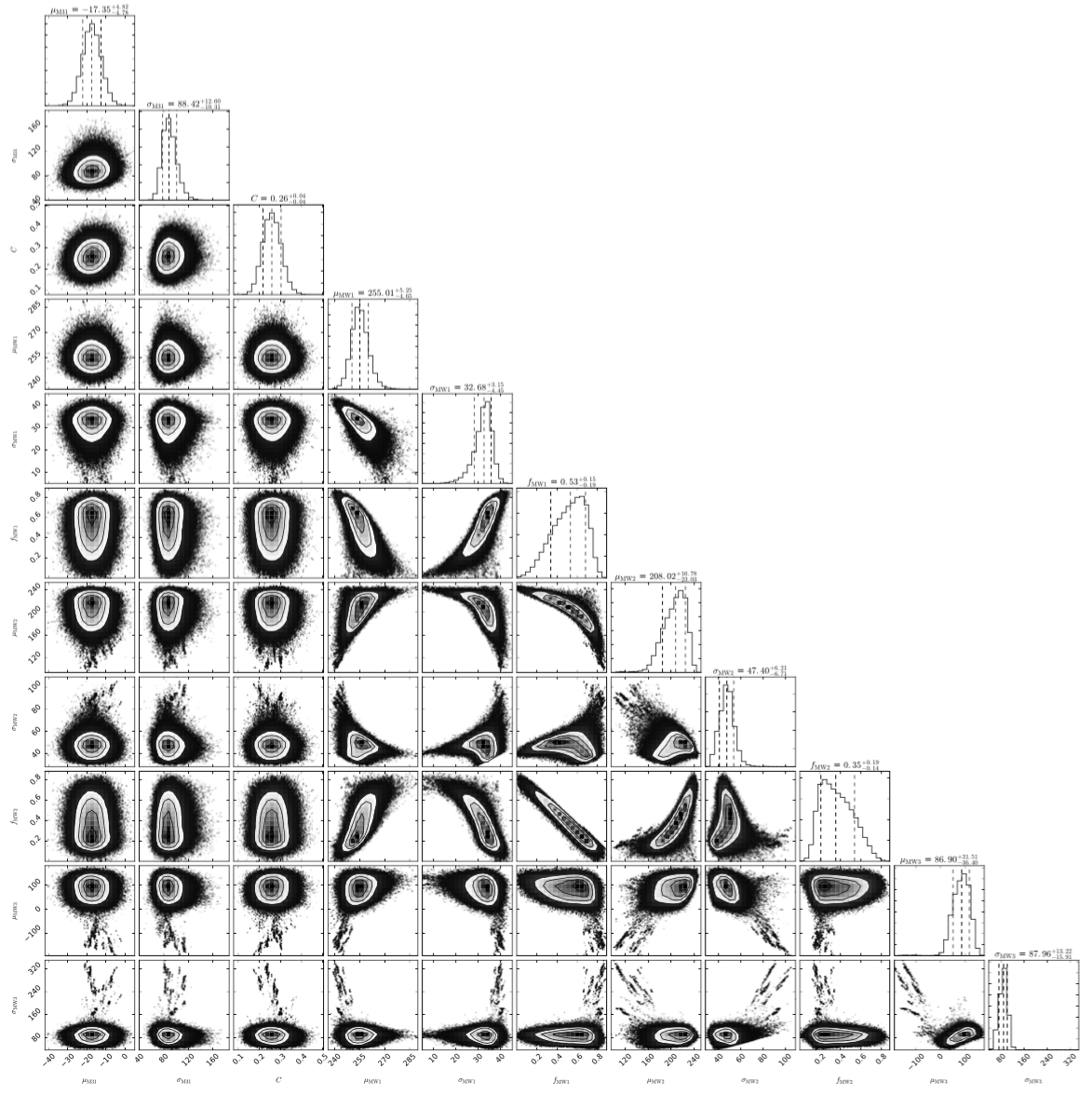}
\caption{Same as Figure~\ref{fig:primary_params_corner_start} for the sixth radial bin ($90\le$\,\rproj\,$<130$~kpc).}
\end{figure}

\begin{figure}
\plotone{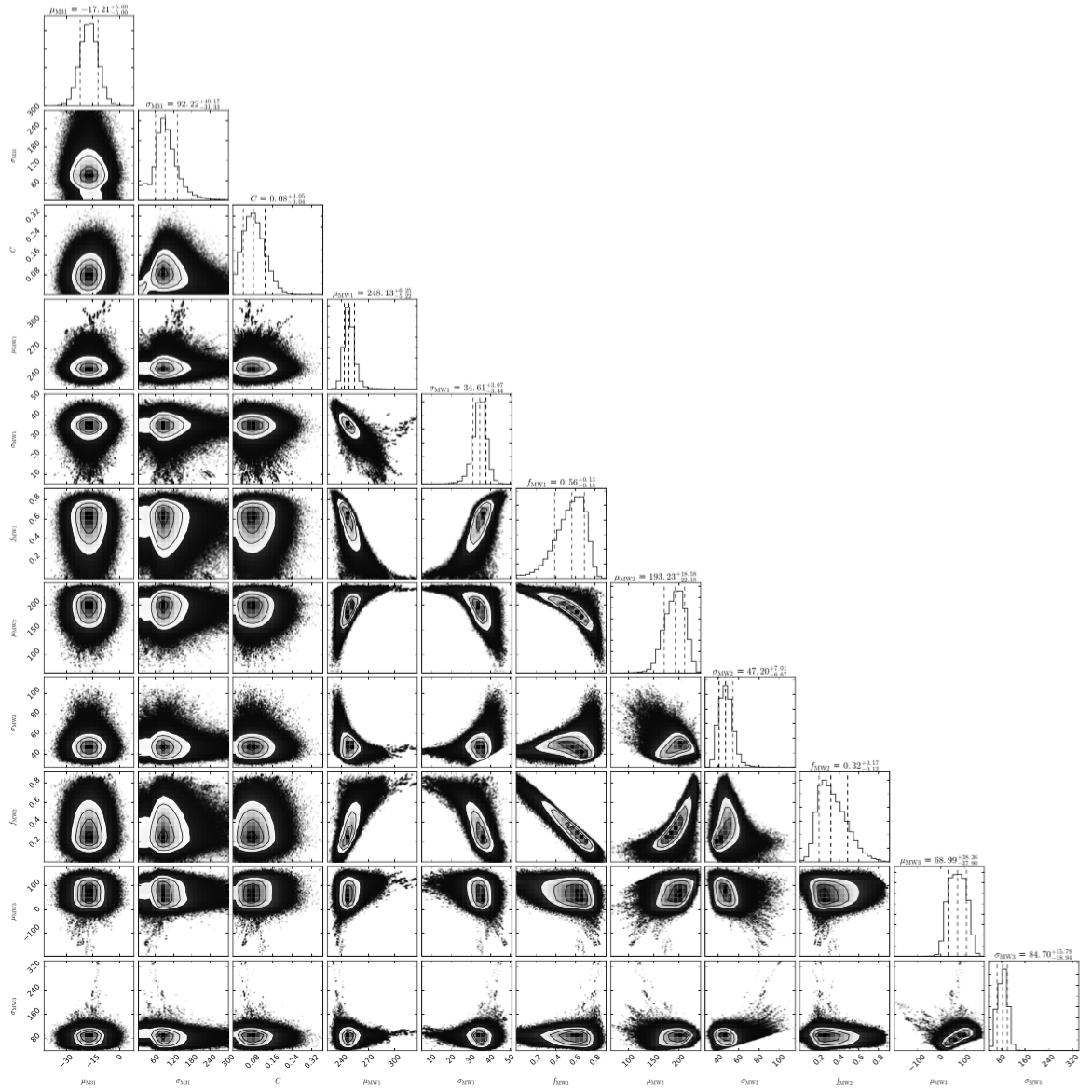}
\caption{Same as Figure~\ref{fig:primary_params_corner_start} for the seventh radial bin ($130\le$\,\rproj\,$<200$~kpc).}
\label{fig:primary_params_corner_end}
\end{figure}

\section{Posterior Probability Distributions of the Field-Dependent Parameters}\label{app:fdep_params}

\begin{figure}
\includegraphics[width=3.5in]{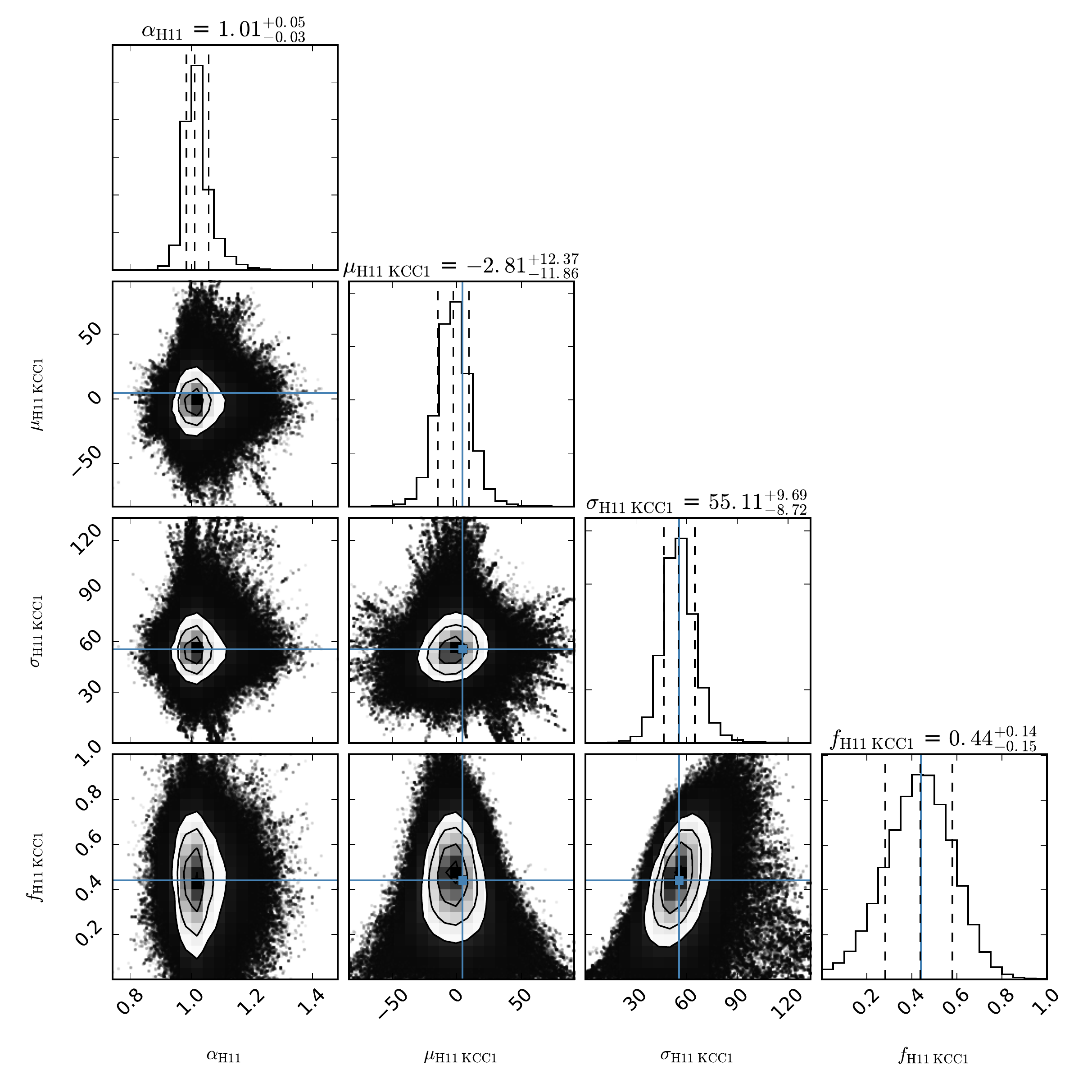}
\includegraphics[width=3.5in]{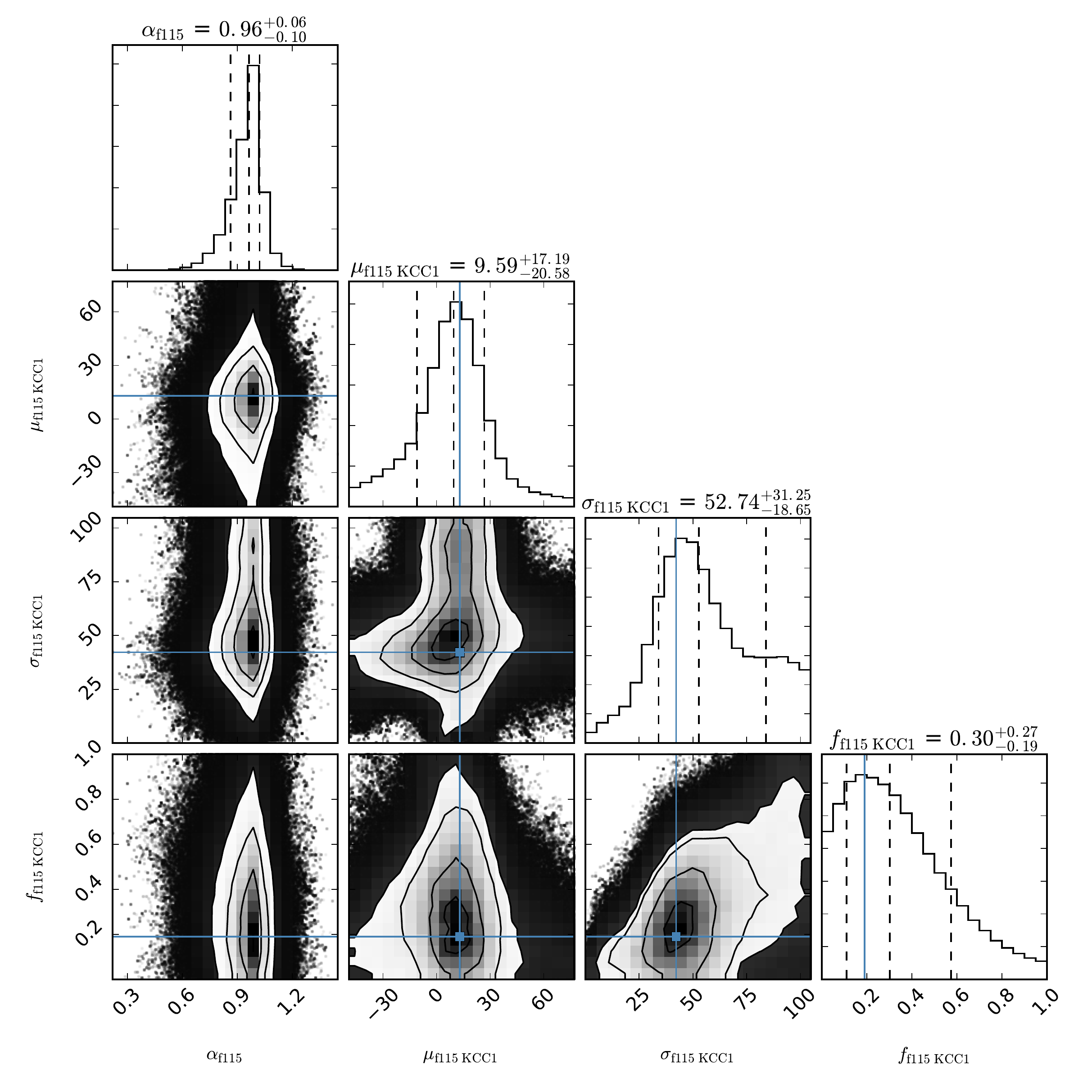}
\includegraphics[width=3.5in]{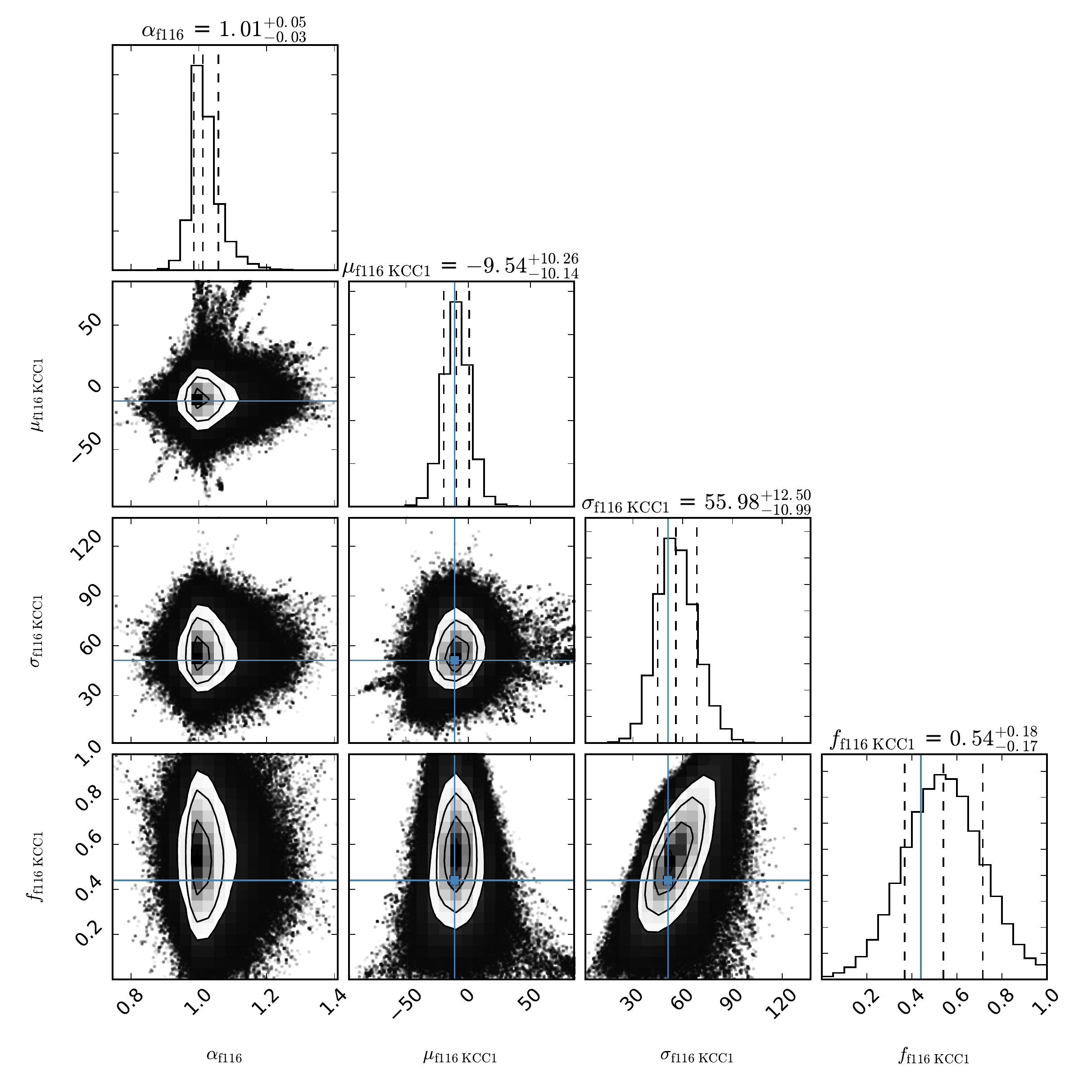}
\caption{Same as Figure~\ref{fig:primary_params_corner_start} for the field-dependent model parameters in the first radial bin ($8.2\le$\,\rproj\,$<14.1$~kpc; Section~A\ref{app:fdep_params}).   The solid blue lines mark the maximum-likelihood values for each component from the literature; the literature measurements form the basis of the prior on the mean velocity and velocity dispersion parameters for each tidal debris feature (Sections~\ref{sec:kccs} and \ref{sec:priors_fdep_params}).  Literature measurements for the mean line-of-sight velocity of each feature were made in the heliocentric reference frame.  They have been transformed to the fit reference frame (Galactocentric with M31's bulk line-of-sight motion removed) using the median RA and Dec of all the stellar sources in the spectroscopic field.
}
\label{fig:subst_corner_start}
\end{figure}

\begin{figure}
\includegraphics[width=3.5in]{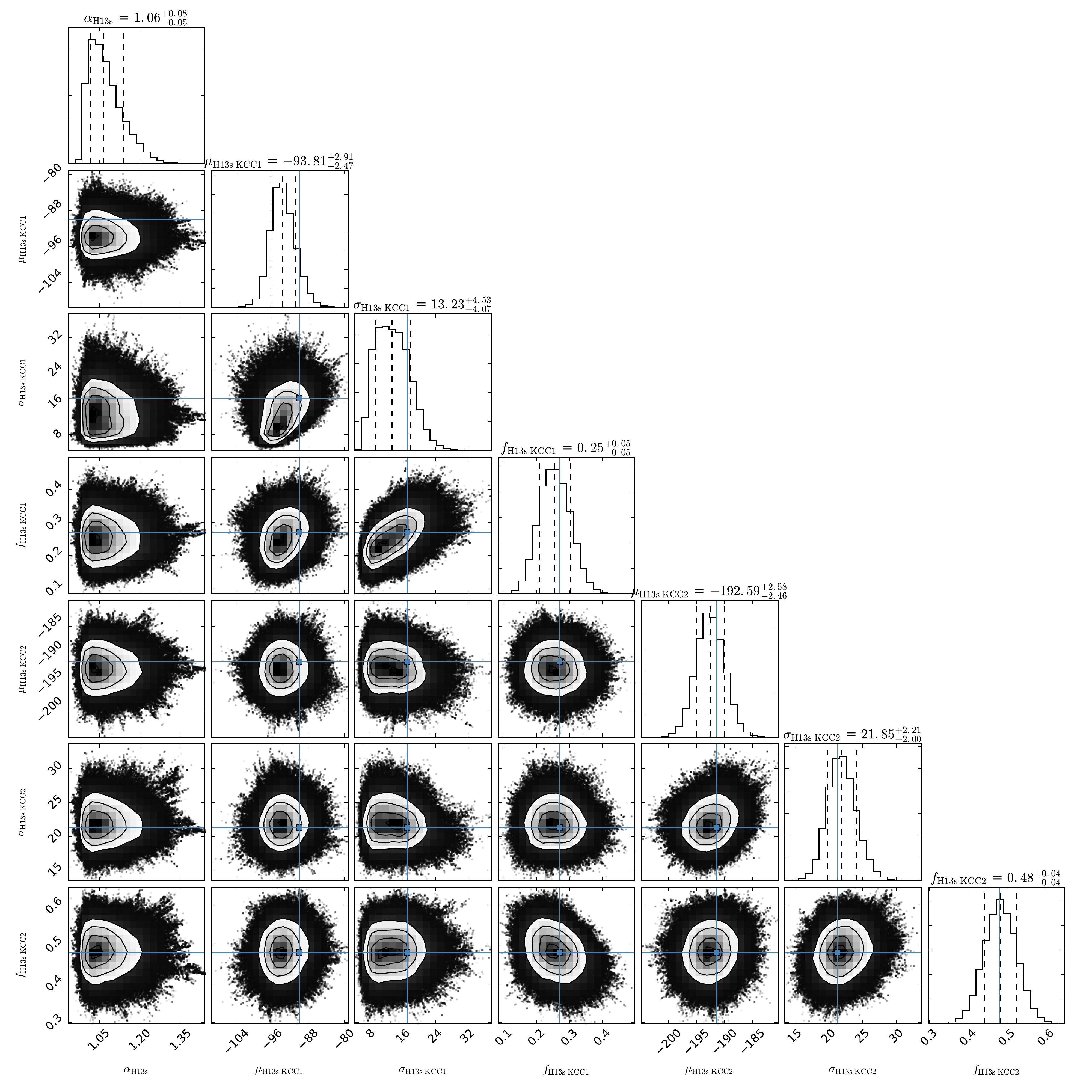}
\includegraphics[width=3.5in]{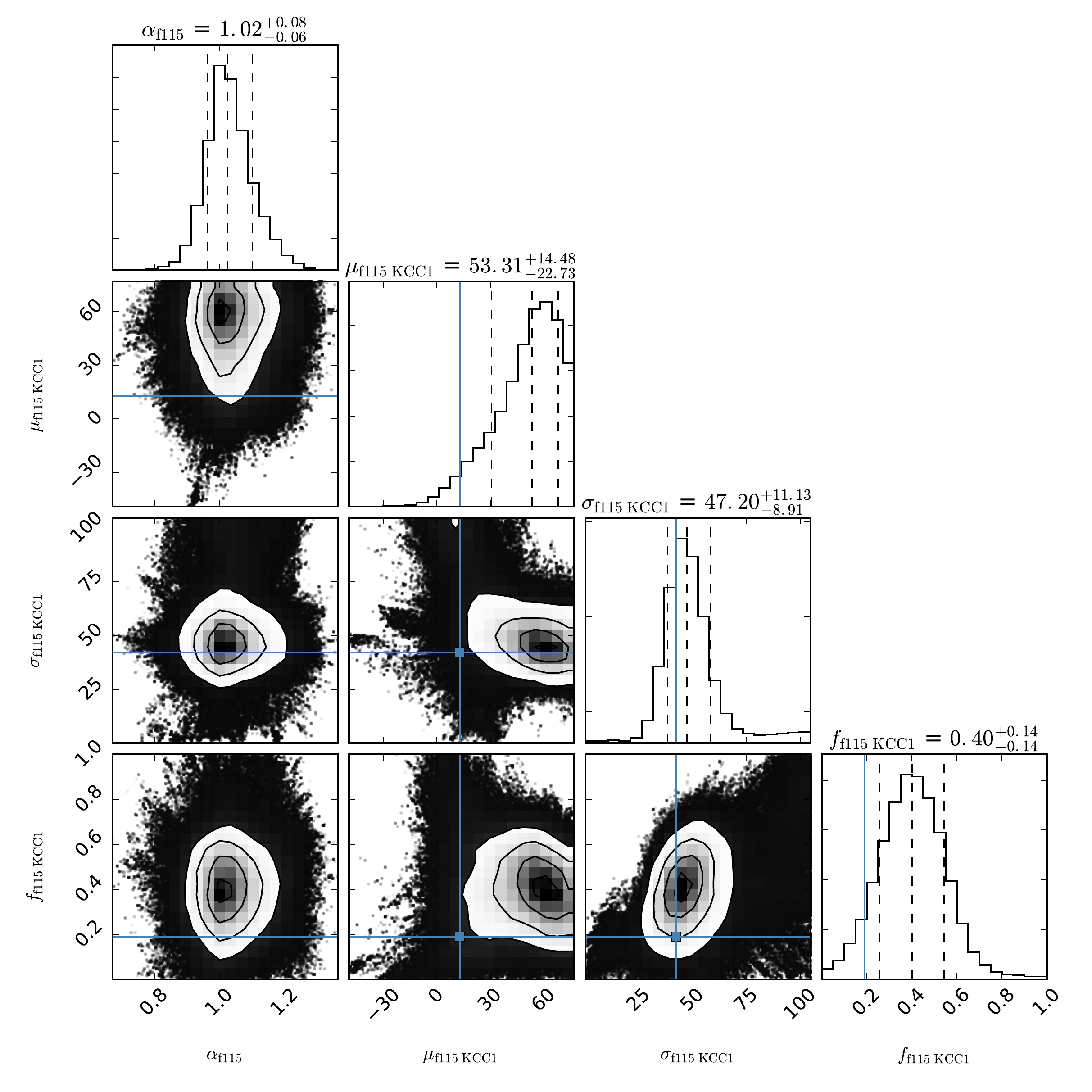}
\includegraphics[width=3.5in]{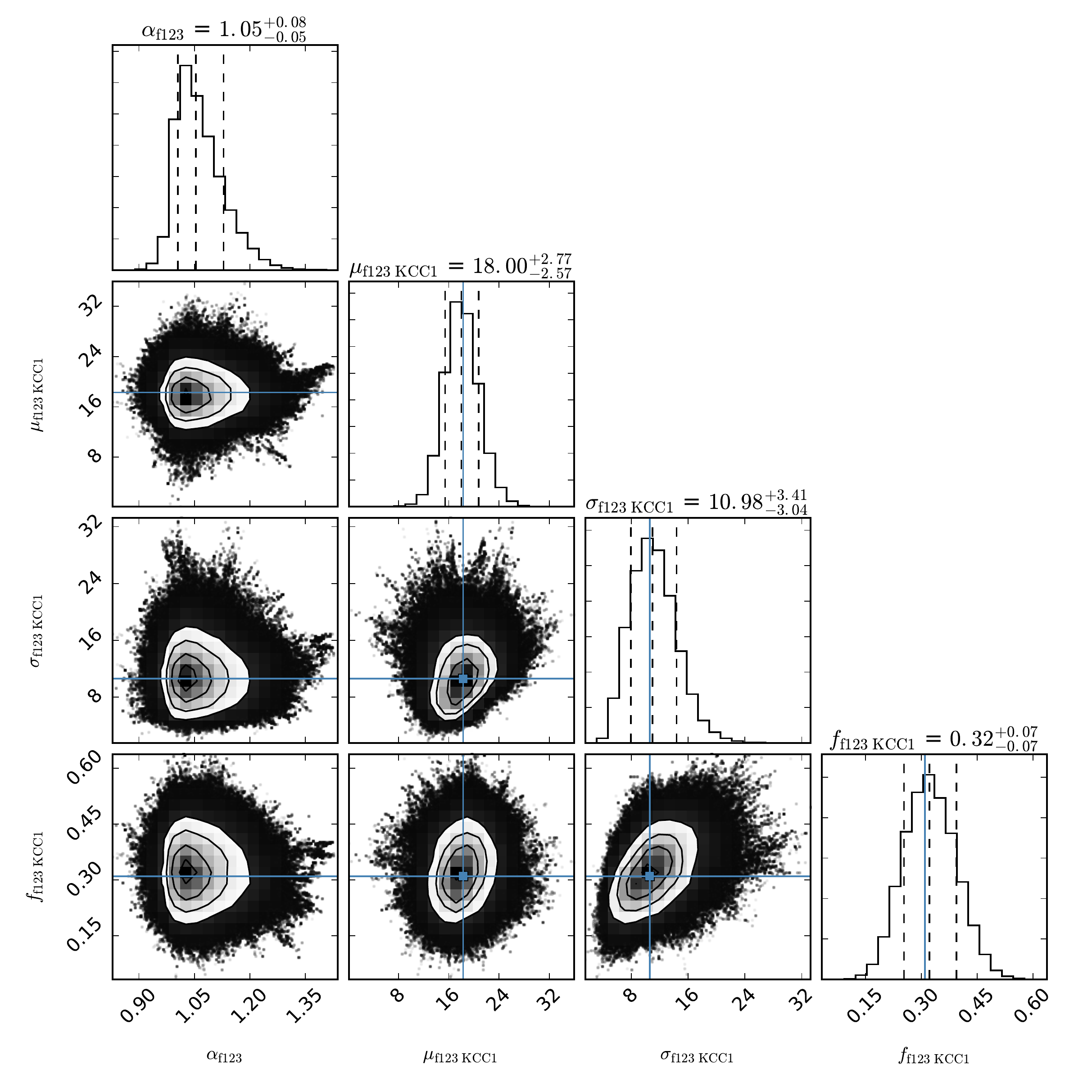}
\includegraphics[width=3.5in]{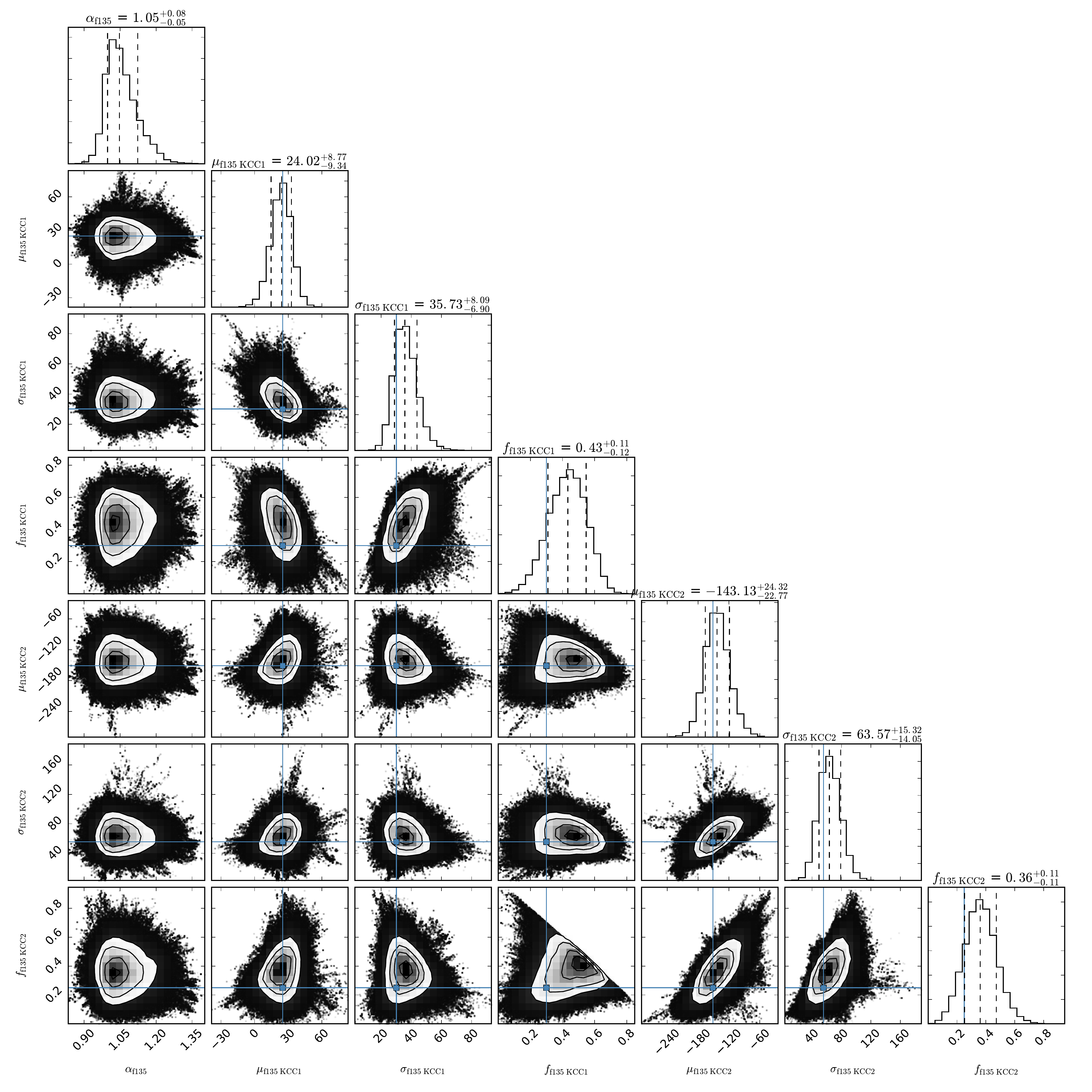}
\caption{Same as Figure~\ref{fig:subst_corner_start} for the field-dependent model parameters in the second radial bin ($14.1\le$\,\rproj\,$<24$~kpc).
}
\label{fig:subst_corner_2a}
\end{figure}

\begin{figure}
\includegraphics[width=3.5in]{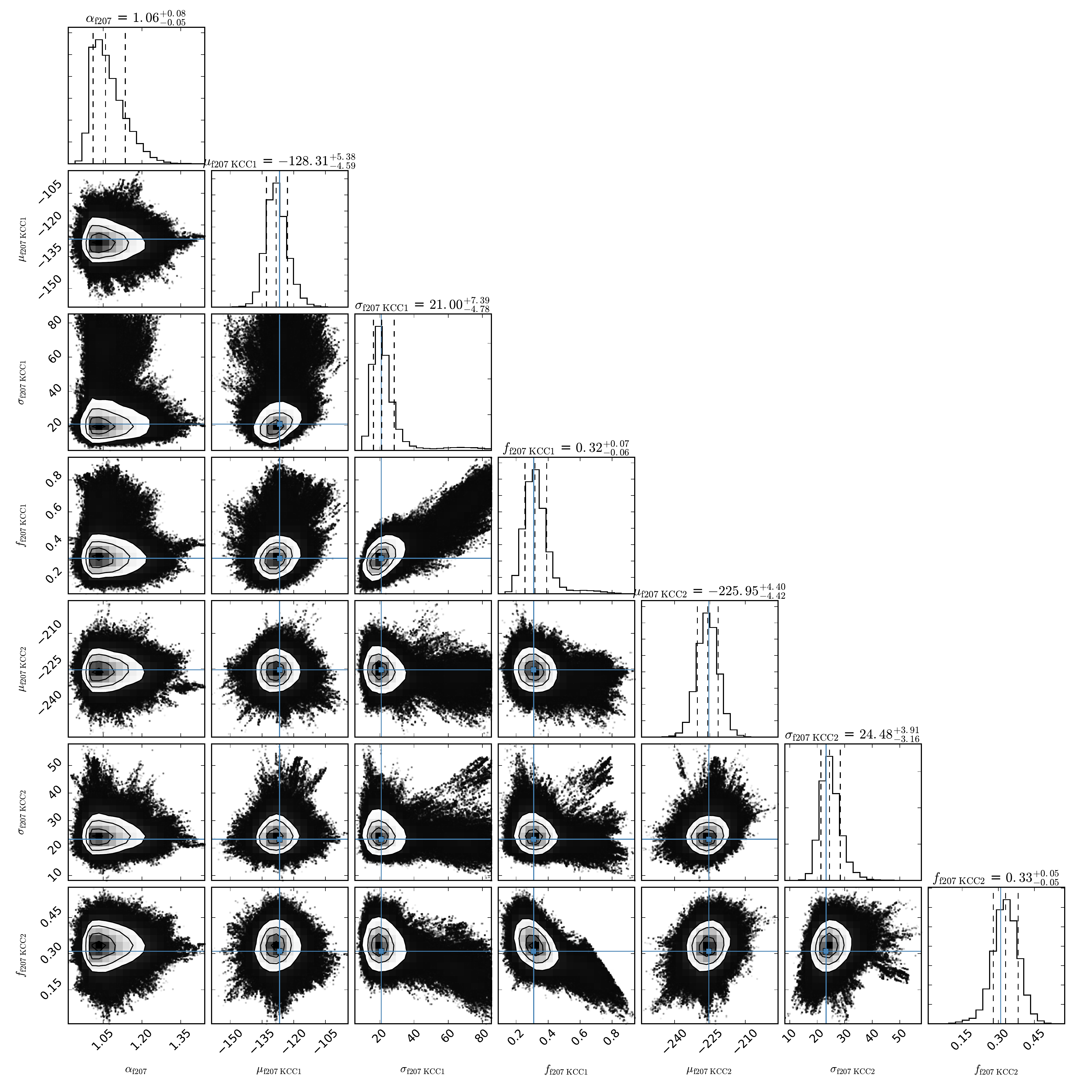}
\caption{Same as Figure~\ref{fig:subst_corner_start} for the field-dependent model parameters in the second radial bin ($14.1\le$\,\rproj\,$<24$~kpc), continued.
}
\label{fig:subst_corner_2b}
\end{figure}

\begin{figure}
\includegraphics[width=3.5in]{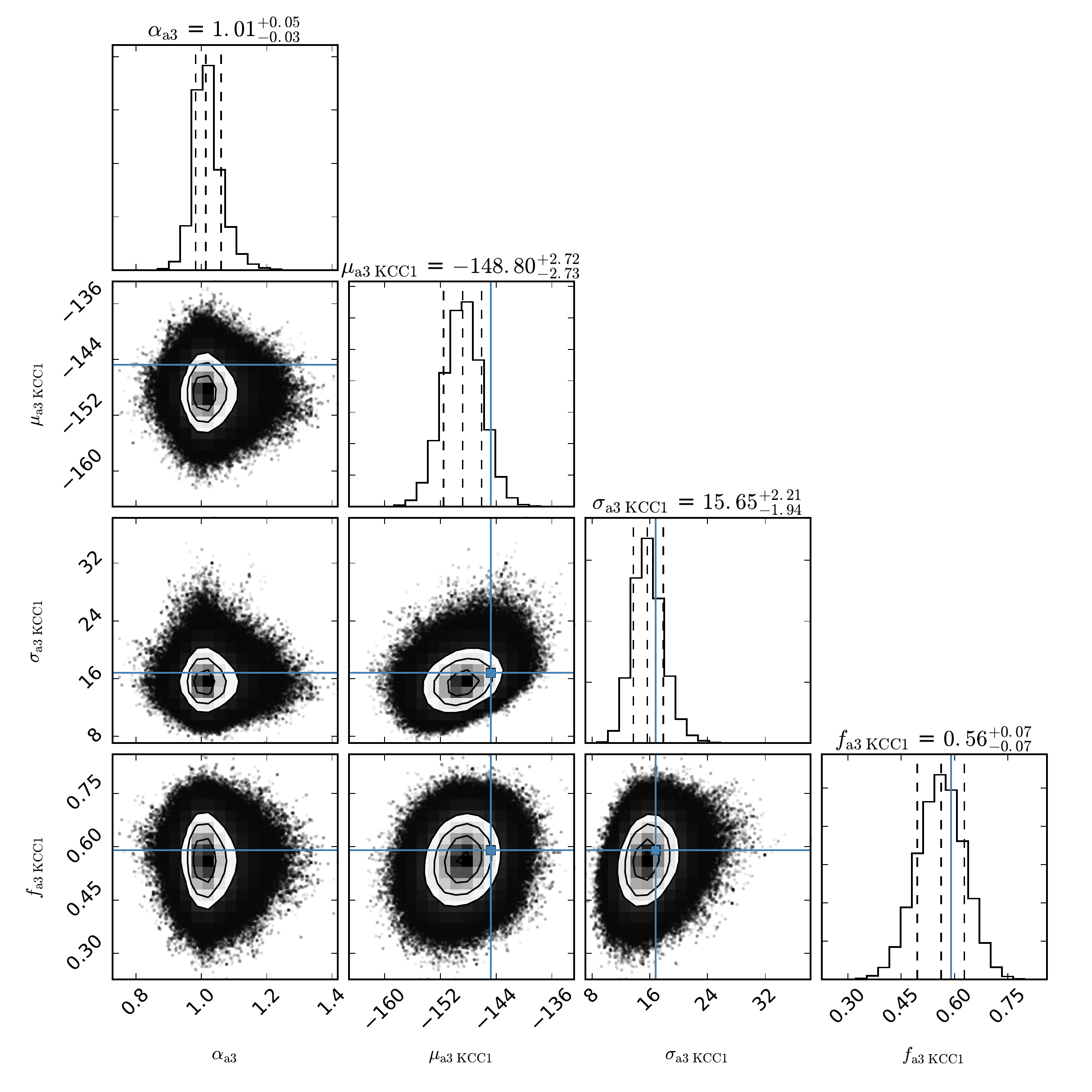}
\includegraphics[width=3.5in]{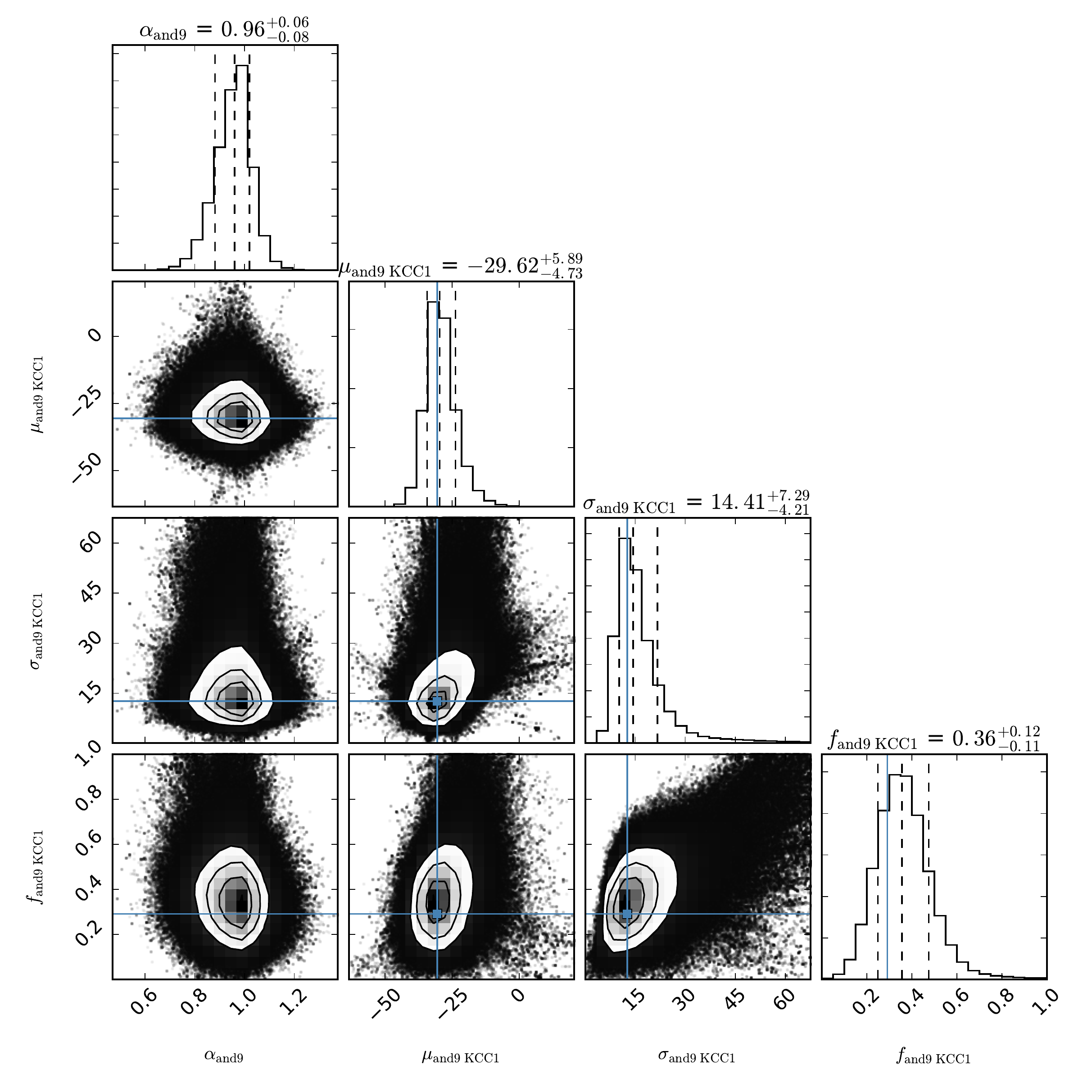}
\caption{Same as Figure~\ref{fig:subst_corner_start} for the field-dependent model parameters in the third radial bin ($24\le$\,\rproj\,$<40$~kpc).
}
\label{fig:subst_corner_3rd}
\end{figure}

\begin{figure}
\includegraphics[width=3.5in]{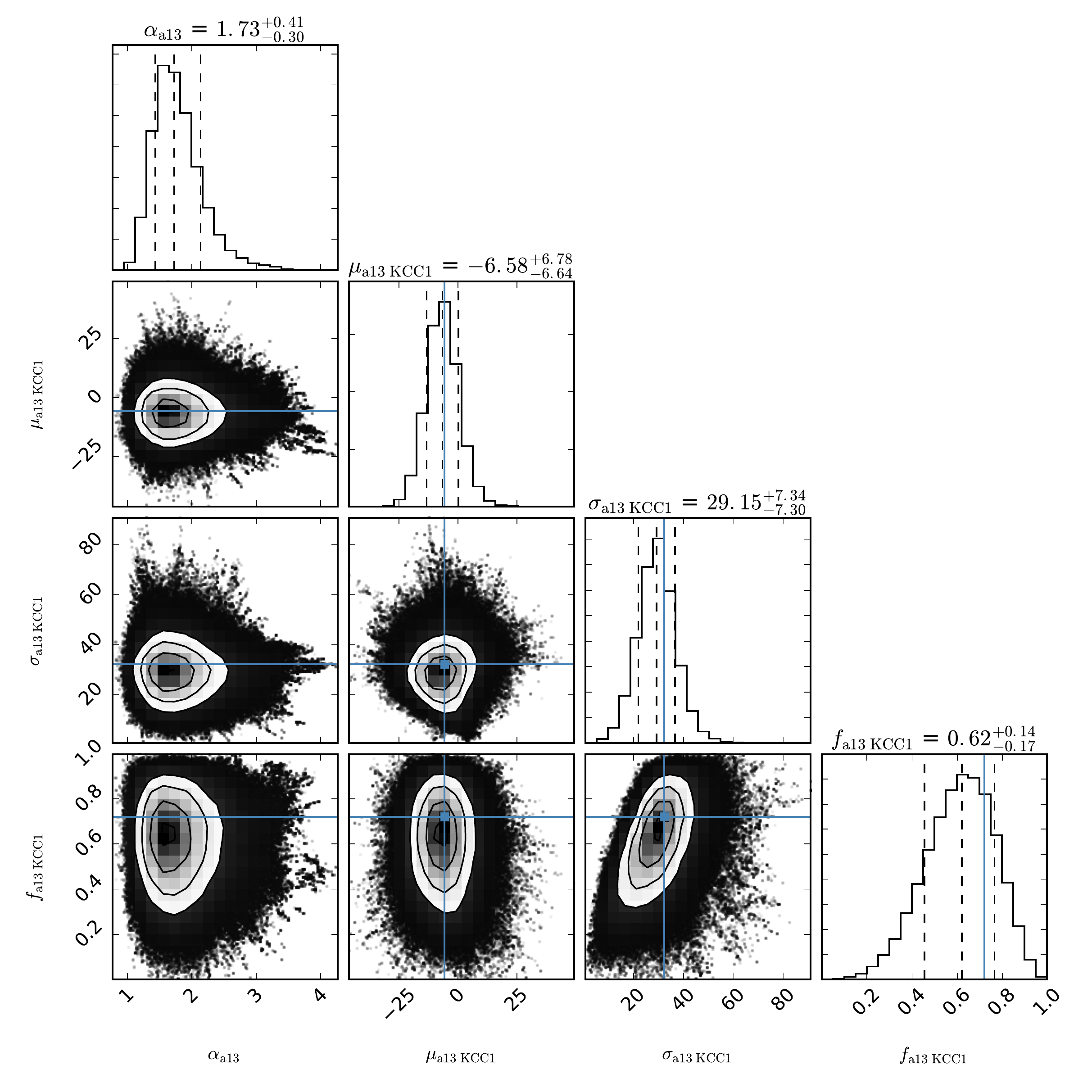}
\includegraphics[width=3.5in]{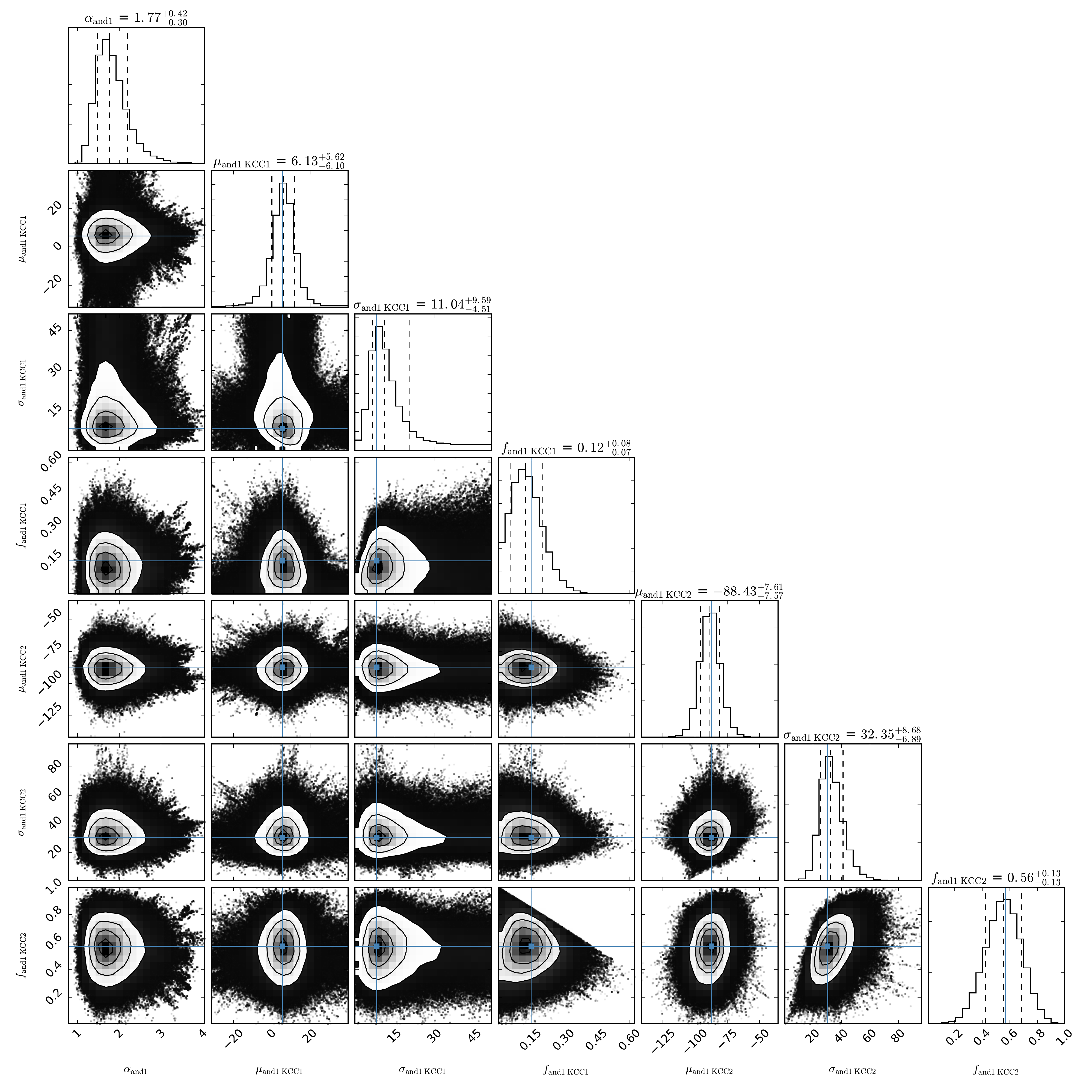}
\includegraphics[width=3.5in]{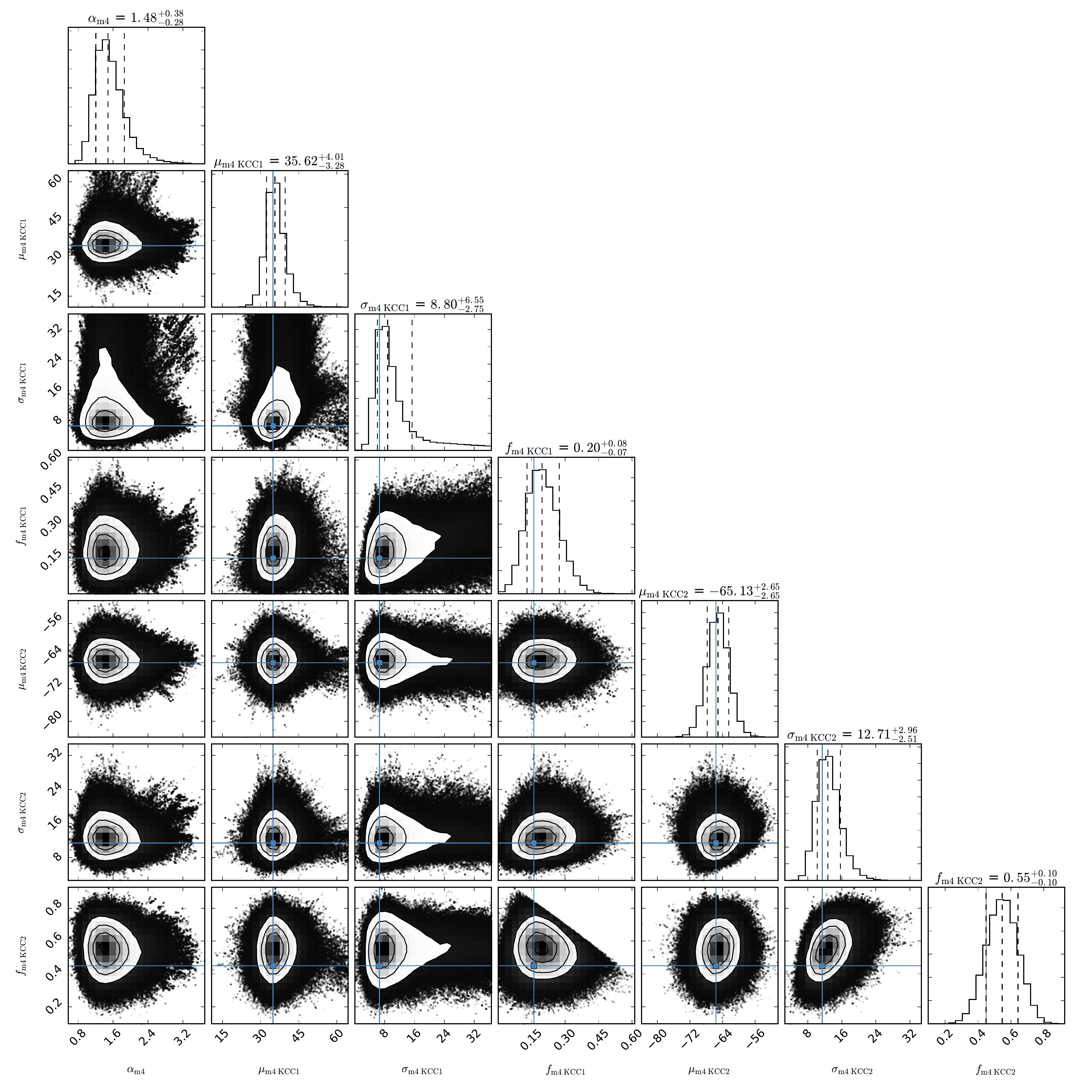}
\caption{Same as Figure~\ref{fig:subst_corner_start} for the field-dependent model parameters in the fourth radial bin ($40\le$\,\rproj\,$<63$~kpc).
}
\label{fig:subst_corner_4th}
\end{figure}

\begin{figure}
\includegraphics[width=3.5in]{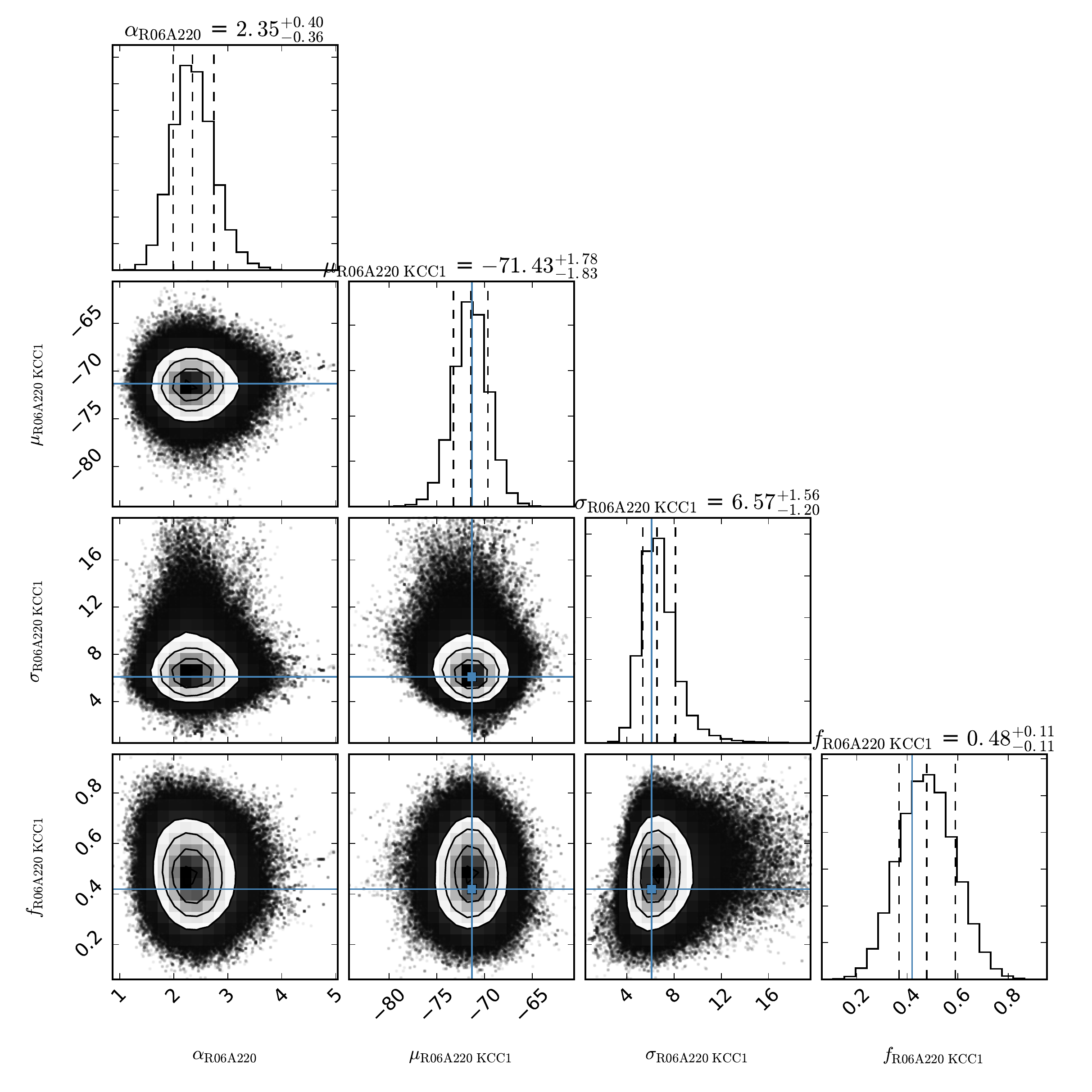}
\caption{Same as Figure~\ref{fig:subst_corner_start} for the field-dependent model parameters in the fifth radial bin ($63\le$\,\rproj\,$<90$~kpc).
}
\label{fig:subst_corner_end}
\end{figure}

Figures \ref{fig:subst_corner_start} through \ref{fig:subst_corner_end} show the marginalized one- and two-dimensional posterior probability distributions for the field-dependent model parameters in each radial bin.  
Each of these features were originally identified and characterized using maximum-likelihood, multi-Gaussian fits to the M31 stars in their respective spectroscopic field (Section~\ref{sec:kccs}).   The results for the mean velocity, velocity dispersion, and fraction of M31 stars in each component are listed in Table~\ref{tab:test_subst_params1}.

Most of the tidal debris features are well constrained when fit within the full model (all M31 plus MW components).   However, a few of the model parameters for individual tidal features show evidence of multi-modality or lower-probability tails in the posterior distributions, hinting that there may be additional kinematical structure in these fields (e.g., fields f207 and m4 in Figures~\ref{fig:subst_corner_2b} and \ref{fig:subst_corner_4th}, respectively).  In a few others, the two-dimensional posterior probability distributions are relatively poorly constrained, indicating that the kinematically-identified feature may not be robust (e.g., the second component in field `f135', Figure~\ref{fig:subst_corner_2a}).

The spectroscopic field `f115' is the only field containing a tidal debris feature that was split between radial bins, with the majority of f115 stars falling in the second radial bin rather than the first.  The original, maximum-likelihood Gaussian fits to the velocity distribution of M31 stars in this field did not require the 
presence of an additional component to the kinematically hot M31 halo component \citep{gilbert2007}.  However,
it is in the middle of the region that \citet{gilbert2007} found to be contaminated by a large shell feature (the Southeast Shelf), which can be connected to M31's most significant tidal debris feature, the Giant Southern Stream \citep{gilbert2007,fardal2007,fardal2012}.  Therefore, we include a single tidal debris feature in the likelihood function for this field, with the prior on the tidal debris feature model parameters based on the multi-Gaussian fit to all of the Southeast Shelf spectroscopic fields \citep{gilbert2007}.    

\begin{deluxetable}{lcccc}
\tablecolumns{5}
\tablewidth{0pc}
\tablecaption{Field-Dependent Model Parameter Results.}
\tablehead{\multicolumn{1}{c}{Component} & \multicolumn{4}{c}{Parameter\tablenotemark{a}} \\
\cline{2-5} 
 & \multicolumn{2}{c}{Mean Velocity} & \multicolumn{1}{c}{Velocity Dispersion} & \multicolumn{1}{c}{Fraction} \\
 & \multicolumn{2}{c}{(\kms)} & \multicolumn{1}{c}{(\kms)} & \\
 & \multicolumn{1}{c}{$\mu_{\rm pec}$\tablenotemark{b}} & \multicolumn{1}{c}{$\mu_{\rm hel}$\tablenotemark{c}}  & \multicolumn{1}{c}{$\sigma$}  & \multicolumn{1}{c}{$f$\tablenotemark{d}}
}
\startdata
\sidehead{{\it 8.2 to 14.1 kpc}}
\hline
f115 KCC 1 &  $   10^{+  17}_{-  21}$ 
 & $ -289$ 
 & $ 53^{+ 31}_{- 19}$ 
 & $0.30^{+0.27}_{-0.19}$ \Tstrut \\ 
\hline 
f116 KCC 1 & $  -10^{+  10}_{-  10}$ 
 & $ -308 $ 
 & $ 56^{+ 13}_{- 11}$ 
 & $0.54^{+0.18}_{-0.17}$ \Tstrut \\ 
\hline 
H11 KCC 1 & $  -3^{+  12}_{-  12}$ 
 & $ -302$ 
 & $ 55^{+  10}_{-  9}$ 
 & $0.44^{+0.14}_{-0.16}$ \Tstrut \\ 
\hline 

\hline
\sidehead{{\it 14.1 to 24 kpc}}
\hline
f207 KCC 1 & $-128.3^{+   5.4}_{-   4.6}$ 
 & $ -427.2$
 & $ 21.0^{+  7.4}_{-  4.8}$ 
 & $0.318^{+0.074}_{-0.062}$ \Tstrut \\ 
f207 KCC 2 & $-226.0^{+   4.4}_{-   4.4}$ 
 & $ -524.9$
 & $ 24.5^{+  3.9}_{-  3.2}$ 
 & $0.331^{+0.052}_{-0.051}$ \Tstrut \\ 
\hline 
f123 KCC 1 & $  18.0^{+   2.8}_{-   2.6}$ 
 & $ -279.7$
 & $ 11.0^{+  3.4}_{-  3.0}$ 
 & $0.322^{+0.073}_{-0.069}$ \Tstrut \\ 
\hline 
H13s KCC 1 & $ -93.8^{+   2.9}_{-   2.5}$ 
 & $ -392.1$
 & $ 13.2^{+  4.5}_{-  4.1}$ 
 & $0.254^{+0.050}_{-0.046}$  \Tstrut \\
H13s KCC 2 & $-192.6^{+   2.6}_{-   2.5}$ 
 & $ -490.9$
 & $ 21.9^{+  2.2}_{-  2.0}$ 
 & $0.483^{+0.042}_{-0.042}$ \Tstrut \\ 
\hline 
f115 KCC 1 & $  53^{+  14}_{-  23}$ 
 & $ -245$
 & $ 47.2^{+ 11.1}_{-  8.9}$ 
 & $0.40^{+0.14}_{-0.14}$ \Tstrut \\ 
\hline 
f135 KCC 1 & $  24.0^{+   8.8}_{-   9.3}$ 
 & $ -274.0$
 & $ 35.7^{+  8.1}_{-  6.9}$ 
 & $0.43^{+0.11}_{-0.12}$ \Tstrut \\ 
f135 KCC 2 & $-143^{+  24}_{-  23}$ 
 & $ -441$
 & $ 64^{+ 15}_{- 14}$ 
 & $0.36^{+0.11}_{-0.11}$ \Tstrut \\ 
\hline 

\hline
\sidehead{{\it 24 to 40 kpc}}
\hline
a3 KCC 1 & $-148.8^{+   2.7}_{-   2.7}$ 
 & $ -444.6$
 & $ 15.7^{+  2.2}_{-  1.9}$ 
 & $0.562^{+0.065}_{-0.067}$ \Tstrut \\ 
\hline 
and9 KCC 1 & $ -29.6^{+   5.9}_{-   4.7}$ 
 & $ -328.8$
 & $ 14.4^{+  7.3}_{-  4.2}$ 
 & $0.36^{+0.12}_{-0.11}$ \Tstrut \\ 
\hline 

\hline
\sidehead{{\it 40 to 63 kpc}}
\hline
and1 KCC 1 & $   6.1^{+   5.6}_{-   6.1}$ 
 & $ -288.7$
 & $ 11.0^{+  9.6}_{-  4.5}$ 
 & $0.125^{+0.079}_{-0.067}$ \Tstrut \\ 
and1 KCC 2 & $ -88.4^{+   7.6}_{-   7.6}$ 
 & $ -383.3$
 & $ 32.4^{+  8.7}_{-  6.9}$ 
 & $0.56^{+0.13}_{-0.13}$ \Tstrut \\ 
\hline 
a13 KCC 1 & $  -6.6^{+   6.8}_{-   6.6}$ 
 & $ -301.6$
 & $ 29.2^{+  7.3}_{-  7.3}$ 
 & $0.62^{+0.15}_{-0.17}$ \Tstrut \\ 
\hline 
m4 KCC 1 & $  35.6^{+   4.0}_{-   3.3}$ 
 & $ -254.1$
 & $  8.8^{+  6.6}_{-  2.8}$ 
 & $0.196^{+0.078}_{-0.067}$ \Tstrut \\ 
m4 KCC 2 & $ -65.1^{+   2.7}_{-   2.7}$ 
 & $ -354.8$
 & $ 12.7^{+  3.0}_{-  2.5}$ 
 & $0.547^{+0.096}_{-0.098}$ \Tstrut \\ 
\hline 

\hline
\sidehead{{\it 63 to 90 kpc}}
\hline
R06A220 KCC 1 & $ -71.4^{+   1.8}_{-   1.8}$ 
 & $ -373.6$ 
 & $  6.6^{+  1.6}_{-  1.2}$ 
 & $0.48^{+0.11}_{-0.11}$ \Tstrut \\ 
\hline 

\enddata
\tablenotetext{a}{Results are the 50th percentile of the one-dimensional posterior probability distribution, marginalized over all other model parameters. The quoted errors are the 16th and 84th percentiles of the posterior probability distribution.}
\tablenotetext{b}{Mean line-of-sight peculiar velocity of the feature (Section~\ref{sec:mcmc}). This is the mean velocity of the feature in the frame of reference used for the MCMC analysis: the observed heliocentric line of sight velocity of each star has been transformed to the Galactocentric frame, and the bulk motion of M31 along the line of sight has been removed.}
\tablenotetext{c}{Mean line-of-sight heliocentric velocity of the feature, calculated using the median RA and Dec of all targets in the spectroscopic field to determine the velocity transformation, which includes the solar motion as well as the line of sight components of both M31's tangential and radial motion.  Heliocentric velocities are provided to enable comparisons to previous estimates of the mean velocities of these tidal debris features.}
\tablenotetext{d}{Fraction of the M31 population in a field in the given component.}
\label{tab:test_subst_params1}
\end{deluxetable}

\bibliography{m31}

\end{document}